\documentclass[10pt,journal]{IEEEtran}
\usepackage{amsmath}
\usepackage{amssymb}
\usepackage{algorithm}
\usepackage{algorithmicx}
\usepackage{algpseudocode}
\usepackage{graphicx}
\usepackage{subfigure}
\usepackage{epstopdf}
\usepackage{cite}
\usepackage{textcomp}
\usepackage{mathrsfs}
\usepackage{color}
\usepackage{booktabs}
\usepackage{cases}
\usepackage{setspace}
\usepackage{bm}
\usepackage{cuted}
\usepackage{booktabs}
\usepackage{stfloats}
\usepackage{makecell}
\usepackage{mathtools}
\usepackage{multirow}
\usepackage{float}
\usepackage{longtable}
\usepackage{supertabular}
\usepackage{array}
\usepackage{bbding}
\begin{document}
	
	\title{\Huge  Breaking the Interference and Fading Gridlock in Backscatter Communications: State-of-the-Art, Design Challenges, and Future Directions} 	
	
	\author{ 
		Bowen~Gu,  Dong~Li, \emph{Senior Member, IEEE}, Haiyang~Ding, \emph{Member, IEEE}, \\Gongpu~Wang,  \emph{Member, IEEE}, and Chintha Tellambura, \emph{Fellow, IEEE}
		\IEEEcompsocitemizethanks{
			\IEEEcompsocthanksitem Bowen Gu and Dong Li are with the School of Computer Science and Engineering, Macau University of Science and Technology, Avenida Wai Long, Taipa, Macau 999078, China (e-mails: gubwww@163.com, dli@must.edu.mo).
			\IEEEcompsocthanksitem Haiyang Ding is with the College of Information and Communication, National University of Defense Technology, Wuhan 430019, China (e-mail: dinghy2003@hotmail.com).
			\IEEEcompsocthanksitem Gongpu Wang is with School of Computer and Information Technology, Beijing
			Jiaotong University, Beijing 100044, China (e-mail: gpwang@bjtu.edu.cn).
	     	\IEEEcompsocthanksitem 	Chintha Tellambura is with the Department of Electrical and Computer
	     	Engineering, University of Alberta, Edmonton, AB T6G 2R3, Canada (e-mail:
	     	ct4@ualberta.ca).
	}
}

	\maketitle
	\thispagestyle{empty}
	\pagestyle{empty}

	\begin{abstract}
		\color{black}
As the Internet of Things (IoT) advances by leaps and bounds, a multitude of devices are becoming interconnected, marking the onset of an era where all things are connected. While this growth opens up opportunities for novel products and applications, it also leads to increased energy demand and battery reliance for IoT devices, creating a significant bottleneck that hinders sustainable progress. At this juncture, backscatter communication (BackCom), as a low-power and passive communication method, emerges as one of the promising solutions to this energy impasse by reducing the manufacturing costs and energy consumption of IoT devices. However, BackCom systems face challenges such as complex interference environments,  including direct link interference (DLI) and mutual interference (MI) between tags, which can severely disrupt the efficiency of BackCom networks. Moreover, double-path fading is another major issue that leads to the degraded system performance. To fully unleash the potential of BackComs, the purpose of this paper is to furnish a comprehensive review of existing solutions with a focus on combatting these specific interference challenges and overcoming dual-path fading, offering an insightful analysis and comparison of various strategies for effectively mitigating these issues. Specifically, we begin by introducing the preliminaries for the BackCom, including its history, operating mechanisms, main architectures, etc,  providing a foundational understanding of the field. Then, we delve into fundamental issues related to BackCom systems, such as solutions for the DLI, the MI, and the double-path fading. This paper thoroughly provides state-of-the-art advances for each case, particularly highlighting how the latest innovations in theoretical approaches and system design can strategically address these challenges. Finally, we explore emerging trends and challenges in BackComs by forecasting potential technological advancements and providing insights and guidelines for navigating the intricate landscape of future communication needs in a rapidly evolving IoT ecosystem.
	\end{abstract}
	\begin{IEEEkeywords}	
	IoT, backscatter communication, performance enhancement,  direct-link interference, mutual interference, double-path fading.
	\end{IEEEkeywords}
	
	\maketitle

\vspace{10mm}

\begin{table} [h]			
	\renewcommand{\arraystretch}{1.1}
	\centering
	\small
	\begin{tabular}{c|l}
		\hline
		Notation & Meaning \\
		\hline
		
		AI& Artificial intelligence \\
								
		ADC & Analog-to-digital converter  \\
		
		AmBC & Ambient BackCom\\
		
		AmIoT & Ambient IoT \\
		
		BackCom & Backscatter communication \\
		
		BER& Backscattered energy recycling \\
		
		BiBC &Bistatic BackCom\\
		
		BLE &Bluetooth low energy\\			
		
		BPSK&binary phase-shift keying\\
		
		CP& Cyclic prefix \\
		
		DC& Direct current\\
		
		DCSK & Differential Chaos shift keying \\
		
		DL & Direct link\\
		
		DLI & Direct-link interference\\
		
		DRFS& Dedicated RF source \\
		
		DSSS &  Direct sequence spread spectrum \\
		
		EH & Energy harvesting\\
		
		FDMA &Frequency division multiple access\\
		
		FS&	 Frequency shift\\
		
		FM& Frequency modulation\\		
		
		IoT&Internet of Things\\
		
		I/Q &In-phase or quadrature\\
		
		LTE& Long-term evolution\\
		
		MI& Mutual interference 	\\
		
		MIMO& Multiple input multiple output   \\
		
		MoBC & Monostatic BackCom\\
		
		
		MS& Modulation silencing\\
		
		MSEH& Multi-source energy harvesting \\
		
		MSMA& Multiple subcarrier multiple access\\
		
		NOMA& Non-orthogonal multiple access \\
		
		OFDM & Orthogonal frequency division multiplexing \\
		
		OFDMA & Orthogonal frequency division multiple access \\
		
		OSIC&Opportunistic SIC\\
		
		OTM& Opportunistic transceiver mechanism\\
		
		PC&Polarization conversion \\
		
		PSK& Phase-shift keying \\
		
		QAM& Quadrature amplitude modulation \\
		
		RF& Radio frequency  \\		
		
		RFID & RF identification\\
		
		RIS& Reconfigurable intelligent surface\\
		
		SDMA&  Space division multiple access\\
		
		SI & Self-interference \\
		
		SIC& Successive interference cancellation\\	
		
		SINR &Signal-to-interference-plus-noise ratio\\
		
		
		TDMA& Time division multiple access\\
		
		THSS& Time-hopping spread spectrum \\
		
		TV& Television \\
		
		UAV& Unmanned aerial vehicle\\
		
		WPCN& Wireless-powered communication network \\
		
		WiFi& Wireless Fidelity\\
		
		WPT& Wireless power transfer \\
		\hline
	\end{tabular}
\end{table}

	\section{Introduction}	
	
\begin{figure}[t]
 	\centerline{\includegraphics[width=3.5in]{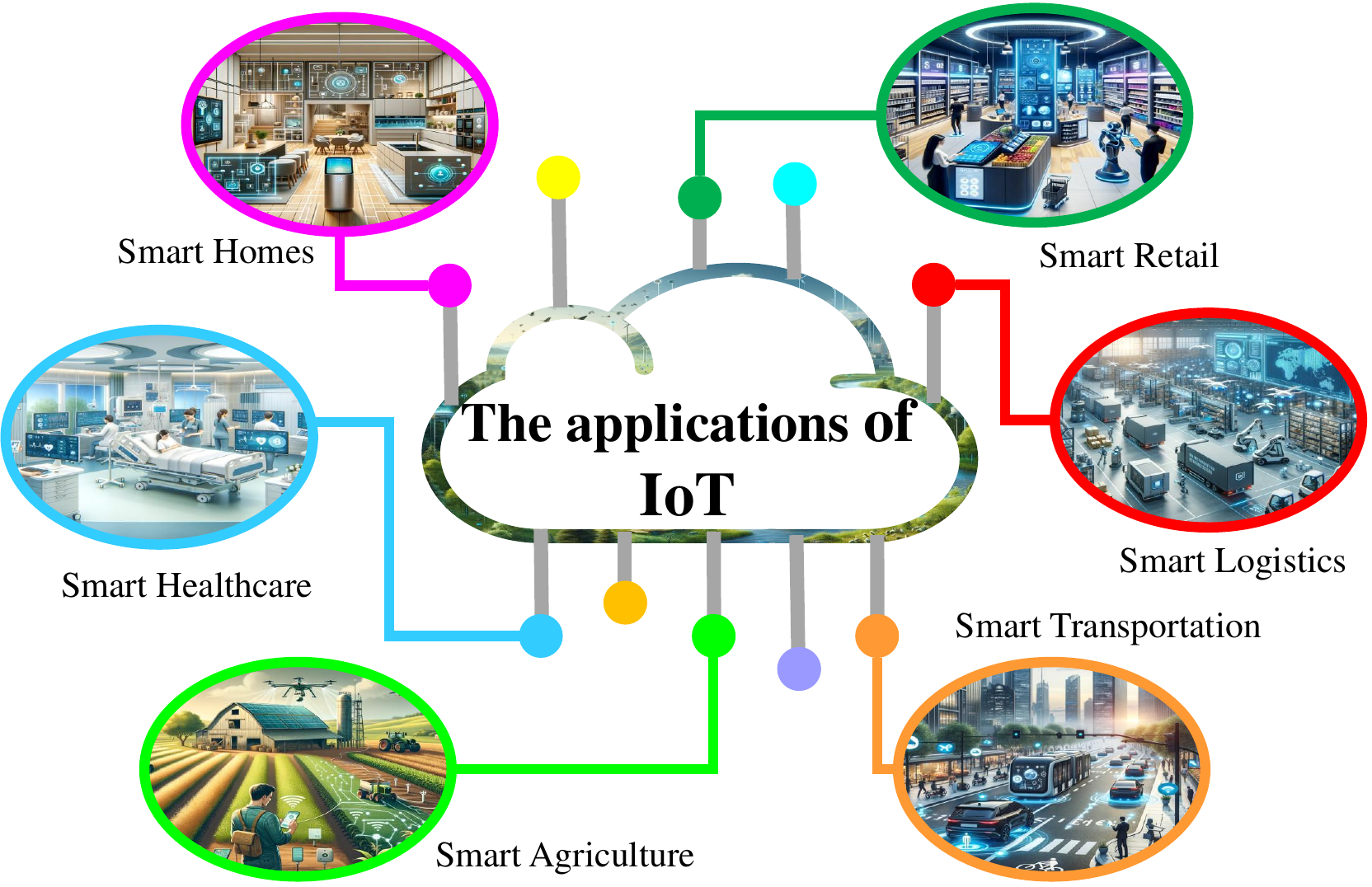}}
 	\caption{The penitential applications of IoT.}
 	\label{fig1}
 \end{figure}  	
	
	{\color{black}Riding on the momentum of wireless communication and electronic manufacture, the Internet of Things (IoT) has made considerable strides in recent years \cite{iot1}, catalyzing a plethora of innovative products and applications tailored to users' diverse needs and preferences, spanning areas such as smart homes, healthcare, agriculture, logistics, transportation, retail, and more, as illustrated in Fig. \ref{fig1}.} In particular, the IoT industry has experienced an exponential increase in connected devices, contributing to an ever-more interconnected world.  {\color{black}Recent market studies \cite{RothmullerBarker} predict that the total number of global active IoT devices will reach 29.7 billion by 2027, a staggering jump from the 16.7 billion connections recorded in 2023, an increase of a whopping 78\%, indicating the deepening integration of intelligent technologies in various aspects of life, reflecting a transformative trend in global connectivity and digital interaction.}

    \subsection{The Energy Bottleneck of IoT}
     
    {\color{black}However, this proliferation is both a blessing and a curse. From the energy consumption perspective, the energy requisite for data transactions and the circuit power demands of an individual IoT apparatus might appear inconsequential. However, when aggregated across the vast landscape of billions of such devices in operation, the accumulated energy requirements become notably tremendous \cite{AnubhaJain}. Transitioning our focus to supply-side challenges,  it is evident that most conventional IoT devices are heavily battery-dependent for their energy needs. Regrettably, the limited capacity of these batteries due to hardware limitations often proves inadequate to support the prolonged operational requirements of these IoT devices. Such limitations necessitate frequent battery charging or replacement, elevating the operating and maintenance costs for IoT devices and exacerbating the environmental issue of electronic waste \cite{LiuLiDuShuHan}.}   Absent effective interventions, these disconcerting trends will likely rise in tandem with the expanding IoT ecosystem. This predicament accentuates the urgent need for innovative, cost-effective, and energy-efficient solutions to reduce the operational and maintenance costs implicated in the expanding IoT ecosystem, ensure long-term sustainability, and mitigate the environmental harm of this technological revolution.
    
    {\color{black}Generally, to overcome the energy and lifecycle constraints of IoT devices, there are two predominant strategies. First, enhancing the energy provision from the supply side to furnish the devices with more abundant energy sources. Second, reducing the inherent energy consumption of the devices themselves to achieve superior energy efficiency.}

	\begin{table*}[ht]
		\centering
		\color{black}
		\renewcommand{\arraystretch}{1.3}
		\small
		\caption{Summary of Energy Harvesting Technologies}
		\label{t1}
		\begin{tabular}{|>{\centering\arraybackslash}m{3cm}|>{\centering\arraybackslash}m{3.5cm}|>{\centering\arraybackslash}m{3.5cm}|>{\centering\arraybackslash}m{4cm}|}
			\hline
			\textbf{Category} & \textbf{Mechanism} & \textbf{Prerequisites} & \textbf{Scenarios}   \\
			\hline
			Solar Energy & Photovoltaic effect & Sufficient sunlight exposure & Bright areas  \\
			\hline
			Thermoelectric Energy & Seebeck effect & Necessary temperature difference & Areas with temperature gradients \\
			\hline
			Kinetic Energy & Vibration harvesters & Continuous motion or vibration & Wind fields, dams, vibration-prone areas  \\
			\hline
			RF Energy & RF conversion & Sufficient RF signal strength & Wi-Fi, cellular, broadcasting zones \\
			\hline
		\end{tabular}
	\end{table*}

	\subsection{Expanding Revenue Sources---Energy Harvesting}

   {\color{black}With the IoT landscape rapidly evolving, it is increasingly evident that traditional power methods like battery replacement or wired charging are unsustainable. This conventional approach places an escalating strain on energy resources and proves impractical for deploying billions of IoT devices envisioned in various industries. Therefore, the exploration of alternative or supplementary energy sources becomes crucial. At this juncture, the energy harvesting (EH) technology has proven its worth. Since its inception, it has garnered widespread attention from researchers and is recognized as a crucial solution for addressing energy supply challenges in IoT applications \cite{LiuLIDaiLIZhang,Pecunia}.

   To be specific, IoT devices possess the capability to harvest and store energy from their surrounding environment through the EH technology, offering a continuous and reliable power source, subsequently extending their operational lifespan, thereby kicking off the explorations and applications for ambient IoT (AmIoT)\footnote{\color{black} A pivotal proposal for AmIoT is the 3GPP TSG RAN\#97e RP-222685 \cite{Amiot1}, which highlights a proposed study titled ``Study on Ambient IoT.'' This initiative marks a crucial step towards addressing the burgeoning need for IoT devices characterized by the low power consumption and the reduced complexity. It underscores the challenges of powering a vast array of IoT devices, particularly those that are battery-dependent and require manual recharging or replacement, posing sustainability and practicality concerns. The standardization process, as proposed by Huawei and HiSilicon, delves into exploring innovative solutions such as energy harvesting from ambient sources, i.e., light, motion, heat, and radio waves, to drive the IoT devices.}\cite{Amiot2, Amiot3, Amiot4}. To date, EH technology taps into a diverse range of potential energy sources, from solar, thermal, kinetic, and radio frequency (RF), as illustrated in Fig. \ref{fig2}.
  
   	  \begin{figure*}[t]
	\centerline{\includegraphics[width=3.6in]{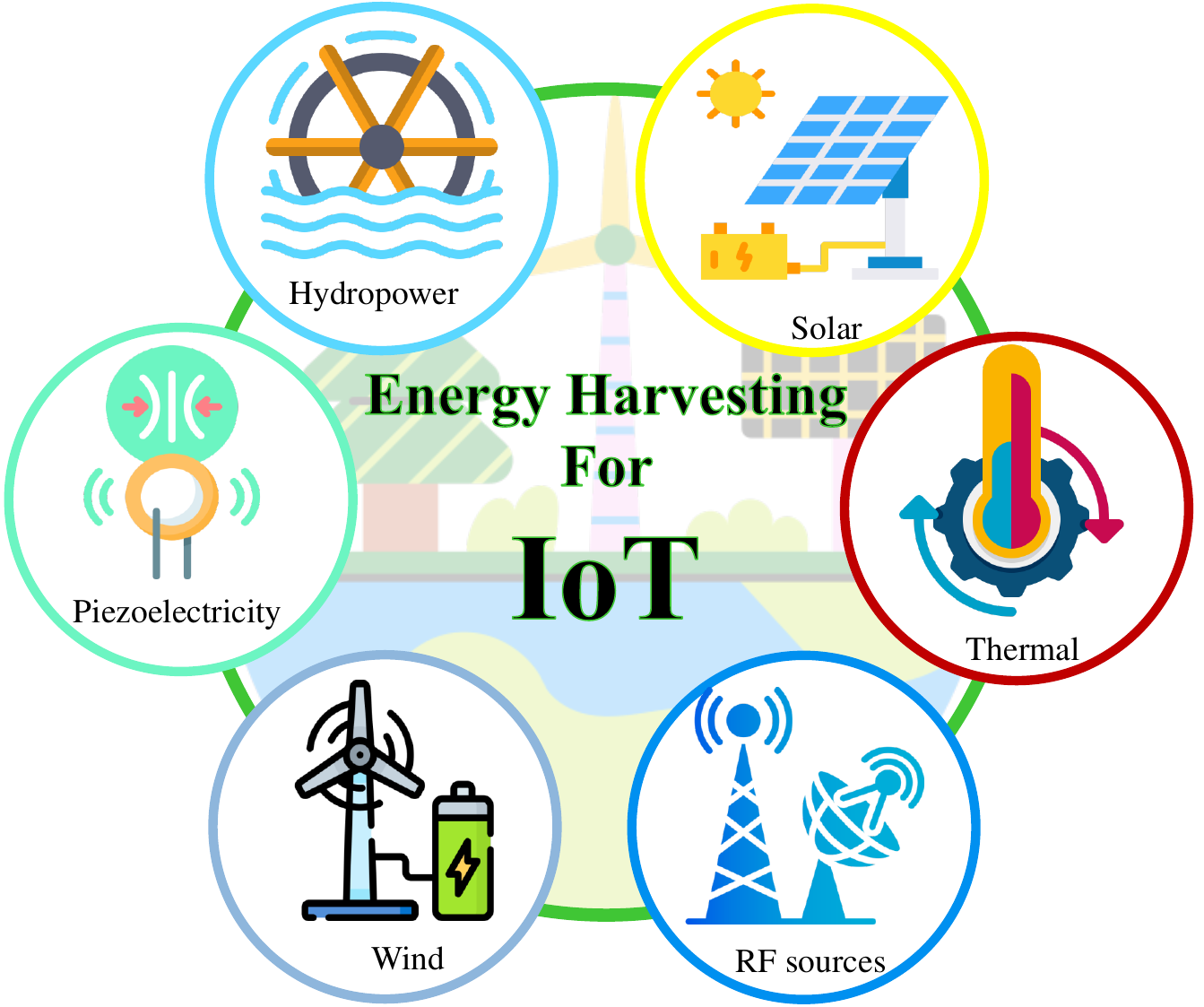}}
	\caption{The potential energy sources for IoT.}
	\label{fig2}
\end{figure*}

Each energy source in EH technology offers unique solutions but also comes with specific requirements and application scenarios. It is prominent to closely evaluate the efficiencies and applicabilities of these sources, as summarized in Table \ref{t1}. 
\begin{itemize}
	\item  	\textbf{Solar Energy}: Solar panels primarily convert outdoor sunlight or indoor artificial light into electricity through the photovoltaic effect \cite{solar}. Despite its remarkable efficiency for IoT devices, its performance drops in dimly lit environments. 
	\item  \textbf{Thermoelectric Energy}:	 Thermoelectric power generation is based on the Seebeck effect \cite{Seebeck}, which can directly convert thermal energy into electrical energy without depending on lighting conditions. It is noted that the effectiveness of this method is closely tied to the presence of a significant temperature gradient in the environment \cite{thermal}. 
	\item  \textbf{Kinetic energy}: Kinetic energy generation converts the motion energy of objects into electrical energy, with standard methods including wind, hydropower, and vibrational EH (such as piezoelectric and electromagnetic) \cite{Piezo}. Notably, wind and hydropower solutions demand substantial investments and expansive infrastructure. In contrast, vibrational EH systems are more constrained, often relying on specific stresses and offering limited conversion efficiency. 
	\item  \textbf{RF energy}:  Within the realm of large-scale IoT applications, RF-powered (a.k.a. wireless-powered) technology shines brilliantly \cite{BiZengZhang}. This innovative approach ingeniously captures omnipresent wireless RF signals from the environment, such as Wi-Fi, cellular, and broadcast signals, and transforms them into usable electrical energy. This approach also aligns well with the principles of AmIoT, harnessing existing environmental resources to power the next generation of smart devices, albeit with the caveat that RF energy is often limited due to controlled power levels of standardized RF sources and significant propagation loss.
\end{itemize}

In the framework of TR 38.848 and the evolving 3GPP systems \cite{Amiot4}, these EH methods are integral to powering a new generation of AmIoT technologies. These technologies are envisioned to support battery-constrained IoT devices or those with energy storage that do not require manual recharging, a critical step towards automation and digitalization in various industries.
}

 \begin{figure*}[t]
	\centerline{\includegraphics[width=4.3in]{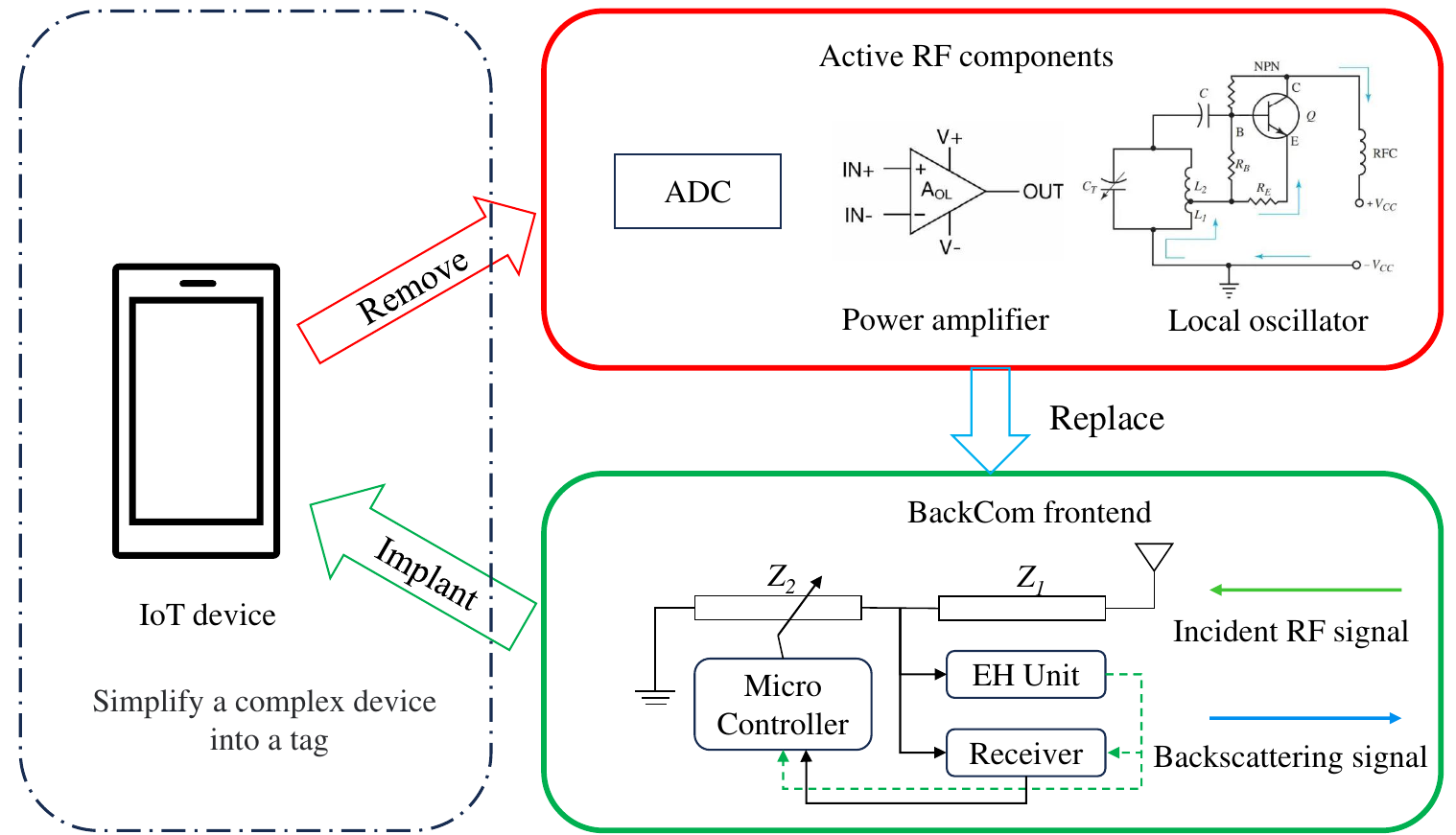}}
	\caption{The concept of BackCom.}
	\label{fig3}
\end{figure*}

\subsection{Reducing Intrinsic Expenses---Backscatter Communication}

{\color{black}Undeniably, existing EH technologies have endowed these devices with supplemental energy, potentially extending their operational lifetime. The silver lining may have a cloud attached, however. It is observed that conventional IoT devices exhibit substantial energy consumption during standard operations \cite{DuhovnikovBaltacci}, particularly in active RF components that integrate crucial details such as analog-to-digital converters (ADCs), power amplifiers, and local oscillators. The inherent high energy consumption still constrains their true long-term operational potential, which deserves to be taken more seriously. Take, for example, traditional EH methods like solar, kinetic, or thermoelectric energy, and there is no denying their impressive performance in many situations. However, these technologies are often tethered to specific environmental conditions, rendering them suboptimal for particular application scenarios. Besides, despite receiving widespread acclaim and attention in recent years, the RF-powered system still faces challenges regarding the low energy conversion efficiency and the weak RF signal strength, leaving room for improvement in its actual EH capabilities \cite{VullersSchaijk}. This implies that  IoT devices may require extended durations to amass requisite energy prior to data transmission, which could hinder their real-time responsiveness and even possibly interrupt their communication process.
	
lIn light of these challenges, a compelling need arises for a more energy-efficient communication strategy. This is where backscatter communication (BackCom) makes its grand entrance, rapidly gaining universal attention across academic and industrial domains. Its salient advantage stems from the clever utilization of load impedance matching to harness ambient RF signals for data transmission, substantially reducing energy consumption throughout the communication process, which means that IoT devices no longer need to expend a significant amount of energy on the active RF chains when transmitting data but merely expend minuscule energy for signal modulation and reflection \cite{BackComS2, BackComS3}, as illustrated in Fig. \ref{fig3}. Especially with the introduction of EH technologies, BackCom is able to maintain its functionality even in scenarios where device batteries are depleted or entirely absent. As a byproduct, this approach paves the way for substantial reductions in IoT device costs and size, enabling a transformation from intricate active devices to simplistic and low-power passive tags \cite{WangZhangYang, LinAhmed}. Consequently, for energy-constrained environments like sensor networks, smart tags, and e-health devices, BackCom unequivocally stands out as an apt solution.}

 \subsection{Motivation and Contribution} 
{\color{black}However, similar to most technological breakthroughs, BackCom exhibits a dual nature, both benefits and challenges. To put it precisely, while its notable energy-saving approach of eliminating active RF components is commendable for low-power communication, BackCom faces significant compromises. Foremost among these is its reliance on signals from nearby RF sources. If these incident signals are weak, it may directly cause a decline in the EH capacity of the tags, subsequently weakening the backscattered signals. Although increasing the strength of the RF source signals may improve the BackCom performance, it could exacerbate interference in RF-dense environments, such as urban or industrial areas, posing a challenge to its widespread adoption.
	
	On the other hand, the effective communication range of BackCom is relatively limited. This limitation stems from the strength of backscattered signals, usually lower than that of directly transmitted active signals. Therefore, in scenarios requiring extensive coverage, such as inter-city communication or remote monitoring, BackCom may not be the most feasible solution. Moreover, compared to traditional active communication technologies, BackCom also falls short in data transmission rate, which could hinder scenarios demanding high-speed data transfer.
	
	To address these challenges, researchers and engineers have proposed various solutions. For instance,  more effective modulation \cite{modulation} and coding \cite{coding} techniques are being explored to enhance the reliability and speed of data transmission. Moreover, spread spectrum communication, exemplified by Millimeter wave \cite{mmwave} and LoRa \cite{Lora}, represents another improvement strategy, which works by expanding the occupancy of the signal in the frequency spectrum, thereby bolstering its robustness. Despite these improvements offering significant performance enhancements in some cases, they have yet to resolve all the challenges faced by BackCom fully. This is especially true in complex interference environments, such as large-scale IoT deployments, where interference from RF sources and other backscatter tags remains a significant issue. Moreover, the problem of double fading of the signal, where it weakens twice during transmission, i.e.,  once at the point of original signal emission and again during backscattering, remains a critical challenge for BackCom technology to surmount.
           
    For the unhindered progress of the BackCom, it is vital to free it from the specific gridlocks mentioned above. Notably, increased research efforts in this field have led to substantial advancements, introducing varied viewpoints and methodologies. This review aims to gather the significant contributions of many researchers to present an exhaustive understanding of present solutions, theoretical accomplishments, and future opportunities concerning complex interference and double fading in BackCom systems. Existing overviews and surveys have already given a thorough overview of the BackCom, clarifying its origin and history, functioning mechanisms, hardware structure, energy supply, and applications across diverse areas. They have, to some degree, already demonstrated what it is, how it works, what it can be employed for, and how it can integrate with existing technologies. To better illustrate the difference between our work and existing work, Table \ref{t2} systematically organizes these works in chronological order of their introduction, offering a clear and comparative visualization of these differences. Our analysis reveals a conspicuous lack of focused, comprehensive summaries of solutions addressing the interference and fading in BackCom systems. This gap emphasizes a unique aspect not discussed in present reviews: how can BackComs efficiently break interference and fading gridlocks? This gap underscores the need for a holistic view that addresses technical challenges and considers the broader implications of BackComs in green communications. Therefore, the incentive for this review is to systematically scrutinize current solutions, offering a lucid, all-inclusive viewpoint that aids researchers in understanding and overcoming these challenges.}
      	
\begin{table*}[t]
	\newcommand{\tabincell}[2]{\begin{tabular}{@{}#1@{}}#2\end{tabular}}
    \color{black}
     \renewcommand{\arraystretch}{1.3}
     \small
	\centering
	\caption{Summary of the Related Surveys for BackComs. \\(Note: I: Fundamentals of BackCom; II: EH Analyses and Summary; III. Potential Sources Summary; IV: Communication Range Summary; V: Interference Analyses; VI: DLI Solutions for BackCom; VII: MI Solutions for BackCom; VIII. Double-fading Solutions for BackCom; IX: Potential Application Summary.\\
	\Checkmark: Detailedly Summarized; ~~~ $\subsetneq$: Only briefly mentioned and not summarized;  ~~~ $\varnothing$: Not involved. }	
	\label{t2}
	\begin{tabular}{|c|c|c|c|c|c|c|c|c|c|c|c|c|}
		\hline
		\textbf{References} & \textbf{Year} & \textbf{I} & \textbf{II}&  \textbf{III} &\textbf{IV}&\textbf{V} &  \textbf{VI}& \textbf{VII}& \textbf{VIII} & \textbf{IX} & \textbf{Main Contribution for BackCom}    \\
		
    	\hline

		{\cite{AmBC}} & 2018& { \Checkmark } & { \Checkmark}  &  { \Checkmark}   &  { \Checkmark}  &   $\subsetneq $  & $\subsetneq $ & $\subsetneq$  & $\subsetneq$ & { \Checkmark}  &  Contemporary survey on AmBC \\
		 
		 \hline
		 
		 {\cite{bcoverview1}} & 2019 & { \Checkmark} & $\subsetneq$ &  $\subsetneq$   &  { \Checkmark}  & $\varnothing$  & $\varnothing$  & $\varnothing$  & $\varnothing$ & { \Checkmark}  &  Overview on BackCom \\
		 
		 \hline
		
		{\cite{GIot}} & 2019& { \Checkmark} & $\subsetneq$ & $\subsetneq$    & $\varnothing$  & $\subsetneq$  & $\varnothing$  & $\varnothing$ & $\varnothing$ & { \Checkmark}  &  Brief survey on AmBC  \\

        \hline
        
       {\cite{bcss3}} & 2019 & { \Checkmark} & $\subsetneq$ &  $\subsetneq$   &  { \Checkmark} & $\subsetneq$ & $\subsetneq$ & $\subsetneq$ & $\subsetneq$  & { \Checkmark}  &  Contemporary survey on BackCom \\
        
        \hline
        
        {\cite{bcss1}} & 2020 & { \Checkmark} & $\varnothing$ &  $\subsetneq$   &  $\subsetneq$ &  $\varnothing$ & $\varnothing$ & $\varnothing$ & $\varnothing$  & { \Checkmark}  &  Overview on BackCom \\
        
        \hline
        
        {\cite{bcss6}} & 2019 &  { \Checkmark} & $\subsetneq$  &   $\subsetneq$   &  $\subsetneq$  &  $\subsetneq$  & $\varnothing$ & $\subsetneq$  & $\subsetneq$  & { \Checkmark}  & Contemporary survey on BackCom \\
        
        \hline	
		
		{\cite{BTTN}} & 2020& { \Checkmark} &  $\subsetneq$ &   $\subsetneq$   & { \Checkmark}  &  $\subsetneq$  & $\varnothing$  & $\varnothing$ & $\varnothing$ & { \Checkmark}  &  Survey on tag to tag communication \\
		
	     \hline
		
	    {\cite{bcss4}} & 2020 &  { \Checkmark} & { \Checkmark}  &  { \Checkmark}   &  { \Checkmark}  &  $\subsetneq$  & $\subsetneq$ & $\subsetneq$  & $\subsetneq$ & { \Checkmark}  &  Contemporary survey on wireless-powered BackCom \\
		
		\hline
		
	   {\cite{bcss}} & 2021 & { \Checkmark} & $\varnothing$ &  $\subsetneq$   &  $\subsetneq$ &  $\varnothing$ & $\varnothing$ & $\varnothing$ & $\varnothing$  & { \Checkmark}  &  Overview on BackCom \\
		
		\hline
		
		{\cite{ACD}} & 2022& { \Checkmark} & $\subsetneq$ & $\subsetneq$   & $\subsetneq$  & $\varnothing$  & $\varnothing$  & $\varnothing$  & $\varnothing$ & { \Checkmark}  &  Survey on chip-level design and system integration   \\
		
		\hline

		{\cite{bcss2}} & 2022 & { \Checkmark} & $\subsetneq$ &  $\subsetneq$   &  $\subsetneq$ &  $\subsetneq$ & $\subsetneq$ & $\varnothing$ & $\varnothing$  & { \Checkmark}  &  Survey on BackCom and sensing \\
		
		\hline
		
		{\cite{BBWS}} & 2022& { \Checkmark} &  $\subsetneq$ &  $\subsetneq$  &  { \Checkmark} & $\subsetneq$  & $\varnothing$   & $\varnothing$ & $\varnothing$ & { \Checkmark}  &  Survey on  wireless sensing \\
		
		\hline
		
	    {\cite{bcss5}} & 2022 &  { \Checkmark} & $\subsetneq$  &  $\subsetneq$   &  $\subsetneq$  &  $\varnothing$  & $\varnothing$ & $\varnothing$  & $\subsetneq$  & { \Checkmark}  &  Survey on link budget \\
		
		\hline

     	 {\cite{bcs2}} & 2022& { \Checkmark} &  { \Checkmark} &  { \Checkmark}  &  { \Checkmark} & $\subsetneq$  & $\varnothing$  & $\varnothing$ & $\varnothing$ & { \Checkmark}  & Survey on Ambient excitation signals \\
     	
     	\hline

     	{\cite{modulation}} & 2022 & { \Checkmark} & $\subsetneq$ &  $\subsetneq$   &  { \Checkmark} & $\subsetneq$ & $\subsetneq$ & $\varnothing$ &$\varnothing$  & { \Checkmark}  &  Review on index modulation \\
     	
     	\hline

     	{\cite{coding}} & 2023& { \Checkmark} & { \Checkmark} &  $\subsetneq$  & $\subsetneq$ & $\subsetneq$  & $\varnothing$   &$\subsetneq$  &  $\varnothing$  &  {\Checkmark} &  Survey on Coding methods \\ 	
     	
     	\hline
     	
     	{\cite{bcss7}} & 2023 &  { \Checkmark} &  $\subsetneq$   &    $\subsetneq$    &   $\subsetneq$   &   $\subsetneq$  &  $\subsetneq$  & $\subsetneq$  & $\subsetneq$  & { \Checkmark}  & Contemporary survey on AI BackCom \\
     	
     	\hline	
     	
     	{\cite{BackComS1}} & 2023 &  { \Checkmark} &  { \Checkmark}   &    { \Checkmark}    &   { \Checkmark} &   $\subsetneq$  &  $\subsetneq$  & $\subsetneq$  & $\varnothing$  & { \Checkmark}  & Contemporary survey on  Battery-Free IoT \\
     	
     	\hline	
		
		{Ours} & --- &{ \Checkmark}& { \Checkmark}&  { \Checkmark}  & { \Checkmark} & {\Checkmark}  & {\Checkmark}  & { \Checkmark} & {\Checkmark} & $\subsetneq$ &  Survey on interference and double fading migration   \\
		
		\hline
	\end{tabular}
	
\end{table*}

         \begin{figure*}[t]
	\centerline{\includegraphics[width=7in]{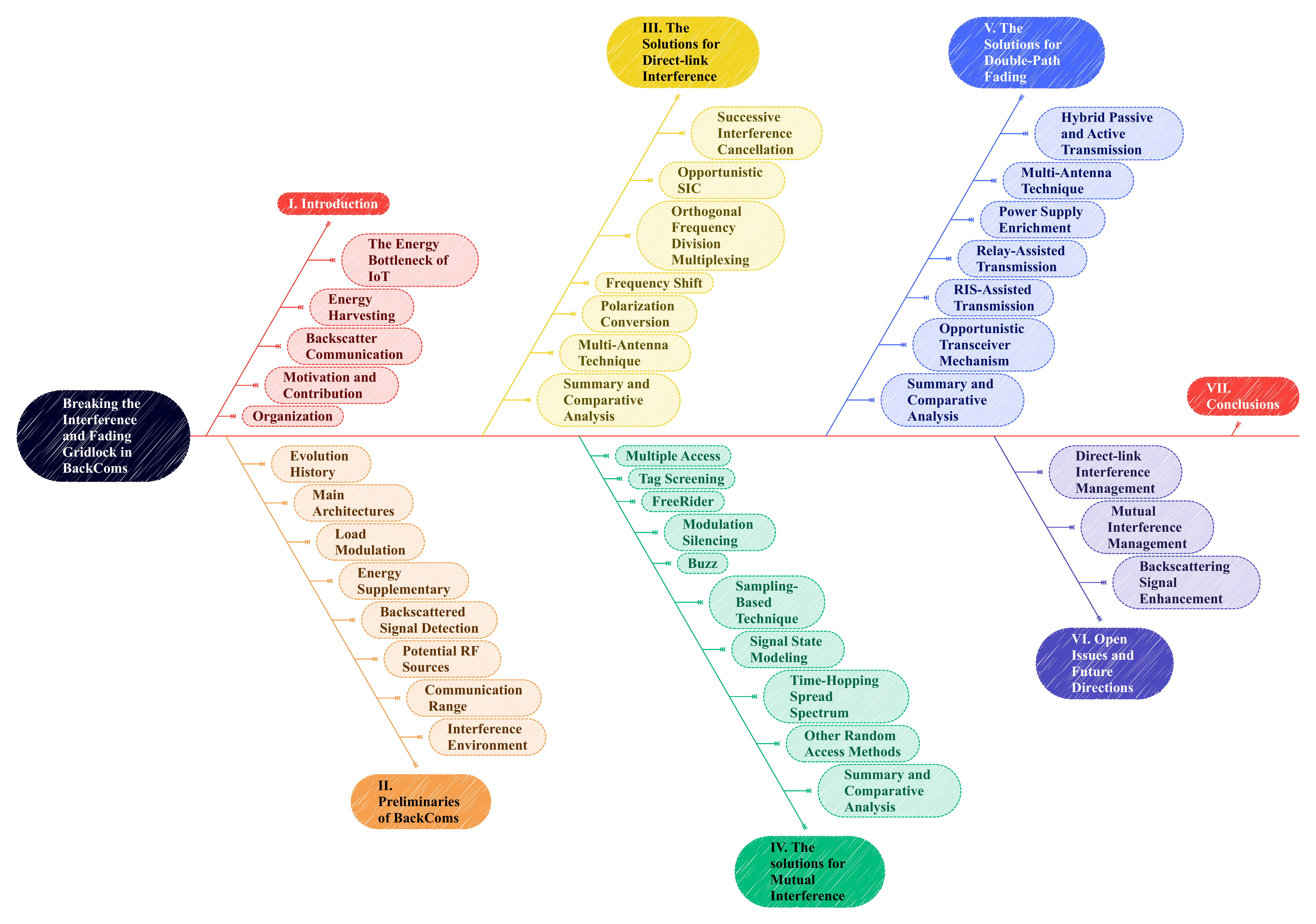}}
	\caption{The taxonomy of this survey.}
	\label{fig5}
\end{figure*}

In a nutshell, the contributions of this survey can be summarized as follows:

\begin{itemize}
	{\color{black}\item Firstly, we provide a systematic presentation of the fundamentals of the BackCom, including its background, concepts, history, architectures, operation mechanism, energy supplementary, potential RF sources, communication range,  and interference environment. This foundational knowledge is crucial for new researchers and seasoned professionals in understanding and advancing BackCom technology.
		\item Following this, our work includes a detailed investigation of existing solutions to the interference issues in BackCom systems. This part of our review summarizes current strategies and identifies challenges and opportunities for innovation, primarily focusing on the direct-link interference (DLI) and the mutual interference (MI) among tags, respectively. A comparative analysis is provided to juxtapose these strategies against each other, providing a clear understanding of their relative merits and limitations.
		\item  Then, we present an in-depth discussion on the solutions for the double-fading issue in BackComs, including exploring emerging technologies and innovative approaches that could revolutionize BackCom systems. Our review rigorously compiles, compares, and assesses the implementation medium, advantages, and challenges of each method, providing a deep dive into technological innovations.
		\item Finally, we explore potential approaches to address the challenges of complex interference and double fading in BackCom systems. Our review also anticipates future technological shifts and user requirements, setting a path for the next generation of BackCom technologies and aiming to inspire ongoing and future advancements in the field.}
\end{itemize}

\subsection{Organization}

The remainder of this survey outlines the rich state-of-the-art in addressing interference and fading within BackCom systems, relying on authoritative citations and succinct summary tables for enhanced clarity and quick reference. By delving deeply into fundamental principles and typical applications, we identify the critical knowledge gaps in existing solutions and critically assess the pros and cons of these methods as well as the trade-offs. This lays the groundwork for addressing open research questions while avoiding common pitfalls and emphasizing specific technological approaches and innovations.

{\color{black}To be more specific, the outline of this survey is shown in Fig. \ref{fig5}, which is designed to visually complement the textual content, providing an intuitive understanding of the structure of this survey. Section I provides the background of BackComs and illustrates the motivation and contribution of this survey. Then, Section II introduces some preliminaries of BackCom systems. Following this, Sections III and IV detail the potential interference solutions for BackCom systems regarding the DLI and MI, respectively. Section V further delves into the solutions of double-path fading in BackCom systems. Section VI outlines promising research directions and open issues in BackCom systems. Finally, Section VII concludes our work and illumines potential research directions in the future.}

{\color{black}\section{\textsc{Preliminaries of BackComs}}

This section provides a comprehensive and updated understanding of BackCom systems, delineating their historical evolution, main architecture, basic modulation principles, energy supplementation, signal detection, various potential RF sources, detailed analyses of communication ranges, and interference environment. The purpose is to set the stage for a comprehensive exploration into the nuanced mechanisms that enable BackCom systems to function efficiently and sustainably. 

\subsection{Evolution History}
	
	         \begin{figure*}[t]
		\centerline{\includegraphics[width=7.6in]{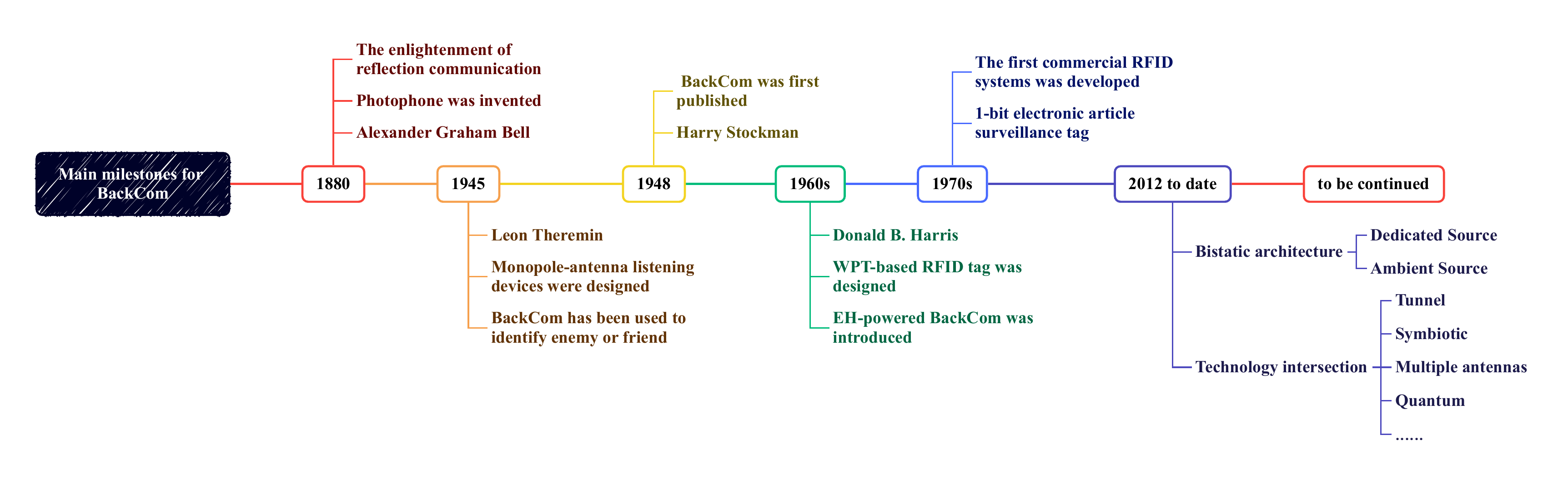}}
		\caption{The main milestones of BackCom.}
		\label{fig5a}
	\end{figure*}  

The story of BackComs was traced back to 1880 when Alexander Graham Bell, in his pioneering endeavor, introduced the world to the photophone \cite{photophone}. This innovative system, transmitting voice via light beams, relied on modulating sound waves onto a mirror, harnessing the vibrations of the reflected light. The culmination of this process allowed a receiver to aptly reproduce the transmitted signal, with successful trials spanning an impressive distance of 213 meters.

Moving to 1945, Leon Theremin, in a covert espionage act, developed a secret listening device ingeniously hidden within the U.S. national emblem \cite{Lero}. This innovative apparatus, a symbol of Cold War subterfuge, empowered the Soviet Union to surveil the U.S. embassy in Moscow clandestinely. The heart of the device lay in its monopole antenna, seamlessly linked to a resonant cavity that housed a sound-sensitive conductive membrane. This configuration deftly modulated backscattered waves as they interacted with ambient sound.

By 1948, Harry Stockman delved deeper into the realms of backscatter, elucidating the communicative prowess harnessed from reflections, particularly those originating from mechanical devices. Such an avant-garde concept, targeting audio transmission via microwave frequencies, resonates with the modus operandi of contemporary tags encapsulating the backscatter ethos \cite{Stockman}. This cascade of innovations solidified the bedrock for the BackCom and catalyzed its journey to commercial prominence.

The 1960s marked the synergistic fusion of backscatter with wireless power transfer (WPT) technologies, an advancement championed by Donald B. Harris \cite{harris}. This collaboration ignited a resurgence in the landscape of  RF identification (RFID) technology \cite{Boyer}. The 1970s epitomized this synergy with radio waves as a deterrent against theft, spotlighting 1-bit electronic article surveillance tags as the flagship embodiment of the RFID paradigm \cite{Finkenzeller}. The introduction of the wireless identification and sensing platform marked a paradigm shift \cite{WISP}, underpinning unprecedented capabilities in environmental sensing and accentuating backscatter-centric research.

Over the years, concerted efforts have been made to advance the BackCom and explore its associated technologies, which are evident in emerging techniques such as bistatic architecture for BackComs, including dedicated sources \cite{BiBC} in 2012  and ambient sources \cite{AmBC} in 2013. 
Then, more technology integration has been explored in succession, such as tunnel backscatter \cite{tunnelscatter}, quantum backscatter \cite{quantum}, symbiotic radio\cite{symbotic}, multi-antenna backscatter \cite{RC1}, and so on.

For convenience, Fig. \ref{fig5a} is included to depict the latest milestones in BackCom during the past years. Each milestone in the history of BackCom has not only marked a leap forward in technology but has also served as a stepping stone for subsequent innovations. Recent technological advancements, particularly in the realm of 5G and 6G networks, have further propelled the capabilities of BackCom systems, offering enhanced bandwidth and reduced latency. This evolution signifies a pivotal step in integrating the BackCom technology with contemporary wireless communication standards.
This multifaceted communication mechanism finds resonance across urban infrastructures\cite{structure}, precision agriculture\cite{agriculture}, smart transport systems\cite{transportation}, digital healthcare \cite{ehealth},  asset tracking \cite{tracking}, underwater communications\cite{underwater}, and so forth, indicating its versatility and adaptability to diverse applications. 

\begin{table*}[t]
	\centering
	\color{black}
	\renewcommand{\arraystretch}{1.3}
	\small
	\caption{Summary of BackCom architectures}
	\label{t2a}
	\begin{tabular}{|>{\centering\arraybackslash}m{1.5cm}|>{\centering\arraybackslash}m{2.5cm}|>{\centering\arraybackslash}m{2.7cm}|>{\centering\arraybackslash}m{4cm}|>{\centering\arraybackslash}m{4cm}|}
		\hline
		\textbf{Category} & \textbf{Source Type} & \textbf{Transceiver Type} & \textbf{Pros.}  & \textbf{Cons.}  \\
		\hline
		MoBC & \multirow{3}{2.5cm}{~~Dedicated source} & Monostatic protocol & Passive communication & Round-trip path loss and near-far effect \\
		\cline{1-1} \cline{3-5}
		BiBC &  & \multirow{3}{2.7cm}{~~~Bistatic protocol} & Higher Flexibility  and lower path loss  & Deployment issue and energy cost  \\
		\cline{1-2} \cline{4-5}
		AmBC &  Ambient source &  &  Ambient source availability and  high energy efficiency & Signal unpredictability and potential interference  \\
		\hline
	\end{tabular}
\end{table*}

  \begin{figure*}[t]
	\centerline{\includegraphics[width=6.0in]{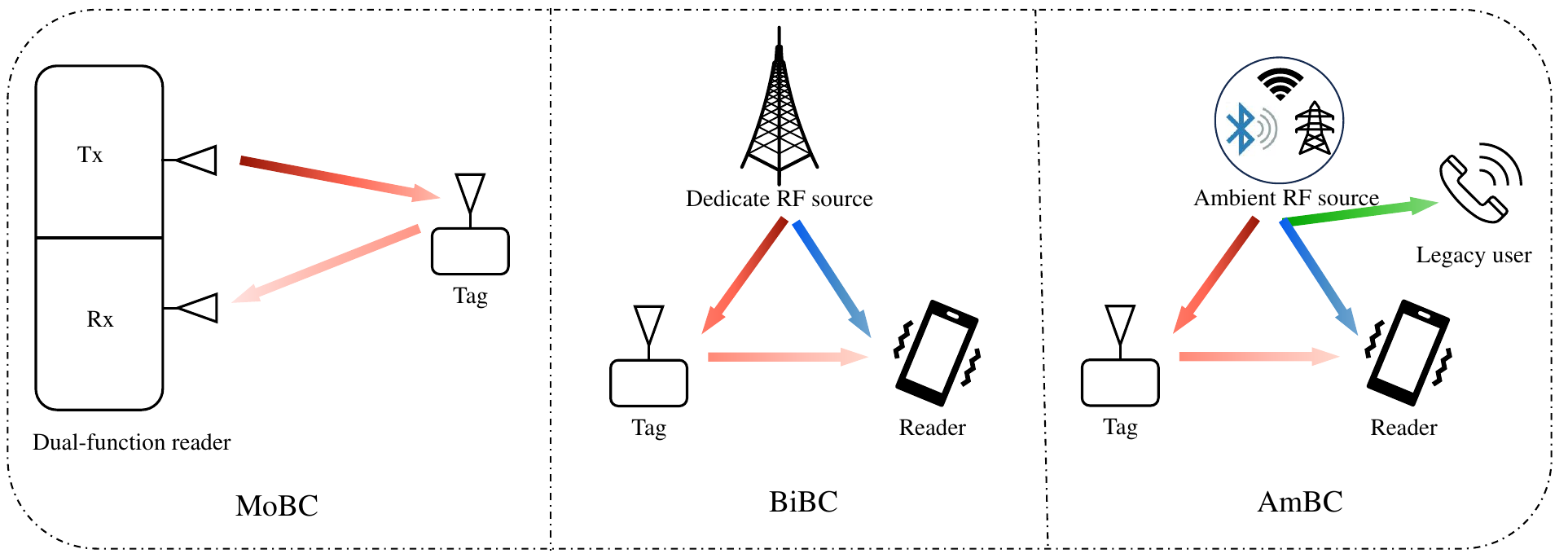}}
	\caption{The architectures of BackCom (Note: The arrow represents the direction of the link, and the gradient of color signifies the strength of the signal.).}
	\label{fig4}
\end{figure*}

\subsection{Main Architectures}

Currently, the architectures for BackCom systems are typically classified into three categories: Monostatic BackCom (MoBC), Bistatic BackCom (BiBC), and Ambient BackCom (AmBC), as shown in Fig. \ref{fig4}, which visually represents these architectures to aid in understanding their operational mechanisms.

\subsubsection{Monostatic BackCom} 
\begin{itemize}
	\item \textit{Technical Merit}: This is the earliest and most mature form of BackCom, often exemplified by technologies such as RFID. In a MoBC setup, the system consists of a tag and a dual-function reader that act as a power source (i.e., Tx) and an information receiver (i.e., Rx). The reader first generates an RF signal to activate the tag. The tag then modulates and backscatters the RF signal back to the reader.
	
	\item \textit{Challenges and Limitations}: In MoBC, it should be noted that the RF signal source and receiver are integrated into a single node. This results in specific challenges, including the round-trip path loss, which can weaken the strength and quality of the modulated signal. Moreover, in MoBC systems, challenges may arise due to the near-far effect, which refers to the variable signal strength and quality depending on the distance between the tag and the reader. When the reader is located far from the tag, there is an increased risk of a higher probability of energy outage and a weakened backscattering signal strength.
	
\end{itemize}

\subsubsection{Bistatic BackCom} 
\begin{itemize}
	\item \textit{Technical Merit}: BiBC presents a more refined protocol, designed to overcome certain inherent limitations associated with traditional MoBC \cite{BiBC}, where the power source and the information receiver are physically separated, which can enhance the flexibility and provide better control over signal propagation, thus reducing the round-trip path loss. Specifically, tags can leverage unmodulated RF signals from nearby power sources to harvest energy and transmit data. By doing so, even tags located far from the reader can maintain a sufficiently high signal strength and reduce the energy outage probability. For instance, BiBC has been successfully implemented in advanced logistics systems for efficient asset tracking.
	
	\item \textit{Challenges and Limitations}: However, deploying and maintaining dedicated power sources in BiBC pose some challenges. First, popularizing BiBC requires additional cost and resource investment, including selecting suitable equipment and considerations for installation and maintenance, which increase the overall cost of the system. Second, extreme environmental conditions pose another obstacle to its implementation. Complex terrain or harsh weather conditions can impact the reliability and stability of the system.
	
\end{itemize}

\subsubsection{Ambient BackCom}
\begin{itemize}
	\item \textit{Technical Merit}: AmBC is garnering attention for exploiting ubiquitous ambient signals from existing RF sources, a concept central to the idea of AmIoT, which is particularly promising in urban environments, where ambient RF sources are abundant. By harnessing the ambient RF energy, AmBC circumvents the constraints and costs associated with deploying dedicated power sources, leading to significant cost and energy savings \cite{AmBC}. Furthermore, by utilizing existing ambient RF sources, AmBC systems eliminate the need to acquire new frequency spectra, obtain licenses, and comply with regulatory requirements. This aspect not only simplifies the deployment process but also accelerates the integration of AmBC into existing communication infrastructures. The accessibility, speed, and energy efficiency of AmBC systems make them an attractive option for a wide range of IoT applications.
	
	\item \textit{Challenges and Limitations}: However, AmBC comes with its own set of challenges. On the one hand, the availability of ambient RF sources can be unpredictable and uncontrollable, posing challenges to the communication reliability. Addressing this issue requires adaptive and resilient system designs capable of coping with varying ambient RF conditions. On the other hand, the presence of backscattering signals from tags in AmBC systems may interfere with traditional/primary receivers, potentially impacting the performance of existing communication systems.
\end{itemize}

In summary, each of these BackCom architectures, i.e., MoBC, BiBC, and AmBC, presents unique technical merits and faces distinct challenges, as briefly summarized in Table \ref{t2a} for a more intuitive perspective. Understanding these different characteristics and limitations of each architecture is crucial for advancing the BackCom technology and tailoring it to specific IoT applications. Besides, recognizing the dynamic nature of these architectures paves the way for innovative advancements, underscoring the importance of continuous research and development in this rapidly evolving field.

\subsection{Load Modulation}

Backscatter tags shift from the traditional archetype of wireless devices that actively produce RF signals. These innovative tags enable communication through a dynamic modification of the reflective properties of incident RF waves, achieved by finely tuning their antenna impedance. This passive modulation paradigm empowers tags to inscribe data upon RF waveforms through a deliberate variation of the load impedance of their antenna. This technique is vital in various BackCom system architectures like MoBC, BiBC, and AmBC. It is increasingly relevant in applications like IoT, where energy efficiency and low-power operation are critical.

The process involves a strategic switch between different load impedances, enabling the tag to modulate its reflection coefficient, which can be represented by \cite{AmBC}
\begin{equation} \label{m1} 
	\Gamma_i=\frac{Z_i-Z_a^*}{Z_i+Z_a},  i\in \{1,2,\cdots M\},
\end{equation}
where $Z_i$ is the load impedance of state $i$ within an $M$-ary modulation scheme \cite{coding}, $Z_a$ is the antenna impedance, and $*$ is the complex conjugate operator. 

In BackCom systems, binary phase-shift keying (BPSK) modulation is commonly adopted, i.e., $M=2$. By varying the sliding impedance $Z_i$, the reflection coefficient can alternate between absorbing and backscattering states, as shown in Fig. \ref{fig5b}. Specifically, RF signals are absorbed in the absorbing state (impedance matching), indicating a binary `0'. In contrast, RF signals are reflected in the backscattering state (impedance mismatching), representing a binary `1'. This load modulation approach, while simple, is highly effective and forms the basis for more complex modulations like $M$-phase-shift keying (PSK) and $M$-quadrature amplitude modulation (QAM), enhancing data transmission capabilities in modern BackCom systems \cite{modulation}.

\begin{figure}[t]
	\vspace{-5mm}
	\centerline{\includegraphics[width=3.2in]{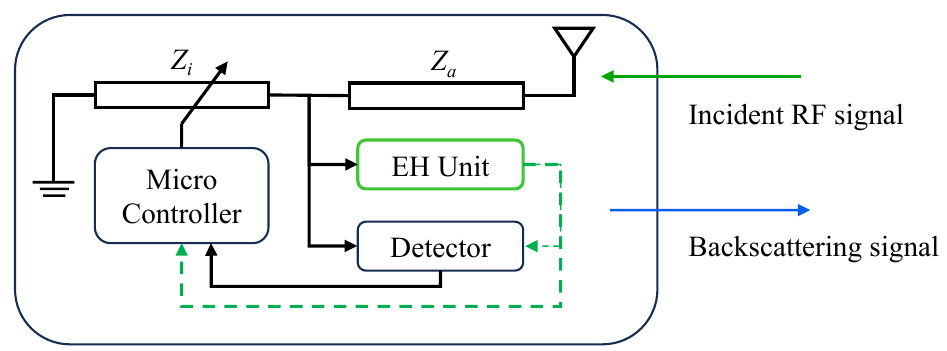}}
	\caption{The architecture of a backscatter tag.}
	\label{fig5b}
\end{figure}

\subsection{Energy Supplementary}

To ensure the stable and prolonged operation of backscatter tags, EH technologies play a crucial role, particularly for those harnessing RF energy, which significantly enhance the capabilities of BackCom systems. These EH technologies are vital to the sustainability and independence of BackCom systems, allowing them to operate in various environments without an external power source.

In particular, tags come with built-in energy harvesters that convert incident RF signals into usable direct current (DC) power, a process crucial for the autonomy of backscatter systems, as depicted in Fig. \ref{fig5b}. An RF energy harvester typically comprises three main components: a capacitor, a voltage multiplier, and an impedance matching circuit \cite{rfeh}, as shown in Fig. \ref{fig5c}. The impedance matching circuit is a resonant circuit that maximizes the power transfer from the antenna to the multiplier. The voltage multiplier, usually constructed with rectifying diodes, converts the RF signal into a DC voltage. At the same time, the capacitor acts as a buffer, ensuring a consistent power supply and serving as temporary energy storage.

\begin{figure}[t]
		\vspace{-5mm}
	\centerline{\includegraphics[width=2.7in]{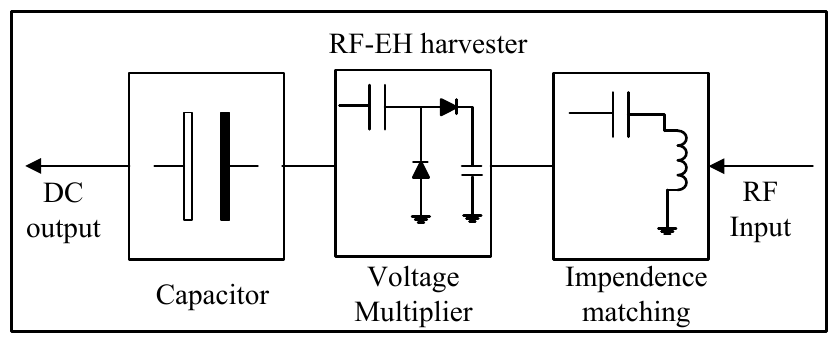}}
	\caption{The architecture of an RF energy harvester.}
	\label{fig5c}
\end{figure} 

The efficiency of energy harvesters, critical to the stability and reliability of BackCom systems, depends on EH sensitivity and energy conversion efficiency. Specifically, EH sensitivity refers to the minimum power level required to activate the circuit, typically around -21 dBm \cite{ehsen}, making it a fundamental parameter for effective operation in low-power environments. Energy conversion efficiency, the ratio of the power output converted to the power input, is influenced by the effectiveness of the antenna, the precision of impedance matching, and the proficiency in converting the RF energy into a DC voltage. Generally, energy conversion efficiency ranges from 22\% to 72\% \cite{ehce}, reflecting the variability in harvester performance under different operational conditions and fabrications. Moreover, the output power of energy harvesters can vary from -25 dBm to -10 dBm \cite{ehputput}, accommodating extensive functional scenarios. This variance in output underscores the adaptability of energy harvesters to diverse environmental energy conditions, making them versatile components in a wide array of applications.

Introducing nonlinear models for EH is a significant advancement, allowing for more accurate prediction of harvester behavior under varying environmental conditions and power levels. These models are essential for developing reliable and efficient BackCom systems, enabling precise predictions of energy harvester performance. The piece-wise linear model provides a segmented approximation of the behavior of the energy harvester \cite{ehmodel0}, and logistic nonlinear models offer a smooth, continuous representation \cite{ehmodel1,ehmode2}. For instance, one typical model is given by 
\begin{equation} \label{e1} 	
	\psi (P_{\text{in}})\text{=}\left\{ \begin{aligned}
		& P_{\text{Sat}}, ~\eta P_{\text{in}}> P_{\text{Sat}}, \\ 
		& \eta P_{\text{in}}, ~P_{\text{Sen}}\le \eta P_{\text{in}}\le P_{\text{Sat}}, \\ 
		& 0,  ~~~~~\eta P_{\text{in}}< P_{\text{Sen}}, \\
	\end{aligned} \right.
\end{equation}
where $\psi ( P_{\text{in}})$  is the actual power that the specific tag can harvest. $P_{\text{Sat}}$ and $P_{\text{Sen}}$ are thresholds representing EH saturation and EH sensitivity for the energy harvester, respectively. $\eta$ denotes the EH conversion efficiency. Besides, $P_{\text{in}}$ denotes the input power of the energy harvester.

  \begin{figure*}[t]
	\centerline{\includegraphics[width=5.6in]{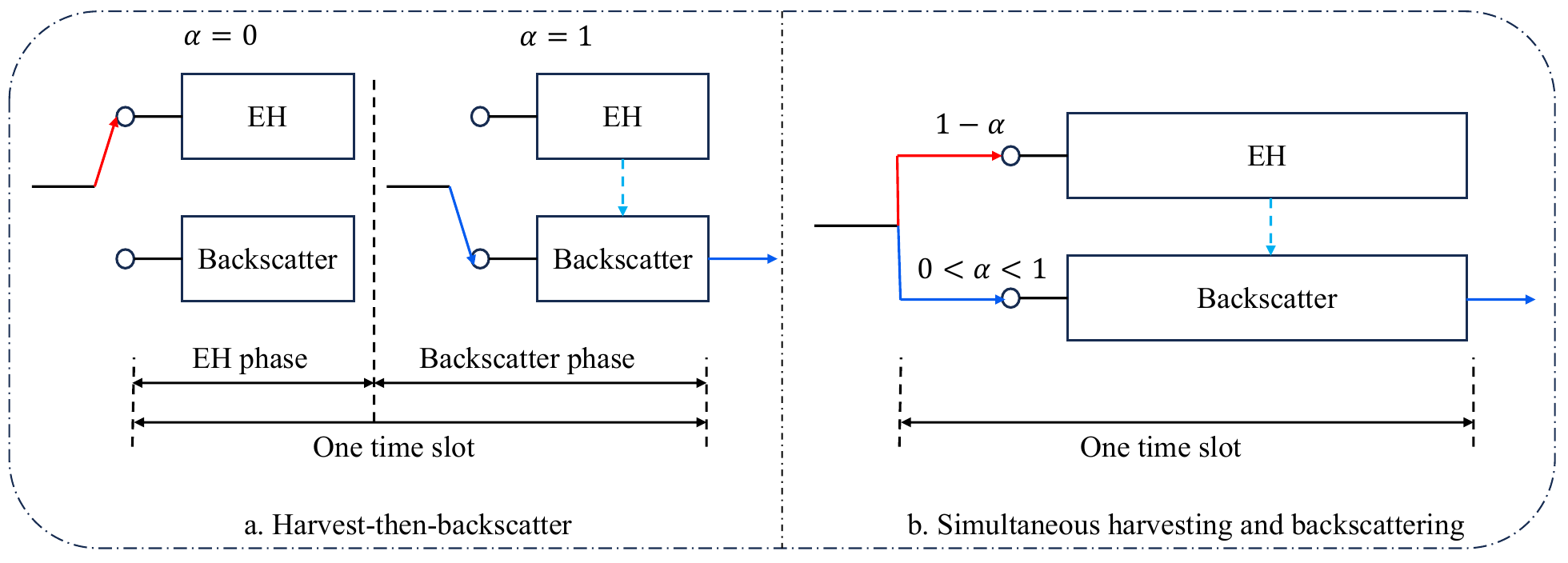}}
	\caption{The EH manners in BackCom.}
	\label{fig5d}
\end{figure*} 

Moreover, BackCom systems employ two primary EH manners: harvest-then-backscatter and simultaneous harvesting and backscattering, shown in Fig. 
\ref{fig5d}, to enhance the adaptability of BackCom systems for different operational requirements and scenarios. In the first manner, the tag can first harvest energy to compensate for the required circuit power at the EH phase (i.e., $
\tau$) and then backscatter information at the backscatter phase (i.e., $T-\tau$). Therefore, the input power and the harvested energy by its energy harvester can be expressed as $P_{\text{in}}^{\text{a}}= P_{\text{R}}$ and $E^{\text{a}}=\tau \psi (P_{\text{in}}^{\text{a}})$, respectively, where $P_{\text{R}}$ denotes the received power of the tag.

In the second manner, the tag can divide a part of the received signal as EH and backscatter the other part of the signal during the whole time slot $T$. Therefore, the input power and the harvested energy by its energy harvester can be expressed as $P_{\text{in}}^{\text{b}}= (1-\alpha)P_{\text{R}}$ and $E^{\text{b}}=T\psi (P_{\text{in}}^{\text{b}})$, respectively, where $\alpha \in [0,1]$ is the power reflection coefficient, denoting the fraction of the power of the received RF signal that is available for backscattering. For its integrity, under the binary modulation, $\alpha$ can be expressed as 
\begin{equation}  \label{e3}
	\alpha=\frac{1}{2}|\Gamma_1-\bar \Gamma|^2+\frac{1}{2}|\bar \Gamma-\Gamma_2|^2=\frac{1}{4}|\Gamma_1-\Gamma_2|^2,
\end{equation}
where $\bar{\Gamma}=\frac{\Gamma_1+\Gamma_2}{2}$.

{\color{black}\subsection{Backscattered Signal Detection}

The detector is a crucial component of passive tags for BackCom systems, as illustrated in Fig. \ref{fig5b}. It plays a vital role in detecting the presence of incident signals, which triggers the backscatter tag to piggyback its data onto the signal. Traditional detectors generally require power-intensive components, making them impractical for low-power tags. The detection methods employed in BackCom systems can be broadly classified into two categories: 

\begin{itemize}
	\item \textbf{Threshold Comparison}: This method relies on the strength of the incident signal and a preset threshold voltage. A typical implementation in Wi-Fi signal detection \cite{wifi1} involves a multistage demodulating logarithmic amplifier and a voltage comparator. This method is characterized by its simplicity and effectiveness in environments with stable signal characteristics. Specifically, the incident signal is first amplified and summed up by a multistage amplifier circuit on the tag. Subsequently, a lowpass filter removes the unwanted noise. The voltage comparator circuit then compares the amplified output ($V_{\text{output}}$) with a threshold voltage ($V_{\text{Th}}$), generating a high voltage output when $V_{\text{output}} > V_{\text{Th}}$, indicating the detection of a Wi-Fi signal. This approach is advantageous for its low power consumption and straightforward implementation, making it suitable for applications where the energy efficiency is crucial.
	
	\item \textbf{Feature Identification}: This method involves leveraging specific features of a signal to detect its presence. A notable example is the detection of LoRa signals, as discussed in \cite{lora1}, which utilizes the distinct features of the LoRa preamble, consisting of ten identical up chirps. This approach is efficient in environments with diverse signal types, as it allows for more precise signal detection by focusing on unique signal characteristics. The detection is achieved by correlating the incoming signal with a pre-stored up chirp, identifying the presence of LoRa signals. Feature identification is more complex than threshold comparison but offers higher accuracy and selectivity, making it ideal for scenarios requiring precise signal discrimination.
\end{itemize}

The choice between these detection methods depends on several factors, including the energy constraints of the tag, the environmental conditions, and the specific application requirements. Feature identification may offer better performance in dynamic environments with fluctuating signal conditions, whereas threshold comparison might be more suitable for scenarios with predictable signal patterns. The development of efficient and accurate detection techniques is critical for the advancement of BackCom technology, as it directly impacts the  overall performance and reliability of the system.}

\subsection{Potential RF Sources}

Recall that, in BackCom systems, harnessing various wireless signals from surrounding RF sources is essential for efficient operation, which aligns perfectly with the principles of AmIoT. These sources act as primary drivers for signal propagation, enabling backscatter tags to modulate and reflect their data. Understanding the diversity and characteristics of these RF sources is crucial to optimizing the performance and applicability of BackCom systems.

\subsubsection{Dedicate sources} 
Dedicated RF sources (DRFSs), specifically designed for BackCom systems, are employed in environments where control over the RF source is crucial. These transmitters can be custom-tailored to provide specific signal strengths and frequencies, meeting the precise operational needs of BackCom systems \cite{BiBC}. They are particularly prevalent in controlled environments like secure communications, underwater communications, and healthcare monitoring systems, where the predictability and stability of the RF source, along with security considerations, are of utmost importance, such as \cite{mbc1,mbc2,mbc3}.

\subsubsection{Ambient sources}
\begin{itemize}
	\item \textbf{TV:} Television (TV) signals, particularly in high-frequency bands, are long-range transmitters covering extensive geographic areas. These signals, capable of penetrating diverse structures, are highly reliable for urban and suburban BackCom applications \cite{tv}. The transition from analog to digital broadcasting has further augmented their applicability for wide-area BackCom systems, offering superior signal quality and stability.

	\item \textbf{FM:} 
	Frequency modulation (FM) signals are renowned for their excellent fidelity and robustness against noise \cite{fm}. They provide consistent and high-quality signals over vast areas, making FM advantageous for BackCom applications that require stable signal propagation, especially in densely populated urban contexts where signal consistency is paramount.
	
	\item \textbf{LTE:} 
	Long-term evolution (LTE) networks, known for their rapid data transmission capabilities, offer a reliable and continuous source of RF energy \cite{lte}. The evolution of LTE, including transitions to 4G and 5G, has broadened its utility as a versatile RF source for BackCom, enhancing its effectiveness in urban and rural settings.
	
	\item \textbf{WiFi:} 
	Wireless fidelity (WiFi) networks, operating under various standards such as WiFi 4/5/6, provide high-frequency wireless connectivity. Their ubiquitous deployment in urban areas renders them ideal for indoor BackCom applications, although the range constrains their utility and potential interference in highly saturated environments \cite{wifi}.
	
	\item \textbf{Bluetooth:} 
	Bluetooth, especially Bluetooth low energy (BLE), is tailored for short-range, low-power wireless communication \cite{FR}. Its significance in personal networks and intelligent environments is notable due to its energy efficiency and optimal operational range.
	
	\item \textbf{Zigbee:} 
	Zigbee operates as a low-power, low-data-rate wireless communication protocol in IoT networks, built on the robust IEEE 802.15.4 standard \cite{zigbee}. Its mesh networking capability significantly enhances coverage and reliability, making it well-suited for indoor and outdoor applications.
	
	\item \textbf{LoRa:} 
	LoRa stands out for its long-range communication capabilities and minimal power requirements, ideal for rural or remote BackCom applications \cite{Lora}. The potential of LoRa in large-scale agricultural monitoring and wildlife tracking is currently under exploration, showcasing its adaptability in diverse environments.

\end{itemize}

The diverse array of RF sources illustrates the broad spectrum of options available for BackCom systems, each bringing distinct attributes to the forefront. From the long-range coverage of TV and FM signals, ideal for urban and suburban settings, to the high-speed data transmission capabilities of LTE networks and the low-power, mesh-networking prowess of Zigbee, these sources cater to a wide array of BackCom applications. Integrating Bluetooth and LoRa further expands potential use cases, from personal networks to extensive rural deployments. By selecting the appropriate RF source, BackCom systems can be meticulously tailored to meet specific operational requirements. As the landscape of RF technologies continues to evolve, so does the potential for innovative AmIoT applications, paving the way for more integrated, efficient, and intelligent wireless communication systems.

\begin{table*}[t]
	\renewcommand{\arraystretch}{1.3}
	\color{black}
	\newcommand{\tabincell}[2]{\begin{tabular}{@{}#1@{}}#2\end{tabular}}
	\small
	\centering
	\caption{Comprehensive Comparison of Communication Ranges in AmBC Systems Using Various RF Sources. \\(Note: $d_{\text{B}}$ and $d_{\text{B}}^*$ denote the  maximum achievable backward distance and maximum achievable rate corresponding to the backward distance.)}	
	\label{t3}
	\begin{tabular}{|c|c|c|c|c|c|c|c|c|c|c|}
		\hline
		\textbf{RF Source} & \textbf{References} & \textbf{Year} & \textbf{Min Cost} & \textbf{$d_{\text{F}}$ } & \textbf{$d_{\text{B}}$} &\textbf{$d_{\text{B}}^*$}& \textbf{Max Rate} & \textbf{Feature} \\

		\hline
		\multirow{3}{1.8cm}	{~~~~DRFS} &\cite{mbc1}&2017&20.4 $\mu$W& 20 m& 20 m& ---& ---&Tunnel diode \\

		\cline{2-9}  &\cite{mbc2}&2022&3004 $\mu$W& 700 m& 700 m& ---&3 Kbps & AllSpark \\

		\cline{2-9}  &\cite{mbc3}&2023& 6 $\mu$W& 300 m& 300 m&--- & 2 Kbps& Underwater \\
		
		\hline
		
		\multirow{3}{1.8cm}{~~~~~~TV} & \cite{tv} &2013& 0.79 $\mu$W& 6.5 mi &2.5 ft &  0.5 ft & 10 Kbps &---\\

		\cline{2-9}& \cite{tv1} &2014& 422 $\mu$W & ---  &---&   7 ft & 1 Mbps& $\mu$\textbf{mo}\\

		\cline{2-9}& \cite{fm4} &2019&  850  $\mu$W&---&9 m& 3.5 m& 10 Kbps& Glaze \\
		
		\hline

		\multirow{4}{1.8cm}	{~~~~~~FM}& \cite{fm} &2017& 11.07  $\mu$W &---& 60 ft  &5 ft & 3.2 Kbps&---\\

		\cline{2-9} & \cite{fm2} &2017&  677 $\mu$W  & 34 km  &5 m&   5 m& 2.5 Kbps&  ---\\

		\cline{2-9}  & \cite{fm4} &2019&  850  $\mu$W&---&10 m& 3.5 m& 10 Kbps& Glaze \\

		\cline{2-9}  & \cite{fm5} &2022& 150 $\mu$W & 4.7 km  & 19.6 m & --- & ---&Tunnel diode\\
		
		\hline
		
		\multirow{2}{1.8cm}{~~~~~LTE} & \cite{lte} &2021& 153 $\mu$W & 40 ft  & 320 ft & 3 ft & 13.63 Mbps &LSscatter \\

		\cline{2-9} & \cite{lte2} &2023& 123 mW & ---  & 36 m & --- &  400 bps &SyncLTE\\
		
		\hline
		
		\multirow{6}{1.8cm}{~~~~~WiFi }  & \cite{wifi} &2016 &  33  $\mu$W& 1 m & 54 m & 34 m& 300 Kbps& HitchHike \\

		\cline{2-9}  & \cite{wifi1} &2018 &  ---& 0.3 m & 14 m & 3 m& 50 Kbps& MOXscatter \\

		\cline{2-9}  & \cite{fm4} &2019&  850  $\mu$W&---&3.5 m& 3.5 m& 10 Kbps& Glaze \\

		\cline{2-9}& \cite{vms} &2021& 32  $\mu$W  & 0,9 m & 34.6 m & 1.5 m & 500 Kbps &VMscatter \\

		\cline{2-9} & \cite{wifiscatter} &2021& 30  $\mu$W  & 5 m&   28 m&---& 500 Kbps &SyncScatter \\

		\cline{2-9} & \cite{wifiscatter1} &2022&  613 $\mu$W  &0.2m&   20 m&---&10 Mbps &SubScatter \\
		
		\hline
		
		\multirow{6}{1.8cm}{~~~~~BLE} & \cite{FR} &2017 & 32 $\mu$W & 1 m  & 12 m & 10 m & 50 Kbps & FreeRider\\

		\cline{2-9} & \cite{bluetooth} &2019 & 198 $\mu$W & 6 m  & 6 m & 1 m & 1 Mbps & NeuroDisc \\

		\cline{2-9} & \cite{blue1} &2020&  37 $\mu$W  &---&   56 m & 1 m &16.6 Kbps& RBLE \\

		\cline{2-9} & \cite{blue3} &2020&  279.5 mW  &0.8 m&   22 m & 2 m &278.4 Kbps& Multiscatter \\

		\cline{2-9} & \cite{blue4} &2020&  77.4 mW  &0.5 m&   23 m & 1 m &662 Kbps& Gatescatter \\

		\cline{2-9} & \cite{blue2} &2021&  53.527 mW  &0.5 m &   20 m & 2m &8.275 Kbps& IBLE \\
		
		\hline
		
		\multirow{3}{1.8cm}{~~~~~Zigbee} & \cite{FR} &2017 & 32 $\mu$W & 1 m  & 22 m & 12 m & 14 Kbps & FreeRider\\

		\cline{2-9} & \cite{blue3} &2020&  279.5 mW  &0.8 m&   20 m & 2 m &26.2 Kbps& Multiscatter \\

		\cline{2-9} & \cite{blue4} &2020&  77.4 mW  &0.5 m&   27 m & 1 m &222 Kbps& Gatescatter \\
		
		\hline
		
		\multirow{5}{1.8cm}{~~~~~~Lora}& \cite{lora} &2017& 9.25 $\mu$W &5 m&   2.8 km & --- &37.5 Kbps& --- \\

		\cline{2-9} & \cite{lora1} &2018& 2.591 mW &---&   1.1 km & 300 m&6.25 Kbps& PLora \\

		\cline{2-9} & \cite{lora2} &2020& 0.3 mW & 1 m&   250 m & 50 m&199.4 Kbps& ALoba \\

		\cline{2-9} & \cite{lora3} &2021& 320 $\mu$W & 1 m&   2.2 km & ---&11.27 Kbps& P$^2$Lora \\

		\cline{2-9} & \cite{lora4} &2022& 93.2 $\mu$W & 10 m&   180 m & 10 m&19.6 Kbps& Saiyan \\
		
		\hline
		
	\end{tabular}
\end{table*}

\subsection{Communication Range}

The communication range is one of the critical metrics for evaluating the performance of wireless systems, playing a significant role in their advancement and implementation. While an extensive communication range is crucial for applications like outdoor IoT, achieving this goal in BackCom systems presents unique challenges due to signal attenuation over distance, a phenomenon more pronounced in BackCom than in traditional active communication systems \cite{bcmag1}.

Specifically, in the BackCom, the tag must first receive the signal from the RF source and then reflect it to the receiver. This process leads to a double-fading effect that considerably increases path loss. Understanding and mitigating this double-fading effect is vital in extending the practical range of BackCom systems. Therefore, when discussing the range issue in the BackCom system, it is essential to recognize that it involves two distinct distances \cite{bcoverview1}. The first is the distance from the RF source to the tag (forward distance, $d_{\text{F}}$), which determines the area within which the tag can be freely deployed while simultaneously receiving the carrier and harvesting energy for its circuit consumption. The second is the distance from the tag to the reader (backward distance, $d_{\text{B}}$), which dictates the range at which the receiver can reliably decode the backscattering data. These distances are interrelated, as the total path loss is a function of attenuation across both paths \cite{bcs2}.

Based on recent research, Table \ref{t3} offers a comprehensive comparison of communication distances achievable by various RF sources to provide a clearer understanding of these dynamics. This table highlights the theoretical possibilities and sheds light on practical implementations in real-world BackCom systems. While summaries in \cite{bcmag1,bcoverview1,bcs2} cover this topic, they may not fully encapsulate the latest advancements in the field. This comprehensive perspective is crucial for researchers and practitioners to bridge the gap between theoretical research and practical application. Furthermore, the impact of different modulation schemes on communication range, as discussed in \cite{modulation}, also plays a vital role in enhancing the overall efficiency and effectiveness of BackCom systems.

\subsection{Interference Environment}

\begin{figure*}[t]
	\centerline{\includegraphics[width=6.0in]{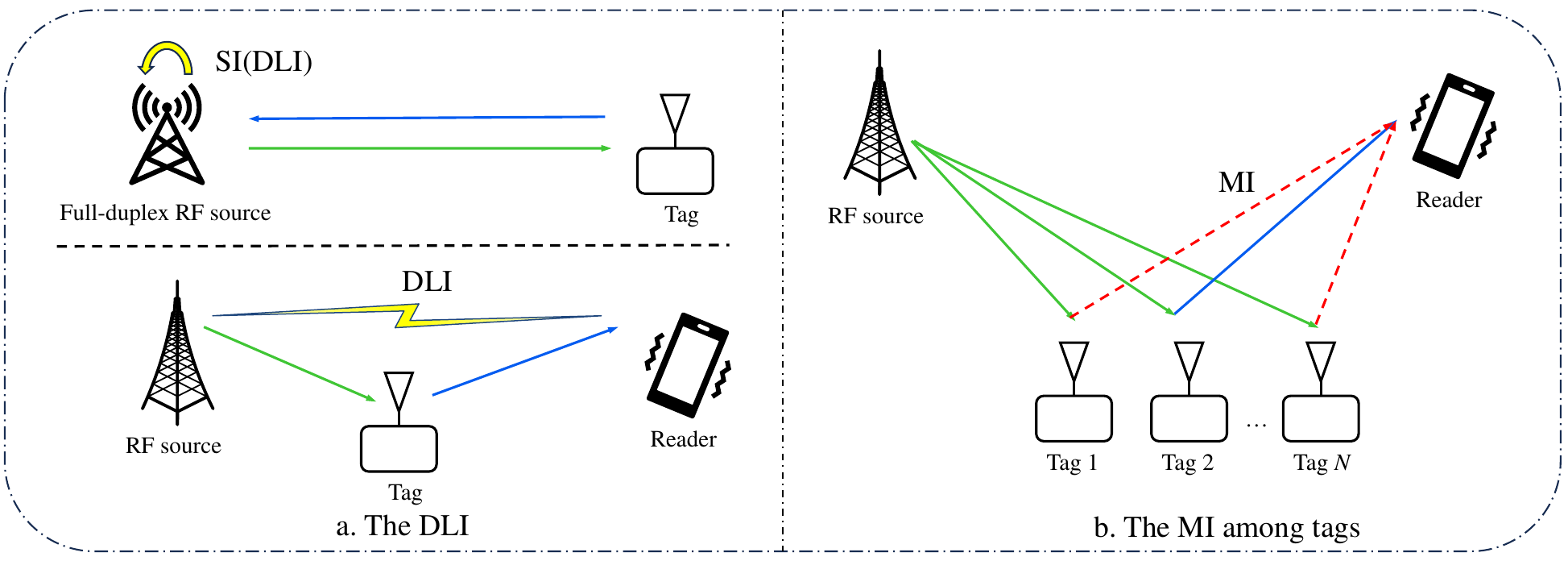}}
	\caption{The complex interference of BackCom.}
	\label{fig6}
\end{figure*} 

The interference, often viewed as a common adversary in wireless communication systems, has the potential to cause substantial degradation in the system performance, which holds particularly true for BackCom systems, where the nature of signal reflection and passive communication amplifies the challenge posed by the interference. In particular, complex interference can significantly hinder their operation. This interference primarily stems from two primary sources that disrupt the delicate balance of signal processing in BackCom systems, as shown in Fig. \ref{fig6}.

\begin{itemize}
	
	\item First, the strong direct-link (DL) signals from the main power source or other external RF sources can create substantial interference. This type of interference is particularly problematic when these strong DL signals directly collide with the weak backscattered signal at the reader. Such DLI can potentially dominate over the desired backscattered signal, leading to a drastic reduction in the signal clarity and readability in the BackCom system. The overpowering nature of the DLI is a critical issue, as it can significantly diminish the effectiveness of the BackCom system, making it challenging to extract valuable data from the backscattered signals. In MoBC systems, since the RF source and reader are integrated, the DLI is also considered as the self-interference (SI).
	
	\item Second, the scenario gets further complicated in environments with multiple tags. When multiple tags operate simultaneously, the MI becomes a significant concern that cannot be overlooked \cite{TC}. This type of interference arises from the simultaneous transmissions of multiple tags. The MI issue is especially pronounced in dense tag environments, where the interplay of multiple backscattered signals creates a convoluted signal environment. When these tags are in close proximity to each other, their backscattered signals can blend, making it difficult for the reader to differentiate and decode them accurately. This blending of signals reduces the accuracy of data retrieval and increases the complexity of signal processing at the reader. This issue is exacerbated when the backscattered signals from different tags have similar strengths or frequencies, further complicating the task of signal separation and identification.
	\end{itemize}}

\section{The Solutions for Direct-link Interference}

{\color{black} Recall that effectively tackling the complex interference is crucial for the sustainability and advancement of BackCom systems. This section is dedicated to exploring various innovative strategies to resolve the challenges the DLI poses. Our goal is to furnish a thorough overview, shedding light on the principles and benefits of each approach while also considering their real-world limitations and hurdles. With this comprehensive analysis, we seek to equip engineers, researchers, and other professionals in the communications field with a lucid understanding of the prevailing solutions for the DLI.

 \subsection{Successive Interference Cancellation}

\subsubsection{Technical Merit}
Successive interference cancellation (SIC) is a pivotal technique in BackCom systems for mitigating the DLI. Its wide adoption, as applied in various studies \cite{ SIC1, SIC3, SIC4}, is a testament to its effectiveness and versatility. SIC is particularly effective in isolating and nullifying interference step-by-step, making it a critical component in complex communication environments. Innovative applications of SIC, such as the low-noise carrier scheme in \cite{ESIC0a}, the front-end architecture in \cite{ESIC0}, the use of sinusoidal RF carriers in \cite{ESIC1}, the differential Chaos shift keying (DCSK) signal approach in \cite{ESIC2},  the direct sequence spread spectrum (DSSS) technique in \cite{ESIC2a}, and the signal reconstruction method in \cite{ESIC3}, highlight its versatility and adaptability. These implementations underscore the capacity of SIC to employ various sophisticated signal-processing mechanisms to mitigate interference effectively.

\subsubsection{Benefits}
\begin{itemize}
	\item Effective DLI Mitigation: The primary advantage is its capacity to significantly reduce carrier leakage and interference, thereby enhancing the clarity and reliability of BackCom systems.
	\item Diverse Implementation Strategies:  The variety of SIC applications, from innovative carrier schemes to complex architectures, demonstrates adaptability to different system requirements.
	\item Advancements in Signal Processing: The advanced techniques employed in SIC, like automatic gain control and phase-locked loops, mark significant progress in interference cancellation technology. These advancements highlight the critical role of SIC in evolving BackCom communication strategies.
\end{itemize}

\subsubsection{Challenges and Limitations}
\begin{itemize}
	\item Residual Interference: A common challenge with SIC is potential residual interference, possibly compromising system performance due to limitations in accurately isolating and canceling interference.
	\item Complexity in Implementation: The sophistication required in the signal processing of SIC, as in \cite{ESIC0a}, \cite{ESIC0}, and \cite{ESIC1}, can be challenging in terms of system design and resource allocation. This complexity necessitates high technical expertise and precision in system configuration.
	\item Precision in System Configuration: The effectiveness of SIC depends on precise system configurations, as seen in waveform design \cite{ESIC2} and DSSS techniques \cite{ESIC2a}.
	\item Dependence on Accurate Signal Reconstruction: Comprehensive signal processing operations, like those proposed in \cite{ESIC3}, demand high technical accuracy and computational resources, impacting system capabilities and stability.
\end{itemize}

\subsubsection{Case Study}

In a recent proof of concept, researchers constructed a prototype system to demonstrate the efficacy of passive DSSS in realistic communication environments \cite{ESIC2a}. The system gateway was implemented using an NI USRP 2922, while the passive DSSS receiver was built with commercial off-the-shelf hardware components. Remarkably, the receiver operated on a mere 166.5 $\mu$W power while robustly demodulating DSSS signals over a 1 MHz bandwidth.

This case study serves as a practical demonstration of the application of SIC in enhancing the communication efficiency, particularly in energy-constrained environments. A critical aspect of this study was evaluating the prototype in environments with RFID and LoRa interference. The researchers conducted extensive stress-testing experiments to determine the system performance regarding the bit error rate under various noise and interference levels. These tests revealed a significant signal-to-interference-plus-noise ratio (SINR) improvement of 16.5 dB for the passive DSSS system compared to conventional receivers used in backscatter devices. This case study demonstrates the technical feasibility and efficiency of passive DSSS systems and highlights their practical application in enhancing wireless communication.

\subsection{Opportunistic SIC}

\subsubsection{Technical Merit}
Opportunistic SIC (OSIC), as applied in BackCom systems, represents a nuanced approach to interference cancellation. This technique, explored in \cite{OSIC}, hinges on using previously decoded signals at the receiver to mitigate residual interference. OSIC demonstrates a deep understanding of signal processing challenges in environments where the characteristics of the ambient RF signals are uncertain or variable. The technical merit lies in its adaptive approach to interference cancellation, tailoring the process based on the quality of the received signals.

\subsubsection{Benefits}

\begin{itemize}
	\item Enhanced Interference Mitigation: OSIC effectively reduces residual interference in systems where RF signal characteristics are known, such as in MoBCs, BiBCs, and some cooperative AmBCs.
	\item Improved System Performance: Utilizing decoded signals opportunistically for interference cancellation enhances signal clarity and reliability, boosting overall performance.
	\item Adaptability to Signal Quality: The adaptability of OSIC to varying signal qualities makes it useful in rapidly changing signal conditions.
\end{itemize}

\subsubsection{Challenges and Limitations}
\begin{itemize}
	\item Dependence on Signal Quality: The efficacy of OSIC heavily relies on the quality of decoded signals. Poor signal quality can significantly diminish its interference cancellation capability.
	\item Complexity in Implementation:  Implementing OSIC requires sophisticated signal processing at the receiver, posing challenges in certain environments.
	\item Variability in Performance: Given its dependence on received signal quality, the performance of SIC can be inconsistent in environments with high signal variability, leading to unpredictability in effectiveness.
\end{itemize}

\subsection{Orthogonal Frequency Division Multiplexing}

\subsubsection{Technical Merit}
Orthogonal frequency division multiplexing (OFDM) in BackCom systems represents a significant stride in addressing the challenge of the DLI. The utilization of this technique, as seen in the three all-digital architectures introduced in \cite{OFDM0}, demonstrates an innovative application of OFDM principles in backscatter modulators. These architectures utilize RF switches and discrete loads to implement digitally controlled single-sideband OFDM backscatter modulators, signifying an advanced level of control and manipulation of signal properties. Additional strategies, such as exploiting the cyclic prefix in OFDM \cite{OFDM1, OFDM1a} and partial spectrum shifting \cite{OFDM2}, further underscore the adaptability and potential of OFDM in diverse signal environments. The technical merit lies in its ability to utilize the inherent properties of OFDM, such as frequency orthogonality and spectral efficiency, to mitigate the DLI.

\subsubsection{Benefits}
\begin{itemize}
	\item Effective DLI Mitigation: The structure and characteristics of OFDM make it an effective tool for mitigating the DLI in BackCom systems. Techniques like exploiting the cyclic prefix and partial spectrum shifting can significantly reduce interference.
	\item Advancements in Signal Processing: The development of architectures that use digitally controlled OFDM modulators demonstrates an advanced level of signal processing, offering more precise control over backscattered signals.
\end{itemize}

\subsubsection{Challenges and Limitations}

A significant limitation of these OFDM-based techniques is their heavy reliance on the waveform of the carrier signal. This reliance can restrict their effectiveness in scenarios where the carrier signal has irregular or incompatible waveform characteristics. Therefore, these methods may not be universally applicable across all BackCom systems, particularly in environments with non-standard or highly variable carrier signals.

 \subsection{Frequency Shift} 
 
\subsubsection{Technical Merit}
Frequency shift (FS) is a powerful technique for interference mitigation in the context of BackCom systems. This approach, initially introduced in \cite{FS1}, has evolved and gained recognition in the field, as evidenced by various studies including \cite{FS2, FS4a, FS5, FS6}. The core concept of FS involves each tag shifting its carrier signal to distinct, non-overlapping adjacent frequency bands. This method showcases a strategic use of the frequency domain to tackle the challenge of interference, a fundamental issue in wireless communication systems. The technical merit lies in its straightforward yet effective nature to segregate signals, thereby mitigating the interference without the need for complex signal processing techniques.

\subsubsection{Benefits}

\begin{itemize}
	\item Effective Interference Mitigation: One of the primary benefits of FS is its ability to segregate backscattered signals from DL signals effectively. This segregation directly addresses the interference issue, enhancing the clarity and reliability of communication.
	\item Simplicity in Implementation: Unlike methods that require complex signal processing, FS offers a relatively more straightforward approach to mitigate the interference. Its implementation mainly involves adjustments in the frequency domain, which can be easier to manage than other techniques.
	\item Enhanced Signal Quality: By ensuring that signals from different tags do not overlap in the frequency domain, FS can significantly improve the quality of received signals, leading to a better overall system performance.
\end{itemize}

\subsubsection{Challenges and Limitations}
\begin{itemize}
	\item Spectrum Resource Consumption: FS relies on the availability of ``white space" or interference-free spectrum segments. This reliance can lead to increased spectrum resource consumption, a critical consideration in limited spectrum availability. In densely populated frequency environments, finding sufficient white space for FS may be challenging, limiting its applicability.
	\item Potential Compromise on Spectrum Efficiency: The segmented utilization of available bandwidth, inherent in FS, may reduce overall spectrum efficiency. This is particularly relevant in scenarios where bandwidth is at a premium, and efficient utilization is paramount.
	\item Energy Requirements: Despite its relatively low energy consumption, FS requires some energy to shift frequencies, which poses a challenge for passive tags typically found in BackCom systems. These tags, designed to operate without an internal power source, might need small-sized batteries or alternative energy-harvesting methods to support the energy demands of FS.
\end{itemize}

\subsubsection{Case Study}

In \cite{FS1}, researchers investigated the interference between FS-Backscatter technology and active radio traffic, particularly Wi-Fi, in the congested 2.4GHz ISM band. This research is crucial for understanding how different communication technologies coexist in the same frequency band.

The first experiment focused on the impact of Wi-Fi interference on FS-Backscatter. The results showed varying FS-Backscatter throughput depending on the proximity of Wi-Fi sources and illustrated the practical implications of signal interference in real-world environments. The second experiment assessed the impact of FS-Backscatter on ongoing Wi-Fi transmissions. These results indicated that FS-Backscatter had a limited interference range and did not significantly impact Wi-Fi throughput.

This case study underscores the importance of considering the dynamic interaction of FS-Backscatter with other wireless technologies in shared spectral environments. It demonstrates that while FS-Backscatter technology can coexist with existing Wi-Fi networks, its proximity and the relative power of signals are crucial factors in the extent of mutual interference. These insights are vital for optimizing the deployment of FS-Backscatter systems to minimize their impact on existing wireless networks.
 
 \subsection{Polarization Conversion} 
 
 \subsubsection{Technical Merit}
Polarization conversion (PC) represents a significant advancement in tackling the DLI in BackCom systems \cite{PC2, PC3}. This technique leverages the inherent properties of electromagnetic waves, particularly their ability to exist in various polarization states such as vertical, horizontal, and circular. By dynamically switching between these polarization states, PC-based approaches can distinctly segregate backscattered signals from the DLI. The innovation of PC lies in its use of physical layer properties to address a pervasive communication challenge, showcasing a deep understanding of electromagnetic wave behavior and its practical application in wireless communications.

\subsubsection{Advantages}

\begin{itemize}
	\item  Effective Interference Mitigation: The primary advantage of PC is its ability to effectively mitigate the DLI, which is a significant challenge in BackCom systems. This is achieved by exploiting the different polarization states to separate the desired signal from the interference.
	\item  Enhanced Signal Clarity: By segregating the backscattered signals from the  DLI, PC can significantly improve the clarity and quality of the received signals, leading to a better communication performance.
	\item Versatility: The use of polarization-based reconfigurable antennas, as studied in \cite{PC2}, demonstrates the versatility of PC approaches. These systems can adapt to different environmental conditions and signal characteristics, making them suitable for various application scenarios.
\end{itemize}

\subsubsection{Challenges and Limitations}
\begin{itemize}
	\item  Dependency on Receiving Antenna: The effectiveness of PC is highly dependent on the ability of the receiving antenna to match the ambient signal polarization. This requirement can limit the PC applicability, especially in scenarios where the antenna of the passive tag cannot be easily modified or controlled.
	\item Complexity in Implementation: Implementing PC, especially in systems like the dual-polarized receiving antenna proposed in \cite{PC3}, can add complexity to the system design. This complexity arises from the need for additional hardware to manage polarization switching and conversion.
	\item Environmental Sensitivity: PC approaches might be sensitive to environmental factors that can alter or affect the polarization state of the signals, potentially impacting the reliability of the interference mitigation process.
\end{itemize}

 \subsection{Multi-Antenna Technique}

\subsubsection{Technical Merit}
Deploying multi-antenna techniques in BackCom systems marks a significant advancement in addressing SINR challenges. This approach leverages the spatial diversity offered by multiple antennas to enhance system performance. Methods like diversity reception, beamforming, and spatial multiplexing employed in these systems are well-established in wireless communications for their ability to improve signal quality and reduce interference. The technical merit of multi-antenna techniques lies in their ability to exploit spatial properties of electromagnetic waves, enabling more robust and efficient communication, as evidenced by studies cited in \cite{MAR0, MAR1, MAR2, MAR2a, MAR2b}.

\subsubsection{Benefits}
\begin{itemize}
	\item Enhanced SINR: The primary benefit of multi-antenna techniques is the significant improvement in SINR. By using multiple antennas, these systems can effectively enhance the desired signal while suppressing interference, particularly the DLI.
	\item Increased System Performance: Techniques such as diversity gain and beamforming can considerably improve the overall performance of BackCom systems, leading to better communication quality.
	\item Adaptability: Multi-antenna systems offer adaptability in various environmental conditions, as they can dynamically adjust beamforming weights for signal reception and interference mitigation.
\end{itemize}
  
 \begin{table*}
 	\renewcommand{\arraystretch}{1.3}
 	\centering
 	\color{black}
 	\small
 	\caption{The solution summary for DLI in BackCom}
 	\label{t4}
 	\begin{tabular}{|>{\centering\arraybackslash}m{2.5cm}|>{\centering\arraybackslash}p{2.0cm}|>{\centering\arraybackslash}m{1.7cm}|>{\centering\arraybackslash}m{4.6cm}|>{\centering\arraybackslash}m{4.7cm}|}
 		\hline
 		\textbf{Solutions} & \textbf{Implementation Medium} & \textbf{References} & \textbf{Main Advantages} & \textbf{Main Challenges} \\
 		\hline
 		SIC & Signal processing & \cite{SIC1,SIC3,SIC4,ESIC0,ESIC0a,ESIC1,ESIC2,ESIC2a,ESIC3} & Effectively reduces interference through sequential isolation; Diverse implementation strategies; Advanced signal processing & Potential residual interference; Complexity in implementation; Precision in system configuration \\
 		\hline
 		OSIC & Signal processing & \cite{OSIC} & Builds upon SIC; Adapts to signal quality variations; Utilizes decoded signals for mitigation & Dependence on signal quality; Complexity in implementation; Performance variability \\
 		\hline
 		OFDM & Digital signal processing & \cite{OFDM0,OFDM1,OFDM1a,OFDM2} & Leverages frequency orthogonality; Effective DLI mitigation; Advanced control over signal properties & Reliance on carrier signal waveform; Limited applicability in certain environments \\
 		\hline
 		FS & Frequency domain & \cite{FS1,FS2,FS4a,FS5,FS6} & Simplifies interference mitigation; Strategic use of frequency bands; Enhances signal clarity & Increased spectrum resource consumption; Potential compromise on spectrum efficiency; Energy requirements for passive tags \\
 		\hline
 		PC & Electromagnetic wave properties & \cite{PC2,PC3} & Utilizes varying polarization states; Effective in segregating signals; Enhances signal clarity & Dependency on receiving antenna's ability; Complexity in implementation; Environmental sensitivity \\
 		\hline
 		Multi-antenna Techniques & Antenna technology & \cite{MAR0, MAR1, MAR2, MAR2a, MAR2b} & Utilizes spatial diversity; Enhances SINR; Improves system performance & Hardware complexity; Increased system cost; Limitations in single antenna settings \\
 		\hline
 	\end{tabular}
 \end{table*}
 
  \subsubsection{Challenges and Limitations}
\begin{itemize}
	\item Hardware Complexity: Implementing multi-antenna techniques often requires sophisticated and advanced hardware configurations. This complexity can be a limiting factor, especially in cost-sensitive deployments or scenarios where space constraints prevent using multiple antennas.
	\item Increased System Cost: The need for additional antennas and the associated signal processing hardware can lead to increased costs in the initial setup and ongoing maintenance.
	\item Limitations in Single Antenna Settings: In environments where only a single antenna is feasible, the benefits of multi-antenna techniques cannot be realized. This limitation can be significant in compact or portable devices where adding multiple antennas is impractical.
\end{itemize}
 
 \subsection{Summary and Comparative Analysis}
 
 \subsubsection{Summary of DLI Solutions}
This section has comprehensively reviewed various DLI solutions in BackCom systems, covering SIC, OSIC, OFDM, FS, PC, and multi-antenna techniques. Each method is distinct in its approach: SIC and OSIC focus on sequential interference reduction and adaptability to signal variations, OFDM utilizes frequency orthogonality for the effective DLI mitigation, FS simplifies interference resolution by frequency band separation, PC uses different polarization states for signal segregation, and multi-antenna techniques enhance the system performance through spatial diversity. While these solutions vary in technical complexity, resource demands, and environmental adaptability, they collectively offer a range of strategies for enhancing the signal clarity and overall system performance in combating the DLI.

\subsubsection{Comparative Assessment of Solution Categories} 

Table \ref{t4} provides a comparative assessment of the DLI solutions discussed, offering a concise overview of each technique, including its type, primary references, advantages, and challenges. This comparative analysis serves as a quick reference guide and aids in understanding the unique features and potential limitations of each solution, thereby facilitating informed decision-making for further research and practical implementation in BackCom systems.}

\section{The solutions for Mutual Interference}

{\color{black}In this section, we venture deep into the complex realm of MI in BackCom systems, a pivotal factor that profoundly influences the efficiency and dependability of wireless communication networks. Our exploration in this domain thoroughly examines a diverse array of mitigation strategies, meticulously offering a nuanced perspective on their potential benefits and inherent limitations. By methodically dissecting these strategies and closely looking at the challenges associated with their implementation, our objective is to present a nuanced landscape of MI mitigation, laying a solid foundation for the evolution of current technologies and inspiring future innovations in BackCom systems.}

 {\color{black} \subsection{Multiple Access} 
  
One solution lies in adopting various multiple access strategies, each uniquely tailored to minimize the interference while maximizing the capacity and performance of the system. In the following subsections, we delve into the specifics of several essential multiple access methods such as orthogonal frequency division multiple access (OFDMA),  time division multiple access (TDMA),  space division multiple access (SDMA), and multiple subcarriers multiple access (MSMA). Each of these strategies offers a distinct approach to managing the shared communication medium, and their careful examination reveals the intricate balance between system complexity, efficiency, and practical feasibility.

\subsubsection{OFDMA}

\begin{itemize}
	
	\item \textit{Technical Merit}:
	OFDMA is an enabled technique designed to mitigate MI in scenarios involving multiple tag backscattering in BackCom systems \cite{OFDMA0,OFDMA1,OFDMA2,OFDMA3}. This technique enhances communication efficiency by allocating distinct frequency bands to the backscattered signal of each tag. The key strength of OFDMA lies in its ability to efficiently use the frequency spectrum, allowing multiple tags to communicate simultaneously without interfering with each other. By dividing the available bandwidth into orthogonal subcarriers, OFDMA ensures that signals are separated in the frequency domain, thereby avoiding the interference in overlapping channels. 
	
	\item \textit{Challenges and Limitations}: 
	The primary challenge of employing OFDMA in BackCom systems is its dependence on the availability and allocation of frequency bands. The effectiveness of this technique is directly tied to the number of available frequency bands within the system and the spectral efficiency of these bands. In environments with limited spectral resources, efficiently implementing OFDMA can be challenging. Besides, there are concerns about system complexity, as managing and maintaining orthogonal frequency allocations in a dynamic communication environment can be technically demanding. These challenges necessitate advanced signal processing capabilities and sophisticated system design to ensure the effective and efficient operation of OFDMA in BackCom systems.
	
\end{itemize} 

\subsubsection{TDMA}

\begin{itemize}
	
	\item \textit{Technical Merit}:
	TDMA is a widely utilized strategy in BackCom systems to facilitate interference-free communication among multiple tags, as referenced in \cite{TDMA4,TDMA5,TDMA6}. Its primary function is to divide the transmission period into discrete time slots, allocating each tag a specific slot for backscattering its signal. This organized structure ensures that only one tag transmits at a given time slot, effectively preventing signal collisions and MI. The ability of TDMA to schedule transmissions in a sequential and non-overlapping manner enhances both the flexibility and reliability of communication within BackCom systems, making it a crucial mechanism for managing interference-free data transmissions. 
	
	\item \textit{Challenges and Limitations}: 
	Despite its benefits, TDMA faces several challenges. A key issue is the need for precise time synchronization among the tags to ensure the correct alignment of time slots. This requirement adds a layer of complexity to the system and can be challenging in environments where tags have limited processing capabilities or external factors affect the timing accuracy. In addition, variable latency caused by the time slot division can be problematic in latency-sensitive applications, where delays in data transmission can affect the overall system performance. Moreover, with fixed time slot durations, TDMA can lead to inefficiencies and potential transmission delays, especially in scenarios involving a large number of tags or in systems with extended time slot durations. These limitations necessitate careful planning and optimization of time slot allocation to balance the need for efficient communication subject to the constraints of the TDMA structure.

\end{itemize}

\subsubsection{SDMA} 

\begin{itemize}
	
	\item \textit{Technical Merit}:
	SDMA is another significant technique for enabling concurrent transmissions in BackCom systems \cite{SDMA1}. The underlying principle of SDMA involves partitioning the available space into several sectors or beams, each dedicated to a distinct tag. This division allows for concurrent communication within the same frequency band. This allocation creates individual spatial communication ``lanes'' for each tag, thus allowing for concurrent communication within the same frequency band. SDMA employs directional antennas to transmit signals to separate sectors to accomplish this spatial separation, consequently reducing the MI among tags. 
	
	\item \textit{Challenges and Limitations}: 
	However, the advantages of SDMA come with their own set of challenges. Directional antennas, which focus on the signal in a specific direction, necessitate advanced beamforming and tracking techniques to steer the signal precisely toward individual tags. This requirement might be a limiting factor in environments where the orientation of the tags changes frequently or is unpredictable. While these techniques are vital for maintaining efficient communication within the assigned sectors, they concurrently add to the complexity of system design and implementation. Furthermore, the coverage area of SDMA systems is generally more confined than systems deploying omnidirectional antennas. The effectiveness of SDMA heavily depends on the deployment of sectorized antennas and the extent of the coverage area of each sector, resulting in coverage limitations. 
	
\end{itemize} 

\subsubsection{MSMA}

\begin{itemize}
	
	\item \textit{Technical Merit}:
	MSMA has gained traction as a promising technique, facilitating concurrent data transmission from multiple tags \cite{MSMA1}. The distinguishing feature of MSMA is the use of dedicated subcarriers produced by digitally controlled RF switches, enabling each tag to modulate its backscattered signal. This strategy effectively increases the transmission efficiency by ensuring that each tag utilizes its unique subcarrier for data transmission, thereby optimizing the spectrum usage. This bypasses the need for complex hardware configurations or intricate scheduling protocols. 
	
	\item \textit{Challenges and Limitations}: 
	Despite its benefits, implementing multiple subcarriers in MSMA can potentially lead to the MI between the subcarriers utilized by the tags and the reader. This necessitates the implementation of appropriate signal processing techniques to mitigate the interference. For instance, in \cite{MSMA2, MSMA3}, MSMA was employed with advanced signal processing algorithms to reduce the interference, demonstrating its practical viability effectively. Further, akin to TDMA, MSMA requires accurate time synchronization among the tags to ensure coordinated and seamless concurrent transmissions.
	
\end{itemize} 

\subsection{Tag Screening}

\subsubsection{Technical Merit}
Tag screening in BackCom systems represents a strategic approach to managing the interference resulting from simultaneous transmissions. The core techniques, namely tag selection and tag clustering, exhibit a sophisticated understanding of network dynamics. Tag selection leverages performance-based criteria to optimize the transmission efficiency \cite{TS1, TS2, TS3, TS4}, while tag clustering utilizes the inherent similarities and correlations among devices to facilitate coexistence and simultaneous transmissions \cite{UC1, UC2, UC3}. These methodologies demonstrate a high level of technical innovation in addressing the complex challenge of interference in dense network environments.

\subsubsection{Advantages}

\begin{itemize}
	\item Efficiency in Transmission: By selectively choosing tags for backscattering, tag screening effectively reduces the potential for interference, thereby enhancing the overall efficiency of the network.
	\item Improved Network Management: Tag clustering simplifies the management of multiple devices by grouping compatible tags together. This organization streamlines communications and maximizes the use of available resources.
	\item Optimized Resource Utilization: Both tag selection and clustering contribute to more effective utilization of time slots and frequency resources, which is crucial in high-density BackCom environments.
\end{itemize}

\subsubsection{Challenges and Limitations}
\begin{itemize}
	\item Potential for Exclusion: The tag selection process may inadvertently exclude certain tags from backscattering, potentially leading to issues with network fairness and reduced throughput for some devices.
	\item Coordination Complexity: Implementing tag clustering requires significant coordination and control overhead, particularly for inter- and intra-cluster communications, adding to the operational complexity of the system.
	\item Dynamic Environment Adaptability: The effectiveness of tag clustering can be compromised in environments where the correlation and similarity between devices frequently change, posing a challenge for maintaining a consistent performance.
	\item Spatial Distribution Impact: Additional considerations, such as the spatial distribution of tags, play a crucial role in the overall effectiveness of tag screening strategies. Studies have shown that factors like node density and cluster size can significantly influence the communication quality, necessitating careful planning and implementation of these strategies.
\end{itemize} 	 	 	

\subsection{FreeRider}

\subsubsection{Technical Merit}
FreeRider, as introduced in \cite{FR}, represents a significant technical innovation in BackCom systems. Its core technique, packet length modulation, where binary values are encoded as the lengths of the excitation packets, is a creative departure from traditional binary encoding schemes. This method showcases a novel use of the physical layer capabilities in BackCom, allowing for a more nuanced and flexible approach to data transmission. The technical sophistication of FreeRider lies in its ability to handle concurrent transmissions from multiple tags by manipulating packet lengths, thereby introducing a new dimension to communication strategies in wireless networks.

\subsubsection{Advantages}
The primary advantage of FreeRider is its capacity to facilitate simultaneous decoding of signals from multiple tags, which is a crucial requirement in dense network environments. This capability potentially enhances network throughput and efficiency, allowing more data to be transmitted and processed within the same time frame. By employing a unique modulation technique, FreeRider can also mitigate some common issues in BackCom systems, such as signal interference and collision, offering a more reliable communication channel. This approach is particularly beneficial in applications where traditional binary encoding may not be sufficient.

\subsubsection{Challenges and Limitations}
Despite its innovative approach, FreeRider faces several challenges and limitations. The most significant one is the risk of packet collisions in scenarios where multiple tags transmit simultaneously. Managing these collisions is critical to maintaining the integrity and performance of the network but adds a layer of complexity to the system. The need to control message transmission to coordinate the tags' activities further complicates the system design and increases computational demands. Besides, the reliance on packet length modulation may limit the types of data that can be communicated, particularly in scenarios requiring high data integrity or strict timing constraints. These limitations suggest that while FreeRider presents a promising approach, its practical application might be best suited to specific scenarios where its unique advantages can be fully leveraged.

\textbf{\textit{Discussion}}: \textit{Centralized access methodologies, like the ones previously discussed, necessitate the centralized management and allocation of wireless resources. These techniques manage resource usage among users, providing a degree of adaptability and flexibility that accommodates a broad spectrum of network requirements. However, these methodologies inherently demand dependence on centralized processing units or BSs for coordination and scheduling, inadvertently increasing the signaling overhead. Besides, coordinating activities, especially among passive tags, introduces challenges due to their inherent passive nature. Moreover, centralized systems are susceptible to single points of failure, with potential widespread repercussions, and limitations dictated by the network topology constrains them. The coverage range and network capacity can experience significant constraints in settings characterized by extensive scale and high-density networks.}

In response to these challenges associated with centralized access methodologies, decentralized access strategies have emerged as compelling alternatives in BackCom systems. Inherently, decentralized approaches do not necessitate a central processing unit or BS for managing and allocating wireless resources. As a result, these strategies substantially lower the signaling overhead and the potential risk associated with a single point of failure. Furthermore, decentralized systems are less hindered by the constraints of network topology, potentially affording a broader coverage and enhanced the network capacity. In what follows, some decentralized strategies are introduced and summarized.

\subsection{Modulation Silencing}

\subsubsection{Technical Merit} Modulation silencing (MS), as presented in \cite{MS1}, represents an innovative approach to mitigate the effects of tag collisions in BackCom systems. The primary mechanism of MS is its carrier collision detection feature, which automatically halts carrier transmission during collision events. This method is particularly beneficial when random access and distributed communication are essential.

\subsubsection{Advantages}

The key strength of MS lies in its ability to respond dynamically to collision scenarios, thereby improving the overall network efficiency and reducing the likelihood of data loss or corruption due to tag collisions.
By enabling random access, MS allows for a more flexible and responsive network, catering to a diverse array of communication requirements.

\subsubsection{Challenges and Limitations}

Implementing MS might necessitate additional hardware, such as specialized carrier collision detection circuits. This requirement could lead to an increased system complexity and higher costs, which might deter cost-sensitive applications.
Integrating this new hardware component poses challenges regarding system design, power consumption, and overall network scalability.

 \subsection{Buzz}
 
  \subsubsection{Technical Merit} Buzz, as detailed in \cite{CS1}, offers a novel solution for enabling the concurrent transmission of multiple tags in BackCom networks. It operates by allowing each tag to transmit within a small, randomly selected subset of collisions. This method generates a sparse and rateless code, which is efficiently decoded using compressive sensing techniques.
 
 \subsubsection{Advantages}
 
 Buzz significantly enhances the communication efficiency and reliability in BackCom networks. Its ability to handle multiple transmissions concurrently is a substantial improvement over traditional methods that often struggle with tag collisions.
 The use of compressive sensing for decoding contributes to the robustness of the network, particularly in environments with high tag densities.

 \subsubsection{Challenges and Limitations}
 However, the implementation of Buzz requires considerable hardware sophistication and computational resources. The need for advanced compressive sensing capabilities might lead to an increased system complexity and higher operational costs.
 The computational burden associated with decoding sparse and rateless codes could be substantial, particularly in large-scale networks with numerous tags and high traffic volumes.}

{\color{black}\subsection{Sampling-Based Technique}

	\subsubsection{Technical Merit}    Some sampling-based techniques, i.e., Laissez-Faire  \cite{LF}, backscatter spike train \cite{BST}, and BiGroup \cite{Bigroup}, were developed to grapple with the MI for BackComs.  These techniques effectively shift the decoding burden from tags to the reader, exploiting the higher sampling rate of the reader to distinguish signals from multiple tags. This is a smart use of available resources and reflects an understanding of the asymmetric capabilities in BackCom systems.
	
	\subsubsection{Practical Implications} By relying on the advanced capabilities of the reader, these methods can potentially enhance system efficiency and reduce the complexity and cost of the tags. This approach is particularly beneficial in scenarios where upgrading the reader is more feasible than upgrading multiple tags.

	\subsubsection{Challenges and Limitations} The heavy reliance on signal stability and distinctiveness in the time or in-phase or quadrature (I/Q) domain is a significant constraint. This dependence might limit the applicability of these techniques in environments with high signal variability or in scenarios where precise signal characteristics are difficult to maintain.

  \subsection{Signal State Modeling}
  \subsubsection{Technical Merit} 
  FlipTracer \cite{Fliptracer} and Hubble \cite{Hubble} were introduced to overcome the limitations associated with solely relying on stable signal features for decoding collided signals. FlipTracer introduced the One-Flip-Graph, a graphical model that captured the transition patterns of the signals, which provided detailed information about the similarity between combined states of the signals, enabling accurate identification of the states. On the other hand, Hubble introduced an interstellar travelling model to capture the bursty Gaussian process of collided signals, which combined burstiness in the time domain and the Gaussian property in the I/Q domain and facilitated the extraction of signal states and the tracing of state transitions.
  \subsubsection{Practical Implications}  The use of such advanced models could significantly improve the accuracy of signal decoding in dense environments. This improvement is crucial for applications requiring high data integrity.
  	\subsubsection{Challenges and Limitations} The complexity of these models might pose challenges in terms of computational requirements and implementation, particularly in resource-constrained environments. Moreover, the effectiveness of these models in diverse operational scenarios needs thorough validation.

\subsection{Time-Hopping Spread Spectrum}
  \subsubsection{Technical Merit}
The Time-hopping spread spectrum (THSS) technique, utilized in \cite{THSS1} for MoBC systems, aimed to minimize the interference and the boost the system reliability and performance by alternating between two TH-SS sequences, each harboring a single random non-zero chip, between the reader and the tag.
 \subsubsection{Practical Implications}  The use of THSS can potentially increase the resilience of BackCom systems to external interference and noise, making it suitable for deployment in crowded spectral environments.
\subsubsection{Challenges and Limitations} The challenges in synchronizing multiple tags and readers are significant, especially in large-scale deployments. Timing discrepancies can degrade the system performance. Besides, the expansive use of bandwidth might lead to the inefficient spectrum utilization and potential conflicts with other wireless systems.

\subsection{Other Random Access Methods}

 \subsubsection{Individual Tag Subframe Selection}
 \begin{itemize}
 	 \item \textit{Technical Merit}: This approach involves each tag individually selecting a subframe to backscatter the continuous wave signal to the reader \cite{RA1}. The reader employs SIC techniques to resolve resultant collisions. This method effectively addresses parallel interference issues.
 	\item \textit{Advantages and Challenges}: 
 		The strategy enables simultaneous transmissions from multiple tags, potentially increasing the system throughput. However, the reliance on SIC requires sophisticated signal processing capabilities at the reader, potentially increasing the system complexity.
 \end{itemize}

  \subsubsection{Channel Sensing with Dual-Backoff Mechanism} 
 \begin{itemize}
 	\item \textit{Technical Merit}: This method combines analog channel sensing with a dual-backoff mechanism, allowing each tag to switch between transmission, reception, and EH states based on the sensed channel state \cite{RA2}.
 	\item \textit{Advantages and Challenges}:  It provides a dynamic and responsive approach, enabling tags to adapt to the current channel conditions, potentially reducing collisions and improving the energy efficiency. However, the need for tags to continuously sense the channel and switch states can increase the operational complexity and energy consumption of tags. 
 \end{itemize}
 
  \subsubsection{Slotted Aloha Protocol}
  \begin{itemize}
 	\item \textit{Technical Merit}: Slotted Aloha is employed in \cite{RA3}, wherein each tag selects a time slot for backscattering its data based on a certain probability. This probabilistic approach aims to avoid significant collisions.
 \item \textit{Advantages and Challenges}:  It is a relatively simple protocol to implement and offers a fair chance for all tags to access the channel. Nowever, the probabilistic nature of the protocol can lead to the ineffective and unpredictable performance, especially in networks with a high density of tags.
  \end{itemize}

\textbf{\textit{Discussion}}: \textit{Although random access offers flexibility and simplicity, it does confront issues related to conflict, inefficiency, and limited control, thereby creating challenges in assuring the system stability and optimality. Due to the random selection and competition among tags, collisions and resource waste can occur, leading to imbalanced resource allocation. These problems may render random access less suitable for certain application scenarios, particularly for those that require more efficient and predictable resource allocation.} 
 
 \subsection{Summary and Comparative Analysis}
 
  \begin{table*}
 	\renewcommand{\arraystretch}{1.3}
 	\centering
 	\small 
 	\color{black}
 	\caption{The solution summary for MI in BackCom}
 	\label{t5}
	\begin{tabular}{|>{\centering\arraybackslash}m{2.5cm}|>{\centering\arraybackslash}p{2.0cm}|>{\centering\arraybackslash}m{1.7cm}|>{\centering\arraybackslash}m{4.6cm}|>{\centering\arraybackslash}m{4.7cm}|}
 		\hline
 		\textbf{Solutions} & \textbf{Implementation Medium} & \textbf{References} & \textbf{Main Advantages} & \textbf{Main Challenges} \\ \hline
 		\multirow{4}{2.5cm}{~~Multiple Access} & OFDMA &  \cite{OFDMA0,OFDMA1,OFDMA2,OFDMA3} & \multirow{2}{4.6cm}{~~~~Interference-free transmission} & Band allocation \& system complexity \\ 
 		\cline{2-3} \cline{5-5} & TDMA & \cite{TDMA4,TDMA5,TDMA6} & & Precise synchronization \& variable latency \\ 
 		\cline{2-5}& SDMA & \cite{SDMA1} & Spatial separation & Beamforming \& limited coverage \\ 
 		\cline{2-5} & MSMA & \cite{MSMA1,MSMA2,MSMA3} & Unique subcarrier per tag & Interference between subcarriers \\ \hline
 		\multirow{2}{2.5cm}{~~~Tag Screening} & Tag selection & \cite{TS1,TS2,TS3,TS4} & Picking one tag that best fits   &  Fairness issue \\ 
 		\cline{2-5} & Tag clustering &\cite{UC1,UC2,UC3} & Grouping compatible tags &  Coordination complexity \\ \hline
 		FreeRider & Packet length modulation & \cite{FR} & Concurrent transmission allowed & Packet collisions \& control transmission needs \\ \hline
 		MS & Collision detection & \cite{MS1} &Dynamically respond to collision & Additional hardware requirements \\ \hline
 		Buzz & Compressive sensing & \cite{CS1} & Manages multiple transmissions & Computationally intensive \\ \hline
 		\multirow{4}{2.5cm}	{~~~Sampling-Based\\~~~~ Techniques} & Laissez-Faire & \cite{LF} & \multirow{4}{4.6cm}{~~~Relies on reader's capabilities} & \multirow{4}{4.6cm}{~~~~~Limited by signal stability} \\ 
 		\cline{2-3} &  Backscatter spike train & \cite{BST}& &\\ 
 		\cline{2-3} & BiGroup & \cite{Bigroup} &  & \\ \hline
 		\multirow{3}{2.5cm}{~~~~Signal State\\~~~~~ Modeling} & FlipTracer & \cite{Fliptracer} & Accurate signal status  identification & \multirow{3}{4.6cm}{~~~~Computationally demanding} \\ 
 		\cline{2-4}& Hubble & \cite{Hubble} & Efficient state extraction and tracking &  \\ \hline
 		THSS & Spread spectrum & \cite{THSS1} & Increases system resilience & Synchronization and bandwidth challenges \\ \hline
 		\multirow{6}{2.5cm}{~~~Other Random\\~ Access Methods} & Tag subframe selection & \cite{RA1} & Parellel transmission allowed & Heavy reliance on SIC \\ 
 		\cline{2-5}& Channel sensing & \cite{RA2}& Flexible state switching &  Operational complexity and energy consumption \\ 
 		\cline{2-5}& Slotted Aloha & \cite{RA3} & Addresses parallel interference & Inefficiencies and unpredictable performance \\ \hline
 	\end{tabular}
 \end{table*}
 
 \subsubsection{Summary of MI Solutions}

In addressing the challenge of MI in BackCom systems, a variety of strategies have been developed, each with unique advantages and limitations. Multiple access strategies like OFDMA, TDMA, SDMA, and MSMA offer solutions ranging from efficient frequency spectrum use to spatial separation. Yet, they grapple with issues like band allocation, synchronization needs, and potential inter-carrier interference. Tag screening strategies enhance the transmission efficiency through selective tag participation but risk complexity in coordination and possible exclusion of tags. FreeRider, utilizing packet length modulation, aims to improve the throughput, balance the risk of packet collisions, and control transmission requirements.

Some decentralized access strategies, such as MS and Buzz, focus on collision mitigation and managing multiple transmissions, respectively, but bring in challenges like additional hardware needs and computational intensity. Sampling-based techniques, including Laissez-Faire and BiGroup, leverage the capabilities of the reader but are constrained by the signal stability. Signal state modeling approaches like FlipTracer and Hubble employ advanced computational models for accurate signal decoding but are computationally demanding. The THSS technique enhances the system resilience but is challenged by the synchronization and bandwidth efficiency. Lastly, other random access methods address parallel interference through sophisticated signal processing, which may be complex to implement.

Overall, selecting an appropriate strategy or a blend of strategies for mitigating the MI in BackCom systems should be informed by the specific operational requirements and environmental conditions of the network.

\subsubsection{Comparative Assessment of Solution Categories} 

To render this intricate information more comprehensible and analyzable, Table \ref{t5} consolidates and summarizes all the essential data and conclusions. This table not only emphasizes the focal points we have discussed but also provides a more comprehensive perspective through comparison and categorization. By examining the table, readers can gain a more intuitive understanding of how these key elements interact with each other, thereby gaining further insight into the core of the subject.}

\section{The Solutions for  Double-Path Fading}

{\color{black}In this section, we delve into the multifaceted challenge of double-path fading, a critical issue that significantly hampers the efficacy of BackCom systems across various configurations, as noted in recent studies \cite{CCH}. Our journey embarks on a comprehensive exploration of innovative strategies designed to counter this pervasive problem in BackCom systems. We meticulously examine a spectrum of solutions, with each characterized by their unique methodologies and advanced technological constructs. These approaches are pivotal in surmounting the inherent limitations of BackCom systems. Through a detailed analysis of these strategies and the intricacies of their implementation, our goal is to impart a holistic understanding of the present state of double-path fading mitigation, fostering an environment ripe for technological evolution.}

\subsection{Hybrid Passive and Active Transmission}

              \begin{figure}[t]
	\centerline{\includegraphics[width=3.1in]{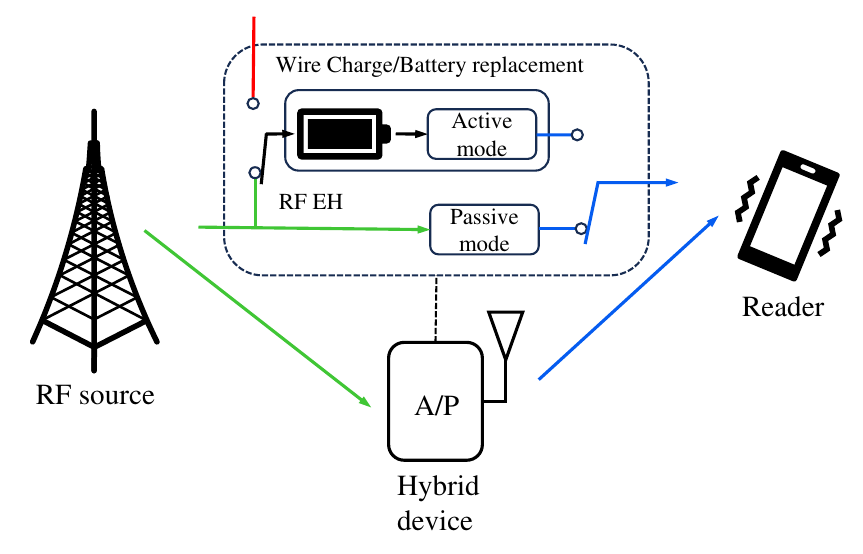}}
	\caption{The hybrid active and passive transmission-based BackCom system.}
	\label{fig7}
\end{figure}

\subsubsection{Technical Merit}

{\color{black}A prevalent strategy in BackCom systems entails the adoption of hybrid active-passive communication, as shown in Fig. \ref{fig7}. This approach intelligently combines the strengths of both active and passive communication technologies within a single device. The technical merit of this strategy lies in its ability to effectively mitigate the issue of double-path loss, which is a major challenge in BackCom systems. By integrating active antennas for extended communication distances and augmented transmission rates, along with passive antennas for reflecting incident signals, the hybrid approach optimizes the communication performance. This dynamic switchability between antennas, based on environmental conditions, showcases a high level of technical sophistication and adaptability in addressing varied communication needs.

The hybrid active-passive communication strategy in BackCom systems encompasses two main schemes, tailored to varying application scenarios and energy needs.
	
	\begin{itemize}
		\item The first is a battery-powered scheme,  where the active communication function of the device is powered by an internal battery \cite{APAS1,APAS1a,APAS2}. Conversely, the device can also participate in communication via backscattering the received signals, in which the harvested energy is primarily used to offset the energy cost of passive communication.
		
		\item The second scheme is based on WPT technology. In this configuration, the device does not rely on an internal battery but is entirely supported by harvesting wireless energy from the environment. In this mode, the device needs to have an efficient EH capability to ensure continuous operation  under the varied energy supply for both active and passive communications \cite{WPBCN5,WPBCN6,WPBCN7,WPBCN4a}.
	\end{itemize}

\subsubsection{Advantages}
\begin{itemize}
	\item Enhanced Communication Range and Quality: The active component of the hybrid system significantly improves the communication distance and signal quality, making it particularly effective in scenarios like urban surveillance and emergency response systems.
	\item Energy Efficiency and Sustainability: The passive component, relying on backscattering, contributes to energy efficiency and sustainability, as it uses the harvested energy for communication, reducing the reliance on external power sources.
	\item Adaptability to Environmental Conditions: The ability to switch between active and passive modes allows the system to adapt to varying environmental conditions, and optimize the communication performance in diverse scenarios.
\end{itemize}

\subsubsection{Challenges and Limitations}
\begin{itemize}
\item Battery Dependence in Powered Schemes: In battery-powered schemes, the reliance on internal batteries introduces concerns about battery life and environmental impact. Regular battery replacement or charging, especially in remote locations, can increase maintenance costs and logistical challenges.
\item Energy Availability in WPT-based Schemes: The effectiveness of WPT-based systems is contingent on the availability of environmental energy sources. In areas with low energy availability, communication capabilities might be limited, affecting the system reliability.
\item Technological Limitations: Current technological limitations in WPT may restrict the use of these systems in high-power communication scenarios, limiting their applicability in certain environments.
\item Complexity in System Design: The integration of both active and passive communication technologies into a single device adds complexity to the system design, which can pose challenges in terms of implementation and operation.
\end{itemize}}

\begin{figure}[t]
	\centerline{\includegraphics[width=3.4in]{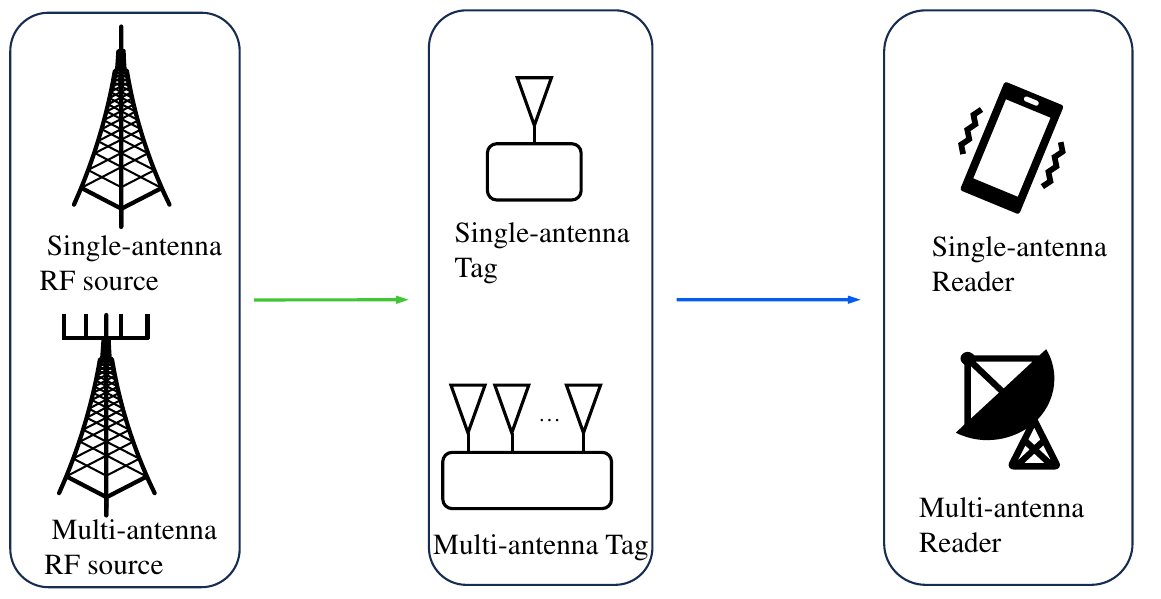}}
	\caption{The multi-antenna-based BackCom system.}
	\label{fig8}
\end{figure}

\subsection{Multi-Antenna Technique}

\subsubsection{Technical Merit}

The integration of multi-antenna technology into BackCom systems serves as a potent strategy to bolster transmission rates, which introduces the concept of an antenna array, where each antenna can independently transmit or receive signals, thus breaking the limitations of traditional single-antenna devices. This approach enhances the ability of the system to handle complex communication scenarios, as demonstrated in recent field studies. {\color{black}Depending on the BackCom network type, this technology is categorized into three: multi-antenna sources, multi-antenna tags, and multi-antenna readers, each illustrated in Fig. \ref{fig8} for clearer understanding.
	
	\begin{itemize}
		\item Multi-antenna sources: The integration of multi-antenna sources in BackCom networks plays a crucial role in mitigating signal attenuation for the forward link \cite{MAS0, MAS1, MAS1a, MAS2}. Specifically, These sources can emit a more directed and focused signal to the tags for both backscattering and EH. This is vital for tags operating in energy-constrained environments or where battery replacement is impractical.
		
		\item Multi-antenna tag: Multi-antenna tags are utilized to receive carrier signals and backscatter information efficiently, which refers to the capability to simultaneously transmit and receive various data signals \cite{MAT1, MAT1a, MAT2, MAT3}. These tags are particularly effective in complex signal environments, enhancing the reliability and quality of BackCom.
		
		\item Multi-antenna readers:  Conversely, readers are equipped with multiple antennas to facilitate reliable data transmission and communication. These antennas, through their advanced signal processing capabilities, have been instrumental in large-scale network implementations where high data throughput is essential \cite{MAR4,MAR5,MAR9}. 
		
		\item Furthermore, multiple input multiple output (MIMO)-based BackCom, composed of multi-antenna source, tags, and readers, has made substantial progress in increasing transmission rates, significantly enhancing the network performance and the user experience \cite{MIMO2,MIMO3,MIMO4}. This approach optimizes available spatial paths, leading to a more robust and efficient communication network.
	\end{itemize}

	\subsubsection{Advantages}
	\begin{itemize}
		\item Enhanced Signal Quality and Efficiency: Multi-antenna systems can significantly improve signal attenuation mitigation, enabling reliable signal transmission even in environments with high interference or physical barriers.
		\item Improved Transmission Rates: The use of multi-antenna technology, especially in MIMO configurations, leads to increased transmission rates, substantially enhancing the network performance.
		\item Advanced Signal Processing Capabilities: Multi-antenna systems can utilize precoding and beamforming techniques, which improve signal directionality and strength, contributing to improved communication performance in dense network environments.
	\end{itemize}

	\subsubsection{Challenges and Limitations}
	\begin{itemize}
		\item Increased Hardware Complexity and Energy Requirements: Implementing multi-antenna systems demands sophisticated hardware and signal processing techniques, leading to higher complexity and energy consumption.
		\item Dependence on Environmental RF Sources in AmBC: The unpredictability of external RF sources in AmBC systems can hinder the effectiveness of transmit beamforming strategies.
		\item Limitations of Passive Tags: In BackCom systems, passive tags may lack the capability to perform complex operations like beamforming, restricting their ability to optimize signal transmission and reception.
		\item Challenges in Network Infrastructure: The complexity of multi-antenna systems extends to the entire network infrastructure, requiring advanced knowledge and resources to fully leverage the advantages of MIMO technology.
\end{itemize} }

   \begin{figure*}[t]
	\centerline{\includegraphics[width=7.0in]{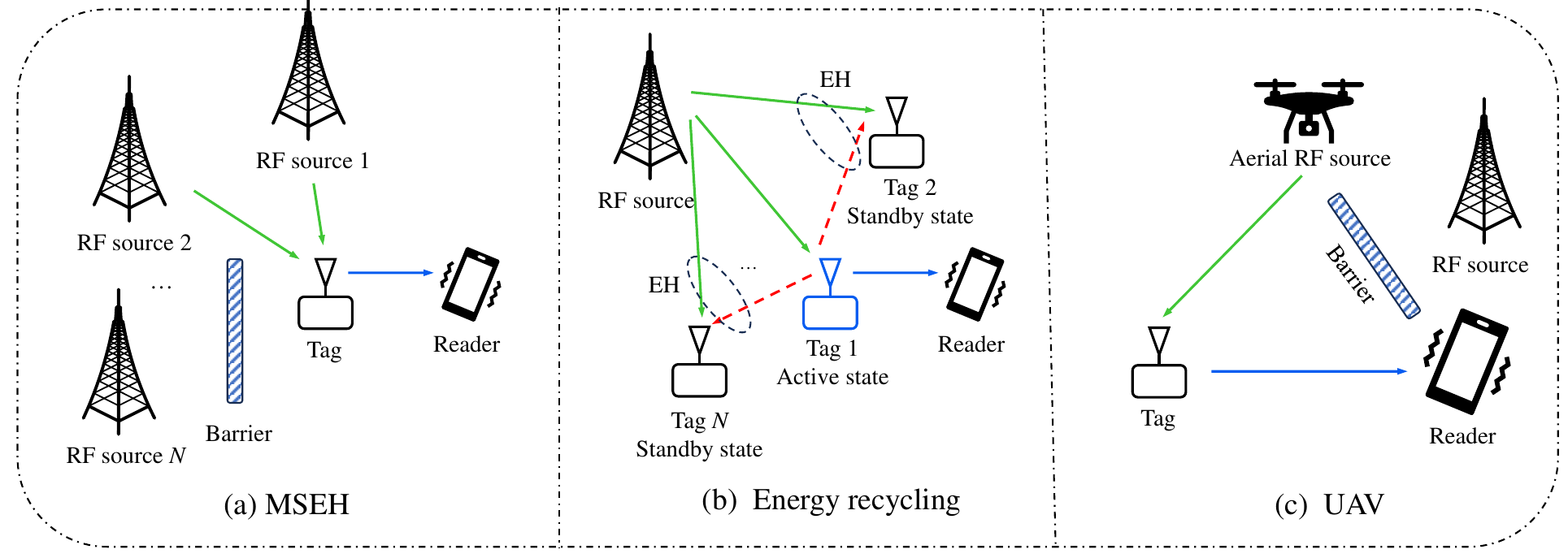}}
	\caption{The enriched power source-based BackCom systems.}
	\label{fig9}
\end{figure*}

 \subsection{Power Supply Enrichment}
 In addition to improving tags and readers, enhancing power source development is also essential to mitigate the double-fading challenge.  Efficient utilization of the power source can directly augment the carrier signal strength, consequently intensifying the backscattered signal. 
 
 \subsubsection{Multiple Energy Sources}
 
 {\color{black} \begin{itemize}
 	\item \textit{Technical Merit}: Multi-source EH (MSEH) is a strategy employed to overcome the issue of inadequate energy from a single RF source in traditional BackCom systems \cite{MSEH0,MSEH1,MSEH2}. This approach has gained traction due to the increasing need for sustainable and reliable energy solutions in the evolving landscape of wireless communications.  This technique allows tags to function for extended durations without solely depending on a single energy source, as shown in Fig. \ref{fig9}(a). With access to a range of power sources, tags can utilize the combined energy supply from these sources, strengthening the carrier signal and effectively counteracting the double fading issue \cite{MSEH2a,MSEH3,MSEH4,MSEH5,MSEH6}.
 	\item \textit{Advantages}:
 	\begin{itemize}
 		\item Extended Operational Durations: The primary advantage of MSEH is its capability to prolong the functional lifespan of tags by diversifying energy sources.
 		\item Strengthened Carrier Signal: Access to multiple power sources allows tags to enhance the carrier signal strength, contributing to more reliable and efficient communication.
 		\item Improved System Robustness: By reducing dependency on a single energy source, MSEH contributes to a more robust and resilient BackCom system, capable of operating under various environmental conditions.
 	\end{itemize}
 	\item \textit{Challenges and Limitations}:
 	\begin{itemize}
 		\item Synchronization and Coordination Difficulties: Implementing MSEH requires the synchronization and coordination of multiple energy sources, which can introduce compatibility and interference challenges.
 		\item Cost and Complexity: Establishing multiple power sources can be a costly and complex endeavor, particularly in terms of installation, maintenance, and management.
 		\item Resource and Expense Management: In realistic environments, managing several energy sources may demand considerable resources and financial investment, which can be a limiting factor for widespread adoption.
 		\item Limited Availability of Energy Sources: In some applications, there may not be enough energy sources available to meet the needs of BackCom systems, limiting the effectiveness of MSEH.
 	\end{itemize}

 	{\color{black}\item \textit{Case Study} 
 	
 	In \cite{MSEH0}, the feasibility of MSEH was explored from water surfaces to power BackCom systems. This novel approach exemplifies the potential of MSEH in utilizing environmental energies, aligning with the trend towards greener and more sustainable communication technologies.
 	
 	\begin{itemize}
 		\item The experiment involved a setup with a water tank, an electric motor to create vibrations, and sensors to capture acoustic and vibration energies. These components were connected to an Arduino MCU and a PC for data collection. The EH was tested at different motor speeds (85 Hz, 170 Hz, 255 Hz) to evaluate the efficiency of energy capture.
 		
 		\item The results demonstrated the ability of this system to operate efficiently even in variable energy conditions, highlighting the adaptability of MSEH. The harvested energy could rapidly charge a 1.5 V rechargeable battery in just 5.7 ms at the lowest speed, allowing continuous operation of the BackCom system without duty cycling.
 		
 		\item This case study successfully demonstrated that MSEH from water bodies was a viable method for powering BackCom systems. It opens up new possibilities for MSEH in aquatic and other unique environments, expanding the scope of BackCom applications. By harnessing environmental energy sources, such as acoustic and vibration energies, it is possible to create more energy-efficient and sustainable communication systems.
 	\end{itemize}}

 \end{itemize}
 }

  \subsubsection{Energy Recycling Strategies}
  
 {\color{black} \begin{itemize}
\item \textit{Technical Merit}: Backscattered energy recycling (BER) introduces an innovative approach to energy management in BackCom systems, as explored in \cite{BER1, BER2}. This strategy allows tags to recycle the backscattered signals from other tags for energy supplementation, thereby reducing their dependence on external power sources, as shown in Fig. \ref{fig9}(b). The technical merit of BER lies in its ability to create a self-sustaining energy ecosystem within the BackCom system. By utilizing the energy inherent in backscattered signals, BER offers a novel way to enhance the strength of these signals without increasing the transmit power of the power source or adding more power sources.

\item \textit{Advantages}:
\begin{itemize}
\item Enhanced Signal Strength: BER provides an additional energy supply to tags, which strengthens the backscattered signals, improving the communication performance and reliability.
\item Extended Operational Duration: In resource-constrained settings, the ability to recycle energy allows tags to operate for longer periods, enhancing the overall sustainability of the system.
\item Reduced Dependence on Power Sources: By leveraging the energy from backscattered signals, BER reduces the reliance on external power sources, making the system more efficient and environmentally friendly.
\end{itemize}

\item \textit{Challenges and Limitations}:
\begin{itemize}
\item Requirement for Non-Overlapping Transmission Times: BER necessitates that the transmission times of tags be non-overlapping, which aligns well with TDMA-based mechanisms. In scenarios where multiple tags need to backscatter simultaneously, the applicability of this method becomes limited.
\item Relatively Small Energy Contribution from Single Tags: The energy contributed by a single tag through recycling is relatively small. Therefore, achieving a significant level of energy supplementation requires the cumulative effort of multiple tags.
\item Complexity in Implementation: Implementing BER strategies requires careful coordination of tag transmissions and sophisticated energy management techniques, which can add complexity to the system design and operation.
\item Dependence on Tag Proximity and Density: The effectiveness of BER is also influenced by the proximity and density of tags within the network, as these factors determine the feasibility of energy recycling among tags.
\end{itemize}
  
   \end{itemize}}

 \subsubsection{Aerial Power Sources}
 
{\color{black} \begin{itemize}
 
\item \textit{Technical Merit}: The use of unmanned aerial vehicles (UAVs) as aerial RF sources in BackCom systems, as shown in Fig. \ref{fig9}(c), introduces a significant advancement in overcoming the limitations of traditional, stationary power sources \cite{UAV1,UAV2,UAV3,UAV4,UAV5,UAV6,UAV7}. UAVs offer the unique capability to dynamically relocate and provide signal enhancement precisely where needed, effectively bridging the distance gaps that fixed power sources cannot manage. This approach significantly mitigates the issue of double channel attenuation and boosts data rates by reducing signal propagation distances. The technical merit of UAVs as aerial power sources lies in their mobility, flexibility, and the potential to ensure high-quality service delivery to distant tags.

\item \textit{Advantages}:
\begin{itemize}
\item Enhanced Network Flexibility: UAVs provide an unparalleled level of flexibility in network configuration, allowing for the adjustment of signal coverage areas as needed.
\item Improved Service Delivery to Remote Areas: By bridging distance gaps, UAVs ensure the delivery of high-quality communication services to remote or otherwise underserved tags.
\item Dynamic Signal Enhancement: The ability of UAVs to swiftly relocate and adapt to changing network requirements allows for dynamic signal enhancement, catering to varying communication demands efficiently.
\end{itemize}

\item \textit{Challenges and Limitations}:
\begin{itemize}
\item Limited Battery Capacity: The flight duration and operational range of UAVs are restricted by their battery capacity, which is a significant concern for long-duration communication tasks.
\item Operational Complexity in Large-Scale Deployments: Deploying UAVs on a large scale introduces challenges in collaborative operation, airspace management, and conflict avoidance, adding complexity to the network management.
\item Maintenance and Cost Considerations: Regular maintenance and operational costs associated with UAVs, especially in large-scale deployments, can be substantial.
\end{itemize}

\end{itemize}}

{\color{black}\subsection{Relay-Assisted Transmission}

   \begin{figure*}[t]
	\centerline{\includegraphics[width=7.0in]{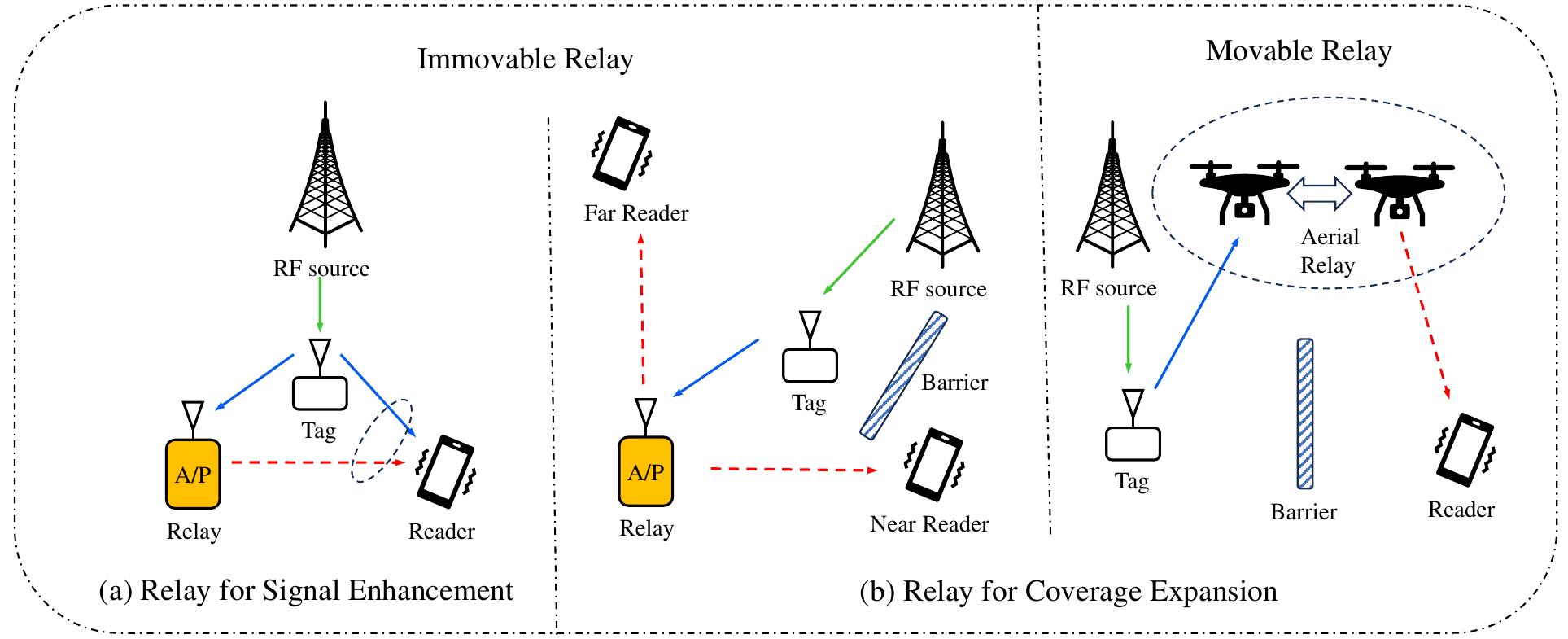}}
	\caption{The relay-based BackCom systems.}
	\label{fig10}
\end{figure*} 	

Whether it involves adopting hybrid active-passive transmission technologies, deploying multiple antenna systems, or enhancing power supply, all these methods point towards a common necessity: the continual advancement and improvement of the internal components of the BackCom system. Integrating relay technology not only addresses signal attenuation but also significantly boosts the overall system performance. This integration strategy aligns with the recent trend of using third-party assistance to boost the performance of BackComs. Specifically, as a critical signal transmission technology in communication networks, the primary responsibility of a relay is to receive signals in transit and amplify or reconstruct them, ensuring their unimpeded transmission to further destinations, as depicted in Fig. \ref{fig10} for enhanced visual understanding.

\subsubsection{Role of Relay for Signal Enhancement} 

Firstly, the signal from the relay, combined with the original reflected signal, can strengthen the backscattered signal at the reader through a multipath approach, as shown in Fig. \ref{fig10}(a). This means that with the assistance of a relay station, the backscattered signal is enhanced before reaching the reader, thereby improving the signal reliability and communication distance while also reducing signal quality degradation caused by path loss.
Presently, this type of system can be classified into three categories.
\begin{itemize}
	\item The first category includes active relay-assisted BackCom systems \cite{ARelay1,ARelay2,ARelay3},  where the relays function in an active mode to relay the backscattered signal. 
	\item The Second is passive relay-assisted BackCom systems \cite{PRelay1,PRelay2,PRelay3,PRelay4}, employing passive nodes or tags as relays that forward signals through backscattering to conserve the energy.
	\item Lastly, hybrid active-passive relay-assisted BackCom systems \cite{HRelay1,HRelay2} allow the EH-powered relays to flexibly switch between active and passive transmission modes, thus enhancing the flexibility and performance of BackCom systems.
\end{itemize}

\subsubsection{Role of Relay for Coverage Expansion}

Secondly, physical obstacles, limited coverage areas, or dual-path fading often lead to link blockages or coverage dead zones, where communication signals struggle to reach effectively. In BackCom systems, dual-path fading particularly causes a significant decrease in the signal strength, leading to more restricted and unstable coverage areas. This adds an extra layer of complexity in maintaining reliable communication links within the BackCom system.

To cope with this, deploying relay nodes is an effective solution, which is able to utilize relay nodes to achieve multi-hop transmission of the signals, effectively conveying the backscattered signals to the reader. In other words, even in complex and variable environments, this multi-hop transmission method can bypass physical obstacles and mitigate the effects of dual-path fading, thereby effectively enhancing the overall coverage area and communication distance. Based on relay characteristics, there are two main system types, as shown in Fig. \ref{fig10}(b).
\begin{itemize}
	\item Immovable relay-based BackComs:   In this system, conventional relays are permanently installed at specific locations to provide stable signal relay services, such as active \cite{ARelay4,ARelay5,ARelay6} and passive \cite{PRelay5,PRelay6,PRelay7,PRelay8} relay-assisted BackCom systems. This setup is suitable for fixed scenarios and environments, such as office buildings, factories, or homes, where communication paths and obstacles are relatively constant, allowing for the optimization of relay node placement for the enhanced coverage.   
	\item Movable relay-based BackComs:  In these systems, the integration of relay technologies with UAV  platforms enables highly flexible and mobile signal transmission, as well as an expanded coverage range, such as \cite{URelay1,URelay2,URelay3,URelay4,URelay5}. This combination allows for the aerial deployment of relay nodes, which can be dynamically positioned to adapt to changing environments and communication needs. For instance, in large public spaces or disaster scenarios, UAV-mounted mobile relays can quickly reposition in response to evolving communication demands and environmental challenges, thereby offering the optimal signal transmission paths. The use of UAVs not only enhances the flexibility in accessing remote or hard-to-reach areas but also effectively overcomes physical barriers, making these systems particularly effective in dynamic or unstable environments.
\end{itemize}

\begin{figure*}[t]
	\centerline{\includegraphics[width=7.0in]{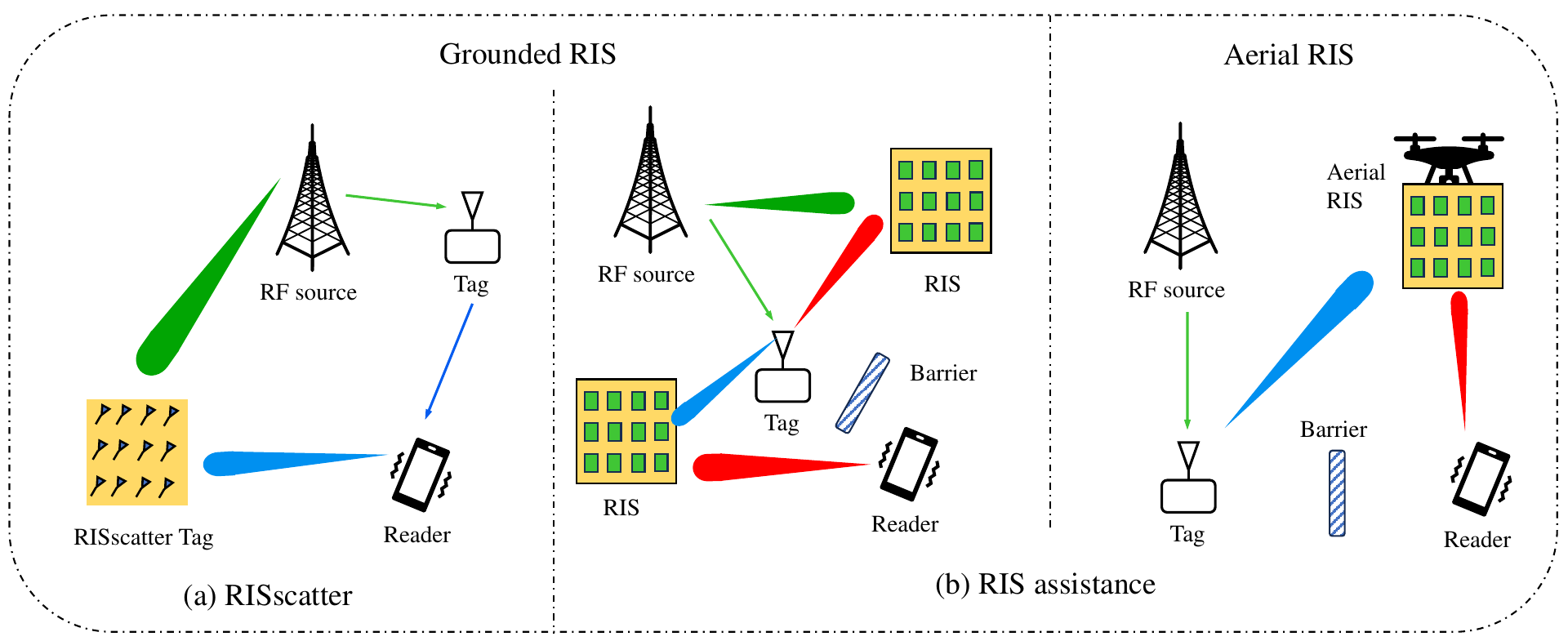}}
	\caption{The RIS-based BackCom systems.}
	\label{fig11}
\end{figure*}

\subsubsection{Challenges and Limitations}

\begin{itemize}
	\item Resource and Power Consumption: Active relay systems, while boosting the signal strength, often struggle with high power consumption and complex signal processing requirements.
	\item Dependence on Source Signal Strength: Passive relay systems' effectiveness largely depends on the strength of the source signal, which can limit their utility in weak signal environments.
	\item Management of Hybrid Systems: Hybrid relay systems, combining active and passive features, require efficient management of energy use and mode transitions, especially in dynamic settings.
	\item Complexity in Movable Systems: Movable relay systems, like UAV-mounted relays, introduce challenges in dynamic network topology management and real-time signal routing adjustments.
	\item Integration with Existing Infrastructure: Ensuring seamless integration of relay systems with existing network infrastructures is crucial for uninterrupted network operations but can be challenging.
	\item Logistical and Cost Considerations: The deployment and maintenance of relay systems, particularly in diverse and dynamic environments, pose logistical and financial challenges.
	\end{itemize}}

\subsection{RIS-Assisted Transmission}

{\color{black}Reconfigurable intelligent surfaces (RISs) represent a cutting-edge development in wireless communication, utilizing surfaces composed of passive elements such as reconfigurable units or antenna arrays. These surfaces can significantly alter the signal propagation environment by dynamically adjusting the phase and amplitude of the elements, effectively manipulating signal characteristics \cite{RIS1,RIS2}. Thanks to its numerous advantages like signal enhancement, interference management, improved spectrum, and energy efficiency, along with its flexibility and cost-effectiveness, RIS has become a focal point of interest in the wireless communication area \cite{RIS3,RIS4,RIS5}.
	
	In BackCom systems, the role of RIS is also increasingly recognized for its exceptional potential in enhancing the signal transmission strength. Based on the deployment characteristics of RIS, there are three categories, i.e., RISscatter, grounded, and aerial RIS-based BackComs \cite{RABC1}.
	
	\subsubsection{RISscatter-based BackCom}
	
	This method, often referred to as ``RISscatter" \cite{RC1}, entails integrating a multiple antenna array RIS directly into the tag, as illustrated in Fig. \ref{fig10}(a). This integration significantly boosts the scattering capacity and strength of the backscattered signal, thereby improving the overall communication efficacy of the system \cite{RC2,RC6,RC7,RC8}. 
	
	\subsubsection{Grounded RIS-based BackComs}
	
	\begin{itemize}
		\item This approach strategically positions RIS within the BackCom system to harness the benefits of multi-path propagation, as shown in Fig. \ref{fig10}(b). Through precise tuning of the phase and amplitude of the reflected signals, RIS effectively optimizes the propagation paths of the backscattered signals. This optimization allows the signals to navigate multiple transmission paths, substantially enhancing their strength \cite{RAB0a,RAB2,RAB3,RAB4,RAB5}. 
		\item Notably, RIS proves invaluable, particularly when forward or backward communication links are compromised by obstacles or similar hindrances. In such cases, RIS is capable of establishing alternative communication channels, thus ensuring continuous and reliable connectivity within the BackCom system \cite{RAB6,RAB7,RAB8}. This feature is incredibly beneficial in complex environments where traditional communication paths are frequently disrupted or weakened. By dynamically reconfiguring signal pathways, RIS offers a flexible and adaptive approach to sustaining communication links, thereby significantly bolstering the resilience and dependability of the BackCom network.
		
	\end{itemize}

	\subsubsection{Aerial RIS-based BackComs}
	
	Despite the substantial enhancements brought about by the grounded deployment of RIS in BackCom systems, certain limitations remain, particularly in terms of the flexibility. Integrating RIS technologies into UAV platforms addresses this by enabling more flexible and mobile signal transmission and coverage range expansion. Such integration permits the aerial deployment of RIS, facilitating dynamic positioning and adaptive coverage. Utilizing UAVs provides greater flexibility in accessing remote or hard-to-reach areas, surmounting physical obstacles, and meeting diverse communication requirements. This innovative approach, as shown in Fig. \ref{fig11}(c) and further discussed in references such as \cite{URIS1}, marks a significant advancement in the deployment and application of RIS in BackCom systems.

	\subsubsection{Challenges and Limitations}
	
	\begin{itemize}
		\item Hardware Complexity: The fabrication and deployment of RIS or RISscatter involve complex engineering and scalability challenges due to the multitude of passive elements.
		\item Signal Optimization: Dynamic signal adjustment via RIS requires sophisticated optimization algorithms, which can be complicated in changing environmental conditions.
		\item Energy Management: Effective energy management is essential, especially in large-scale deployments, to maintain the system efficiency and performance.
		\item Aerial RIS Limitations: Integrating RIS with UAV platforms introduces challenges such as the need for advanced stabilization, complex flight control algorithms, and comprehensive power management to ensure prolonged operations and consistent communication under varying conditions.
\end{itemize}}

\begin{table*}
	\renewcommand{\arraystretch}{1.3}
	\centering
	\small
	\color{black}
	\caption{The solution summary for the double-path fading in BackCom}
	\label{t6}
	\begin{tabular}{|>{\centering\arraybackslash}m{2.5cm}|>{\centering\arraybackslash}p{2.0cm}|>{\centering\arraybackslash}m{1.7cm}|>{\centering\arraybackslash}m{4.6cm}|>{\centering\arraybackslash}m{4.7cm}|}
		\hline
		\textbf{Solutions} & \textbf{Implementation medium} & \textbf{References} & \textbf{Main Advantages} & \textbf{Main challenges} \\
		\hline
		\multirow{3}{2.7cm}{Hybrid Passive and Active Transmission} & Battery assistance & \cite{APAS1, APAS1a, APAS2} & \multirow{3}{4.7cm}{~~~Adaptive active and passive\\ ~~~~switching; Improved system\\~~~~~~~~~~~~ flexibility} & Battery maintenance; Increased system complexity \\
		\cline{2-3} \cline{5-5} & WPCN & \cite{WPBCN2, WPBCN4a, WPBCN6, WPBCN7, WPBCN5} & & EH inefficiency; Integration challenges \\
		\hline
		\multirow{6}{2.5cm}{~~~Multi-Antenna\\~~~~~ Technique} & Multi-antenna sources & \cite{MAS0, MAS1, MAS1a, MAS2} & \multirow{6}{4.6cm}{Improved spatial reuse and signal quality; Multiplexing, Diversity,\\~~~~~~~~~ Beamforming} & \multirow{6}{4.6cm}{Increased hardware complexity;\\ Signal synchronization and calibration issues;\\ Limitations from ambient sources and passive tag} \\
		\cline{2-3} 
		&Multi-antenna tags & \cite{MAT1, MAT1a, MAT2, MAT3}  & & \\
		\cline{2-3} 
		& Multi-antenna readers & \cite{MAR4, MAR5, MAR9} & & \\
		\cline{2-3} 
		& MIMO & \cite{MIMO2, MIMO3, MIMO4, MIMO5} & & \\
		\hline
		\multirow{5}{2.5cm}{~~~Power Supply \\~~~~Enrichment} & MSEH & \cite{MSEH0, MSEH1, MSEH2, MSEH2a, MSEH3, MSEH4, MSEH5, MSEH6} & Power source expansion and selectivity & {Energy transfer efficiency issues; Implementation cost, Interference complexity; } \\
		\cline{2-5}
		& BER & \cite{BER1, BER2} & Self-energy supply within the system; Improved network sustainability & TDMA-based protocol only; Limited energy contribution from single tags \\
		\cline{2-5}
		& Aerial power source & \cite{UAV1, UAV2, UAV3, UAV4, UAV5, UAV6, UAV7} &  Flexible energy supply; Enhanced coverage in remote areas & Regulation and control Design; UAV stability and safety issues \\
		\hline
		\multirow{7}{2.5cm}{~~~Relay-Assisted\\~~~ Transmission} & Active relay  & \cite{ARelay1, ARelay2, ARelay3, ARelay4, ARelay5, ARelay6} & \multirow{5}{4cm}{Signal amplifying/forwarding;\\ Backscattering signal relaying and/or replenishing; \\Enhanced adaptability in dynamic environments} & High power consumption; Scheduling and coordination complexity  \\
		\cline{2-3}  \cline{5-5}
		& Passive relay & \cite{PRelay1, PRelay2, PRelay3, PRelay4, PRelay5, PRelay6, PRelay7, PRelay8} &  & Source signal strength dependency; Weak relaying signal  \\
		\cline{2-3}  \cline{5-5}
		& Hybrid relay & \cite{HRelay1, HRelay2} &  & Mode switching control; Operational complexity \\
		\cline{2-4}  \cline{5-5}
		& Aerial relay & \cite{URelay1, URelay2, URelay3, URelay4, URelay5} & Flexible and movable relaying; Enhanced coverage in dynamic scenarios  & Real-time routing adjustments; UAV flight control challenges \\
		\cline{2-3}  \cline{5-5}
		\hline
		\multirow{5}{2.5cm}{~~~RIS-Assisted \\~~~Transmission} & RISscatter & \cite{RC2, RC6, RC7, RC8} & \multirow{3}{4cm}{Altering the signal propagation environment; Backscattering signal replenishing; Enhanced scalability and flexibility} & Dynamic signal optimization issues; Hardware complexity  \\
		\cline{2-3} \cline{5-5} & Ground RIS & \cite{RAB0a, RAB2, RAB3, RAB4, RAB5, RAB6, RAB7, RAB8} & &  Dynamic Optimization; Environmental variability challenges \\
		\cline{2-4} \cline{5-5} & Aerial RIS & \cite{URIS1} &Flexible and movable RIS; Dynamic coverage adaptation  & Regulation and control; Computation complexity; UAV integration challenges \\
		\hline
		OTM &  Tag to tag &  \cite{OSB} & Adaptive role switching; Improved network capacity and resource utilization & Transceiver dependency; Network coordination and management difficulties \\
		\hline
	\end{tabular}
\end{table*}

\subsection{Opportunistic Transceiver Mechanism}

{\color{black}\begin{itemize}
	\item \textit{Technical Merit}:
	The opportunistic transceiver mechanism (OTM), introduced in the work in \cite{OSB}, demonstrates a significant technical advancement in backscatter tag-tag networks. This approach is grounded in the adaptability of tags to changing channel conditions, determining their roles as either transmitters or receivers based on the strength of the forward link. The technical merit of this mechanism lies in its dynamic response to the varying energy and channel conditions prevalent in BackCom systems. It signifies an intelligent and flexible approach to maximizing the efficiency and reliability of communication in these networks.
	
	\item \textit{Advantages}: 
	\begin{itemize}
	\item 	Adaptive Role Assignment: The mechanism enhances network efficiency by adaptively assigning the roles of tags as transmitters or receivers based on their current channel conditions.
	\item Improved Energy Utilization: By enabling tags with stronger forward links to act as transmitters, the mechanism optimizes the use of available energy resources within the network.
	\item	Resilience to Channel Variability: This approach mitigates the impact of double fading on tags in less favorable channel conditions, improving the overall resilience and reliability of the BackCom system.
	\end{itemize}
	
	\item \textit{Challenges and Limitations}:
		\begin{itemize}
		\item Limited Impact on Backscatter Tags: While the mechanism assists tags in unfavorable conditions by shifting their roles, it does not substantially alleviate fading effects for tags continuing to backscatter, which can hinder their performance.
		\item Operational Complexity: The continuous monitoring and analysis required for dynamic role determination add complexity to the operation and management of the system.
		\item Risk of Network Imbalance: There is a potential for creating an imbalanced network load, where certain tags are consistently burdened with transmitter roles while others predominantly serve as receivers, leading to uneven wear and energy depletion across the network.
	\end{itemize}
		
\end{itemize}

\subsection{Summary and Comparative Analysis}

\subsubsection{Summary of Double-Path Fading Solutions} In the realm of BackCom systems, various innovative strategies have been developed to tackle the double-path fading, each offering distinct advantages for different scenarios. These include hybrid active-passive transmission for fluctuating energy environments; multi-antenna techniques enhancing the signal quality in urban areas; MSEH providing robust energy for tags in remote locations; BER for sustainable network scalability; Aerial power sources like UAVs for dynamic coverage; Relay-assisted transmission for extended coverage in large-scale settings; RIS-assisted transmission for signal control in environments like smart homes; and OTM for adaptive tag role assignment. While these methods collectively mitigate the double-path fading and improve the system performance, they also bring unique challenges, highlighting the need for continuous research and innovation in BackCom technology.

\subsubsection{Comparative Assessment of Solution Categories}

To better understand and compare these solutions, Table \ref{t6} presents a summarized assessment, highlighting the distinct features, benefits, and challenges of each method. This comparison is not only instrumental in showcasing the relative strengths and weaknesses of each approach but also in pinpointing the most suitable strategies based on application requirements and environmental conditions.}

    \section {Open Issues and Future Directions}
    
{\color{black}In this pivotal section, we embark on a forward-thinking exploration of some open issues and potential solutions that could shape the future landscape of BackCom systems. Our focus is on identifying the critical challenges that currently limit the full realization of the potential of BackComs and we propose innovative approaches that could revolutionize this field. We delve into various topics, including the DLI management, the MI management, and the backscattering signal enhancement.

\subsection{The DLI Management}

\subsubsection{Open Issues}

The realm of DLI management in BackCom systems is rife with challenges that necessitate innovative solutions and thoughtful consideration. Addressing these open issues is about enhancing current technologies and rethinking our approach to deal with the DLI and its potential impacts on the system performance. Here, we delve into some of the most pressing concerns that are currently shaping the research and development in this field:

\begin{itemize}
	
	\item \textit{How to Enhance the Cancellation Performance?}
	
	The effectiveness of current techniques like SIC and OSIC is limited by their inability to completely eliminate interference, leading to residual effects that can degrade the system performance. These techniques traditionally rely on sequentially cancelling interference signals, which can be less effective in complex signal environments or when dealing with multiple sources of interference. Furthermore, the reliance of OSIC on the quality of previously decoded signals means its performance can vary significantly in different conditions, particularly where the signal quality is compromised.

	\item \textit{How to Balance the Complexity and the Cost?}
	
	Advanced signal processing methods in OFDM and multi-antenna systems offer the improved performance but at the expense of the increased system complexity and cost. This issue arises because these techniques often require sophisticated hardware and computational resources, making them less feasible for resource-constrained environments. The challenge is to find a middle ground where the benefits of advanced signal processing can be harnessed without incurring prohibitive costs or overly complicating the system design.
	
	\item \textit{How to Enhance the Spectrum Utilization?} 
		
	While the FS method is effective in separating signals to reduce interference, it typically does not use the full potential of the available spectrum.  This inefficiency stems from the fact that the FS method requires different frequency bands to be allocated to different signals. Consequently, the overall spectrum utilization might be compromised, especially in scenarios with multiple signals. For instance, in environments with several interfering signals, these signals would occupy a portion of the spectrum, thereby reducing the performance of BackComs as the spectrum allocated to them diminishes. This can be a significant drawback, particularly in high-demand communication environments or situations where spectrum resources are scarce.
	
	\item \textit{How to Ensure the Adaptability and Robustness?} 
	
	While innovative in utilizing the polarization properties of electromagnetic waves for interference mitigation, PC techniques are susceptible to environmental factors. These factors can include physical obstructions, atmospheric conditions, and changes in signal propagation paths, all of which can alter the effectiveness of polarization-based methods. Developing systems that can adapt to these environmental changes and maintain the robust performance is a crucial challenge.

	\item \textit{Is it Possible to Make the DLI Beneficial?} 
	
	The conventional view of DLI as a detrimental factor to be mitigated or avoided overlooks the potential for utilizing this interference constructively. The challenge here lies in identifying scenarios and developing methods where DLI can be transformed from a hindrance into an asset. This requires a radical shift in thinking about the DLI, exploring ways it can be harnessed to enhance the communication rather than impede it.
\end{itemize}

\subsubsection{Future Directions}
Looking ahead, the future of the DLI management in BackCom systems is both promising and demanding. It calls for a multi-faceted approach that not only addresses current limitations but also anticipates emerging trends and technological advancements. The following points outline some of the key areas where future research and development efforts could lead to significant breakthroughs and improvements in how we manage and utilize the DLI in BackCom systems:
\begin{itemize}
	
	\item \textit{Deep Learning for the Interference Cancellation:}
	\begin{itemize}		
		\item Deep learning algorithms have shown remarkable proficiency in processing large and complex datasets to detect patterns in interference cancellation, leading to more accurate predictions and effective mitigation strategies \cite{DL}. These algorithms can be incorporated into BackCom systems as an advanced processing layer, significantly enhancing the systems' existing capabilities to cancel the DLI.

		\item However, deploying deep learning for interference cancellation comes with several challenges. The need for extensive training data to effectively teach the algorithms can be a significant hurdle. The high computational demands of deep learning models and the risk of overfitting, where models become too tailored to the training data and fail to generalize new data, are key considerations.
		
		\item To address these challenges, several approaches can be adopted. Utilizing transfer learning can reduce the amount of training data required by leveraging pre-trained models on similar tasks. Developing lightweight neural network models can help manage the computational demands by reducing the complexity of the models without significantly sacrificing the performance. Employing edge computing for on-site data processing alleviates the load on central systems and reduces the latency. Furthermore, exploring methods to develop learning models based on insufficient data can offer practical solutions in scenarios where collecting extensive datasets is challenging or impractical. In this regard, the work in \cite{YuLi} may be a living proof.
	\end{itemize}

	\item \textit{Developing Low-Power, Cost-Effective Solutions:}
	
	\begin{itemize}
		\item  The implementation of low-power solutions is essential for enhancing the longevity and reducing the operational costs of tags, thereby making the technology more accessible and sustainable. Integrating such solutions into existing BackCom systems can significantly improve their energy efficiency, which is crucial for large-scale deployments, without compromising the performance. This approach helps make BackCom technology more eco-friendly and economically viable, especially in scenarios requiring extensive network coverage.
		
		\item A key challenge lies in balancing the low power consumption and the improved performance, particularly in environments with complex interference. This balance is critical to ensure that energy-saving measures do not negatively impact the functionality of the BackCom system or its ability to handle the DLI effectively.
		
		\item To overcome these challenges, innovations in both hardware and software are necessary. On the hardware side, developing energy-efficient chipsets and components can drastically reduce the power requirements of tags. In terms of software, enhancing algorithmic efficiency, especially in signal processing, is vital. Besides, implementing techniques such as dynamic power scaling, which adjusts the power usage based on real-time performance needs, can effectively manage the power consumption while ensuring that the system remains responsive and capable of handling the demands of complex network environments.
		
	\end{itemize}

	\item \textit{Advancing Cross-Layer Design and Optimization:}
	
	\begin{itemize}
		\item  Cross-layer design is a strategy that enhances the system operation by leveraging interactions between different protocol layers, which leads to optimized resource allocation and improved system performance. Integrating this design approach into existing BackCom systems offers a more responsive and adaptable solution for managing the DLI. This strategy can significantly enhance the overall efficiency of BackCom systems by ensuring that various layers work in harmony, thereby optimizing the communication and reducing the interference.
		
		\item However, the complexity of coordinating and optimizing interactions across multiple layers presents a significant challenge. Besides, this approach can potentially increase the overhead of the system due to the need for more sophisticated management and control mechanisms across the layers.
		
		\item To address these challenges, the development and implementation of intelligent, decentralized algorithms are crucial. These algorithms should be capable of operating autonomously at each layer while synchronizing with other layers periodically. This approach helps in maintaining the system performance and minimizing the overall overhead. By enabling each layer to handle its specific tasks independently while still being a part of a coordinated whole, these algorithms can effectively balance the need for cross-layer optimization with the desire to keep the system efficient and manageable.
	\end{itemize}

	\item \textit{Cultivating Intelligent Adaptive Systems:}
	
	\begin{itemize}
		\item Adaptive systems, especially those incorporating cognitive radio technologies, can significantly improve the ability of BackComs to manage the DLI. These systems can dynamically adjust operational parameters in response to the evolving radio environment, enhancing the system flexibility and responsiveness. Integrating cognitive radios and other adaptive technologies with existing BackCom systems enables them to adjust more effectively to environmental changes and the dynamics of interference.
		
		\item However, developing real-time, energy-efficient algorithms for these adaptive systems poses a complex challenge. These algorithms must be capable of making intelligent decisions based on environmental inputs, requiring a delicate balance between the rapid responsiveness and the energy conservation.
		
		\item To overcome these challenges, leveraging advancements in machine learning, particularly reinforcement learning, is a promising approach. Reinforcement learning algorithms can be designed to efficiently adapt transmission strategies based on real-time feedback from the environment. Besides, implementing distributed processing across the network can help in reducing the energy load on individual nodes. This approach allows for a more scalable and energy-efficient way of managing the DLI, ensuring that BackCom systems remain responsive and effective even in dynamically changing environments.
	\end{itemize}

	\item \textit{Harnessing the Constrictive Interference:} 
	
	\begin{itemize}
		\item  This approach involves reimagining the DLI as an advantageous resource rather than a hindrance. By doing so, it opens up innovative methods to boost the signal strength and data throughput in BackCom systems \cite{CI1}. This requires a significant shift in perspective but can be effectively integrated into existing systems with advanced signal processing and thoughtful system design. By turning this traditional challenge into an advantage, BackCom systems can achieve an enhanced performance in terms of the signal reliability and data transmission rates.
		
		\item One of the major challenges in harnessing the constructive interference is the unpredictable nature of the DLI and the complexity involved in accurately manipulating it for beneficial outcomes. The variable and often chaotic nature of interference patterns makes it difficult to consistently transform them into constructive elements within the system.
		
		\item To address these issues,  advanced signal processing techniques are essential. Techniques like phase alignment and signal amplification strategies can be pivotal in converting the DLI into beneficial components. Besides, a broader exploration of the DLI from various perspectives can provide deeper insights into how these interference patterns can be utilized to serve BackCom systems. Different signal considerations, such as timing, frequency, and spatial factors, can play a crucial role in understanding and exploiting the DLI, thereby enhancing the overall effectiveness and efficiency of BackCom systems.
		
	\end{itemize}
	
\end{itemize}

In conclusion, by addressing these open issues and actively exploring these innovative future directions, significant advancements are anticipated in the DLI management for BackCom systems. This progress is expected to establish a robust foundation for efficient, adaptable, and reliable wireless communication networks in an ever-evolving technological landscape.

\subsection{The MI Management}

\subsubsection{Open Issues}

Although critical for the system performance and reliability, managing the MI in BackCom systems encompasses several unresolved challenges. Here, we delve deeper into each open issue, analyzing their root causes and complexities:

\begin{itemize}

	\item \textit{How to Decode Signals in Dense Environments?}
	
	The challenge stems from the sheer number of tags communicating simultaneously in dense BackCom networks. This density leads to a crowded signal environment, where distinguishing individual tag signals becomes increasingly tricky. Traditional decoding techniques may fall short in effectively separating these overlapping signals, leading to increased error rates and reduced the overall system performance. The need for advanced decoding methods that can cope with this high-density signal environment is imperative, particularly as the IoT and other technologies continue to add IoT devices to these networks.
	
	\item \textit{How to Manage the Dynamic Spectrum?}
	
	The scarcity of spectrum resources is a fundamental issue exacerbated by the growing number of passive tags. The challenge lies in dynamically allocating this limited resource to maximize the utilization and minimize the interference. Current static spectrum allocation methods are insufficient in rapidly changing network environments, leading to the inefficient spectrum use and intensive MI. Developing dynamic spectrum management strategies that can adapt in real-time to changing the network demands and conditions is crucial.
	
	\item \textit{How to Enhance the Energy Efficiency in Decentralized Systems?}
	
	Decentralized BackCom systems distribute the processing and communication load across multiple nodes, reducing reliance on central units. However, this distribution often means that individual nodes, which may be resource-constrained, face the increased energy demand. The challenge is to develop energy-efficient mechanisms that can sustain these nodes without compromising the decentralized nature of the system. This involves optimizing the energy usage while maintaining the effective communication and processing capabilities.
	
	\item \textit{How to Ensure the Scalability and Flexibility?}
	
	The scalability issue arises from the need to extend the MI management techniques to accommodate a growing numbers of tags without loss of effectiveness. Besides, the flexibility is critical to adapt these techniques to different network conditions and topologies. The challenge is to design the MI management strategies that are not only scalable but also versatile enough to handle various network scenarios, including the changes in the device density, mobility patterns, and communication protocols.

	\item \textit{Are There Any Alternative Solutions to Cancel the MI?}
	
	Traditional methods for managing the MI in BackCom systems are effective historically, but they exhibit significant limitations, particularly when handling a large number of tags. This situation necessitates exploring alternative or supplemental strategies to overcome these inherent challenges in BackCom systems. These new approaches should be capable of efficiently managing the increased levels of interference and complexities brought about by the proliferation of tags, ensuring robust and reliable network performance even in highly congested environments.

	\item \textit{How to Make the MI Constructive?}
	
	Typically, the MI is seen as a detrimental factor that needs to be minimized or eliminated. However, there is potential in rethinking this approach by exploring ways to use the MI constructively. This paradigm shift requires a profound understanding of the underlying signal-processing mechanisms and innovative architectural designs. The challenge is identifying scenarios where the MI could enhance the signal strength or reliability and develop techniques to harness this potential. This approach could lead to more robust and efficient BackCom systems, especially in the complex and dynamic environments.	
	
\end{itemize}

\subsubsection{Future Directions}

The future of the MI management in BackCom systems, especially in the context of the growing IoT ecosystem, demands innovative and multifaceted solutions. The following enhanced future directions not only address current challenges but also lay down a pathway for harnessing emerging trends and technological advancements:

\begin{itemize}
	\item \textit{Development of Advanced NOMA Strategies:}
	\begin{itemize}   
		\item Non-orthogonal multiple access (NOMA) strategy significantly enhances spectral efficiency in communication networks \cite{NOMA,NOMA3}, especially in BackCom systems. This approach is more efficient than traditional orthogonal access methods like TDMA and FDMA. It can be combined with other technologies to form hybrid architectures like TDMA-NOMA and Aloha-NOMA, optimizing performance.

		\item However, addressing the unique challenges posed by weak backscattered signals in BackCom systems is critical. Unlike traditional communication systems, where transmitted signals are typically strong and distinct, BackCom signals are inherently weaker and more prone to the interference.
		
		\item  This situation necessitates a combination of innovative signal processing techniques and algorithms specifically designed for the unique aspects of BackCom environments. There is a significant need for advanced algorithms aiming at enhancing the ability of the reader to distinguish between these weaker, closely-spaced signals.     
	\end{itemize}
	
	\item \textit{Adaptive AI-Driven Interference Management:}
	
	\begin{itemize}
		\item Integrating artificial intelligence (AI) and machine learning for interference management in BackComs may significantly enhance the network efficiency \cite{DL1}, which aids in predicting and adapting to complex interference patterns, optimizing network performance. AI algorithms analyze historical and real-time data to mitigate the MI proactively, ensuring the high network quality.
		
		\item The main challenge for AI-driven management is processing the vast data from tags in a real-time manner while continuously adapting to changing network conditions and user behaviors. Efficient, real-time data processing is vital for reliable networks.
		
		\item To tackle these challenges, distributed AI systems at the network edge are proposed. These edge-based AI systems reduce the load on central systems by processing data locally, decreasing latency, and providing faster responses. This approach enhances the network scalability and flexibility, offering customized solutions for diverse network areas.
	\end{itemize}

	\item \textit{Blockchain-Enabled Spectrum Management:}
	
	\begin{itemize}
		\item Implementing the blockchain technology in BackCom systems offers a solution for transparent, secure, and equitable management of spectrum resources \cite{Blockchain}, which is especially pivotal in decentralized network environments. The decentralized nature of blockchain aligns well with BackCom systems, providing a framework for efficiently and securely, and transparently managing the spectrum resources utilized by numerous tags.
		
		\item However, incorporating blockchain technology into existing network infrastructures presents scalability challenges. Blockchain systems, while robust and secure, can be complex and resource-intensive, which may pose difficulties when integrating with the existing, potentially less complex BackCom systems.
		
		\item To address these challenges, the development and use of lightweight blockchain models are essential. These models would reduce the computational and resource burden of blockchain technology, making it more feasible for integration with BackCom systems. Besides, creating integration frameworks that allow for the seamless incorporation of blockchain technology into existing BackCom infrastructures is crucial. These frameworks would ensure that blockchain, and can be effectively utilized without disrupting the existing operations and performance of BackCom systems, thereby providing a secure and efficient method for spectrum resource management.
	\end{itemize}

	\item \textit{Energy-Efficient Decentralized Coordination:}
	
	\begin{itemize}
		\item Prioritizing energy efficiency and decentralized coordination in BackCom systems can greatly improve their sustainability and scalability. These approaches are particularly beneficial in environments where tags are constrained by their energy capacity or rely on EH. By focusing on low-energy designs and decentralized management, BackCom systems can operate more effectively in a wide range of scenarios, including those with numerous distributed tags, by minimizing their environmental footprint and operational costs.

		\item However, achieving a balance between the energy efficiency and the performance presents a significant challenge. Besides, ensuring effective coordination among a large number of tags,  with its own energy constraints and operational parameters, adds complexity to the system management.
		
		\item To address these issues, the development of advanced EH technologies is crucial. These technologies can provide a sustainable power source for tags, reducing their reliance on external power supplies. Furthermore, algorithms designed for efficient decentralized management are essential. These algorithms should be capable of autonomously adapting to variations in network conditions and the energy levels of individual tags. Such adaptive, decentralized systems can optimize the energy usage across the network while maintaining high levels of the system performance and the communication quality.
		
	\end{itemize}
	
	\item \textit{Scalable Solutions for Massively Connected Networks:}
	\begin{itemize}
		\item In large-scale BackCom networks, scalable solutions are key to effectively managing the synchronization and coordination, particularly as these systems are anticipated to support an immense number of interconnected tags. Scalability ensures that the network can handle an increasing number of tags without a significant drop in the performance or efficiency. This is crucial for the growing demands of modern networks, where the number of connected tags and the volume of data traffic are rapidly increasing.
		
		\item A major challenge in such massively connected networks is ensuring the network stability and performance even as the number of tags increases significantly. As the network expands, maintaining efficient communication and coordination among a vast number of tags becomes challenging.
		
		\item To tackle this challenge, the development and implementation of advanced optimization algorithms are essential. These algorithms should be capable of dynamically adjusting network parameters to optimize resource allocation and communication pathways, thereby maintaining the network stability and performance. Besides, incorporating game theory-based models can provide a framework for strategic decision-making within the network. These models can help manage the interactions between numerous network entities, ensuring that individual actions lead to the overall network optimization and mitigating potential conflicts or inefficiencies in the resource usage.
		
	\end{itemize}

	\item \textit{Cross-Layer and Quantum-Enhanced Approaches:}
	
	\begin{itemize}
		\item Cross-layer and quantum-enhanced strategies offer comprehensive MI management by integrating insights from various communication layers and harnessing the unique capabilities of quantum mechanics. Cross-layer approaches are particularly beneficial for BackCom systems, as they provide a holistic understanding of network dynamics, which is essential for the effective MI management. The use of quantum technologies introduces advanced capabilities in managing and securing data\cite{Quantum, Quantum1}, potentially revolutionizing the MI management in BackCom systems.  
		
		\item However, challenges arise in integrating quantum technologies with existing BackCom infrastructures and maintaining coherence across different communication layers. Quantum technologies are at the forefront of communication research, but their practical application in existing network systems, especially in BackCom, requires careful consideration of compatibility and integration issues.
		
		\item To overcome these challenges, the development of standardized protocols and interfaces for cross-layer communication are necessary. These protocols would facilitate the smooth interaction among different layers of the communication network, ensuring efficient data transfer and MI management. Simultaneously, investing in research to advance quantum technologies is crucial. The goal is to make these technologies more accessible and compatible with current communication systems, allowing for their seamless integration into the BackCom infrastructure and unlocking new potentials in the network performance.
	\end{itemize}

	\item \textit{Constructive Interference Utilization:}
	
	\begin{itemize}
		\item Leveraging the MI constructively in BackCom systems can significantly enhance the spectral efficiency and improve the overall system performance, particularly in dense network environments. Given that managing interference is a fundamental challenge in BackCom systems, using MI in a constructive manner is a highly relevant and beneficial approach.
		
		\item However, identifying specific scenarios where the MI can be used constructively and devising techniques to manipulate and exploit interference effectively are complex tasks. These challenges stem from the need to understand the intricate dynamics of interference in varying network conditions and to develop methods to control and utilize this interference positively.
		
		\item Addressing these challenges involves evaluating various performance indicators to gain a deeper understanding of interference dynamics within BackCom systems \cite{CI2,CI3}. Innovative signal processing techniques play a crucial role in this context. They need to be developed to transform the MI from a disruptive factor into a valuable resource that enhances the communication efficiency. These techniques would enable BackCom systems to not only cope with but also capitalize on the presence of interference, thereby improving their functionality in high-density settings.
	\end{itemize}
\end{itemize}

By addressing these enhanced future directions, we can anticipate significant advancements in the MI management for BackCom systems. These efforts are expected to lead to more robust, efficient, and adaptable communication networks capable of meeting the demands of an increasingly connected and complex world.

\subsection{The Backscattering Signal Enhancement}

\subsubsection{Open Issues}

Enhancing backscattering signals in BackCom systems is pivotal for their efficiency and effectiveness. However, several open issues currently impede optimal performance and broader application. These challenges must be addressed to realize the potential of BackCom systems fully in various environments. The key open issues include:

\begin{itemize}
	
	\item \textit{How to Enhance the EH Efficiency?}
	
	The primary cause of this challenge lies in the inconsistent and often low levels of RF energy in various environments, which BackCom systems rely on for power supply. This inconsistency is particularly pronounced in areas with fluctuating energy availability, like urban settings with varying RF noise levels or rural areas with limited RF sources. An efficient EH leads to an adequate power supply for tags, compromising their operational stability and communication efficacy.

	\item \textit{How to Balance the Scalability and Deployment Costs?}
	
	Scalability issues arise from extending BackCom systems over larger areas, requiring more sophisticated infrastructure like RIS panels or relay nodes. The complexity increases with the need for the system robustness and the consistent performance across the expanded network. Large-scale deployment of these technologies incurs significant costs, both in terms of hardware and the maintenance required to manage a more complex network. This is a barrier to the widespread adoption of BackCom systems, especially in resource-limited settings.

	\item \textit{Are There Any Alternative Solutions for the Double-Fading Issue?}
	
	The double-fading issue in BackComs primarily results from signal attenuation due to both the initial transmission and the subsequent backscattering process. This issue is exacerbated in environments with physical obstructions or significant distances between communication nodes. Current solutions might not be sufficient in specific scenarios, prompting a need for alternative or supplemental strategies to enhance the backscattering signal strength for BackCom systems.
	
	\item \textit{How to Ensure the Security and Privacy Demands?}
	
	The increasing complexity of BackCom systems and their integration into diverse and often public networks raises significant security and privacy concerns. The data transmitted through these systems can be sensitive, making them a target for cyber threats. Inadequate security measures could lead to data breaches, unauthorized access, and privacy violations, undermining the trust and reliability of BackCom systems.
	
\end{itemize}

\subsubsection{Future Directions}

Looking forward, addressing these open issues paves the way for several promising directions in the field of BackCom systems. The future of backscattering signal enhancement lies in innovation and the adoption of emerging technologies, which could significantly improve the performance and applicability of BackCom systems. The potential future directions include:

\begin{itemize}
	
	\item \textit{Advanced EH Technologies:}
	
	\begin{itemize}
		\item  Advanced EH technologies have the potential to enhance the energy efficiency of BackCom systems significantly. By harnessing innovative energy sources such as triboelectric nanogenerators or cutting-edge photovoltaic cells,  tags can function autonomously for prolonged periods. This reduces the reliance on external power sources, extending the lifespan of tags and minimizing their environmental impact. These advanced EH methods allow BackCom systems to operate more sustainably and efficiently, which is a top priority in large-scale deployments and remote applications.
		
		\item However, the development of these technologies faces substantial challenges. High costs associated with researching and developing new materials and technologies are a major hurdle. Besides, miniaturizing these technologies for integration into existing BackCom frameworks without losing efficiency presents a significant technical challenge. The balance between making these technologies both small enough to fit within current tags and efficient enough to be effective is crucial.
		
		\item To address these challenges, the focus should be on developing cost-effective and environmentally sustainable materials for EH. Research and innovation in this area should aim to reduce the costs while maintaining or improving the performance of these technologies. Furthermore, efforts are needed to design miniature versions of these EH technologies. Ensuring the integration into existing BackCom systems without requiring extensive modifications is essential for their widespread adoption. This will allow BackCom systems to benefit from the enhanced energy efficiency and independence, contributing to their broader applicability and sustainability.
	\end{itemize}

	\item \textit{Low-Cost, Scalable Deployment Models:}
	
	\begin{itemize}
		\item   Implementing scalable deployment models is key to affordably expanding BackCom networks. These models are essential for facilitating the deployment in diverse environments, including areas with economic constraints. By adopting modular design principles, these models can allow for easy network upgrades and expansions, minimizing disruptions and costs associated with the network growth. This approach is particularly valuable for ensuring that BackCom technology is accessible and beneficial in a wide range of settings, from urban centers to remote and underserved regions.
		
		\item However, scaling up BackCom networks poses risks, such as potential declines in the system performance and increased complexity in the network management. As the network grows, ensuring consistent performance and efficiently managing the expanded infrastructure becomes more challenging, which could impact the overall functionality and reliability of the network.
		
		\item To effectively address these challenges, the adoption of adaptive network management software is critical. This software should be designed to dynamically adjust to changes in network size and complexity, ensuring optimal performance and efficient resource utilization. Besides, incorporating cloud-based data management solutions can offer a cost-effective and efficient way to handle the increased volumes of data generated by larger networks. Cloud-based solutions provide scalable storage and processing capabilities, which can significantly enhance the management and analysis of data in expanding BackCom networks, supporting their growth and evolution.
	\end{itemize}

    \item \textit{Tunnel Diode-Enhanced Backscatter:}
    
    \begin{itemize}
     \item The primary potential of tunnel diode tag lies in its low energy consumption and microwatt-level operating power. This significantly reduces the overall energy consumption of its circuit, allowing the more incident signal to be backscattered, which in turn improves BackCom performance. Besides, its ability to effectively reflect or absorb signals offers stable and efficient communication, making it highly suitable for the long-term operation in battery-free or EH-based systems \cite{fm5, Tunnel}.
     
    \item  While the tunnel diode strategy offers numerous advantages, it also faces challenges such as the limited communication range due to lower radiated power, the potential instability of the emitted signal under energy-limited conditions, and the susceptibility to external interference in certain environments.
    
    \item To overcome these challenges, several approaches could be implemented. Improving signal processing algorithms to enhance the detection and handling of weak signals is one method. Optimizing the design of tunnel diode tags to improve performance across various environments is another. Additionally, combining the use of other technologies or devices, such as high-sensitivity receivers or deploying auxiliary devices near the tags, can effectively extend the communication range and enhance the overall system performance.
    
    \end{itemize}
	
	\item \textit{Movable Antenna-Enhanced Backscatter:}
	\begin{itemize}
		
		\item The movable antenna technology, a novel development, capitalizes on the spatial variation of wireless channels within a confined area by allowing the local movement of the antenna \cite{MA1}. This technology enables dynamic adjustment of antenna positions at the transmitter and/or receiver, which can lead to the enhanced channel conditions and the improved BackCom system performance. By adapting to environmental variables that typically impact the signal quality, BackCom systems can be more reliable.

		\item However, integrating movable antenna technology into existing BackCom systems may add to their operational complexity. This integration would require advanced signal processing capabilities to handle the dynamic nature of antenna positioning and the associated variations in the communication channel.
		
		\item The development of adaptive signal processing algorithms is crucial for this technology. These algorithms should be capable of intelligently determining antenna positions and optimizing spatial diversity. They need to be responsive to environmental changes, continuously adjusting to maintain optimal antenna placement and spatial configurations for effective signal transmission and reception. This adaptability is key to leveraging the full potential of the movable antenna technology in enhancing the BackCom system performance.
	\end{itemize}
	
	\item \textit{Millimeter-wave-Enhanced Backscatter:}
	
	\begin{itemize}
		\item Utilizing the millimeter-wave spectrum in BackCom systems presents an opportunity for high-speed data transmission, opening up new application possibilities \cite{MMwave1}. This approach could provide increased bandwidth, catering to applications that require higher data rates and supporting more data-intensive tasks within BackCom environments.
		
		\item However, this shift to higher frequencies brings several technical challenges. Key among these is the development of suitable antennas and circuitry that can efficiently operate at millimeter-wave frequencies. Besides, addressing the inherent propagation losses characteristic of these frequencies is crucial to ensure the effective communication.
		
		\item Ongoing research and development in millimeter-wave technology are essential. The focus should be on designing components that are not only cost-effective and energy-efficient but also compatible with the specific requirements of BackCom systems. Advancements in this area could significantly enhance the capabilities and applications of BackCom technology, making it more versatile and powerful.
	\end{itemize}
	
	\item \textit{Active RIS-Enhanced Backscatter:}
	
	\begin{itemize}
		\item Active RIS marks a significant breakthrough in signal enhancement, with the ability to actively amplify signals during reflection \cite{ARIS}. This capability could revolutionize signal propagation in BackCom systems, offering more robust and extended communication capabilities in various environments.

		\item However, a major challenge with active RIS lies in their power consumption and the need to manage the amplified noise that might be introduced into the system. The balance between signal amplification and power efficiency is a critical aspect that needs careful consideration.
		
		\item Therefore, it is essential to design active RIS components that are energy-efficient. In conjunction with this, integrating advanced noise cancellation technologies becomes imperative. Such technologies are crucial for minimizing the impact of noise, ensuring that the benefits of signal amplification do not compromise signal quality. The development of these technologies is vital for maximizing the effectiveness of active RIS in BackCom systems, maintaining high-quality signal transmission while managing power and noise challenges.
	\end{itemize}
	
	\item \textit{Enhanced Security Protocols:}
	
	\begin{itemize}
		\item While enhancing the strength of backscattered signals, it is crucial not to overlook the security aspects of the communication system. Security in communication systems can be achieved in two ways: secure communication aiming at protecting information data, and covert communication intended to conceal transmissions. Both methods are effective in safeguarding the communication security.
		
		\item The use of artificial noise is a key strategy to ensure the communication security. This approach exploits the differences in interference cancellation capabilities between eavesdroppers and intended users. By adding controlled noise to the communication channel, it becomes more challenging for eavesdroppers to decipher the transmitted information. However, a challenge arises in accurately adjusting and controlling the noise level to ensure the effective security without overly impacting the communication quality of the intended recipient. The solution lies in adopting adaptive algorithms that dynamically adjust the noise level to maintain a balance between the security and the communication quality.
		
		\item  Switching between active and passive modes in the tags is another method to enhance the security. In the presence of eavesdroppers, the tag can switch to passive communication to reduce the strength and visibility of its signal, making it harder to intercept. Conversely, in a secure environment, active transmission can be utilized to boost the transmission efficiency. The challenge here is to accurately identify potential eavesdroppers and switch to passive mode when necessary, without compromising the overall efficiency of the system. Advanced monitoring and analysis tools can be developed to detect potential security threats and automatically adjust the transmission mode. This dynamic switching mechanism offers a flexible way to balance the security and the performance in BackCom systems.

	\end{itemize}
\end{itemize}

Addressing these open issues and pursuing these future directions will be key to advancing the field of BackCom systems, particularly in overcoming the challenges posed by the double-path fading.}

		\section{Conclusions}

		{\color{black} This survey begins by revealing the profound impact of strong interference and fading on the performance of BackCom systems, which elucidates how these factors contribute to  reduced data rates, shorter communication distances, and overall system instability, thereby emphasizing the urgent need for dedicated research in this area. Following this, the survey presents a thorough review of the current state-of-the-art solutions that target these specific challenges, where insightful analyses and comparisons of various strategies are provided, offering clarity on effective methods for mitigating issues like  the DLI, the MI, the and double-path fading. 
		Progressing further, the survey shifts its focus to exploring open issues and potential directions for BackComs related to these challenges and solutions, which aims at addressing the technical obstacles inherent in BackCom systems, thereby facilitating  their unobstructed progression and efficient development within the field.
			
        In conclusion, this survey underscores the critical importance of continuous innovation and research in overcoming the technical challenges faced by BackCom systems. This relentless pursuit of advancement is vital for their seamless integration into the expanding AmIoT ecosystem. While BackCom offers a promising pathway in the AmIoT era, its ultimate success hinges on our collective ability to effectively tackle the inherent challenges of interference and fading. By proactively addressing and resolving these issues, BackCom has the potential to unlock a more sustainable and interconnected world, representing a significant leap forward in the realm of IoT technologies.}


\begin{thebibliography}{1}
	
	\bibitem{iot1}
	Y. Liu, D. Li, B. Du, L. Shu, and G. Han, ``Rethinking sustainable sensing in agricultural Internet of Things: From power supply perspective," \textit{IEEE Wireless Commun.}, vol. 29, no. 4, pp. 102--109, Aug. 2022.
	
	\bibitem{RothmullerBarker}
{\color{black}S. Sinha, ``State of IoT 2023: Number of connected IoT devices growing 16\% to 16.7 billion globally,'' Website, May. 2023.  https://iot-analytics.com/number-connected-iot-devices/.}

\bibitem{AnubhaJain}
{\color{black}S. Griffiths, ``Why your internet habits are not as clean as you think,'' Website, Mar. 2020. https://www.bbc.com/future/article/20200305-why-your-internet-habits-are-not-as-clean-as-you-think.}

\bibitem{LiuLiDuShuHan}
{\color{black}C. Glickman, ``Green IoT: The shift to practical sustainability,'' Jul. 2023.  https://cio.economictimes.indiatimes.com/news/internet-of-things/patents-need-to-be-part-of-cutting-edge-products-rajaraman/101456241.}

\bibitem{LiuLIDaiLIZhang}
{\color{black}D. Ma, G. Lan, M. Hassan, W. Hu, and S. K. Das, ``Sensing, computing, and communications for energy harvesting IoTs: A survey,"  \textit{IEEE Commun. Surveys Tut.}, vol. 22, no. 2, pp. 1222--1250, 2nd Quart. 2020.}



\bibitem{Pecunia}
V. Pecunia, L. G. Occhipinti, and R. L. Z. Hoye, ``Emerging indoor photovoltaic technologies for sustainable Internet of Things,'' \textit{Adv. Energy
Mater.}, vol. 11, no. 29, pp. 1--31, Aug. 2021.

{\color{black}
	
\bibitem{Amiot1}
3GPP RP-222685, ``Study on Ambient IoT,'' Huawei, HiSilicon, RAN\#97e, Sept. 2022.
	
\bibitem{Amiot2}	
3GPP RP-232408, ``Discussion on feasibility assessment and required functionalities for Ambient IoT," Huawei, HiSilicon, RAN\#101, Sept. 2023.

\bibitem{Amiot3}
3GPP S2-2310033, ``Study on Architecture support of Ambient power-enabled Internet of Things," OPPO, SA2\#160, Nov. 2023.


\bibitem{Amiot4}
3GPP TR 38.848, ``Study on Ambient IoT (Internet of Things) in RAN,'' Sept. 2023. https://www.3gpp.org/ftp/Specs/archive/38\_series/38.848/.
}
	



\bibitem{solar}
C. Yang, J. Gummeson, and A. Sample, ``Riding the airways: Ultrawideband ambient backscatter via commercial broadcast systems,'' in \textit{Proc. IEEE INFOCOM’17},  May 2017, pp. 1--9.

\bibitem{Seebeck}
 K. Uchida, S. Takahashi, K. Harii, J. Ieda, W. Koshibae, K. Ando, S. Maekawa, and E. Saitoh, ``Observation of the spin Seebeck effect,'' \textit{Nature}, vol. 455, no. 7214, pp. 778--781, 2008.
 
 \bibitem{thermal}
 H. Solar, A. Beriain, A. Rezola, D. del Rio, and R. Berenguer, ``A 22-m operation range semi-passive UHF RFID sensor tag with flexible thermoelectric energy harvester," \textit{IEEE Sensors J.}, vol. 22, no. 20, pp. 19797--19808, Oct. 2022.
 
 \bibitem{Piezo}
 H. S. Kim, J. -H. Kim, and J. Kim, ``A review of piezoelectric energy harvesting based on vibration,'' \textit{Int. J. Precision Eng. Manuf.}, vol. 12, no. 6, pp. 1129--1141, Dec. 2011.

\bibitem{BiZengZhang}
S. Bi, Y. Zeng, and R. Zhang, ``Wireless powered communication networks: An overview," \textit{IEEE Wireless Commun.}, vol. 23, no. 2, pp. 10--18, Apr. 2016.



\bibitem{DuhovnikovBaltacci}
S. Duhovnikov, A. Baltaci, D. Gera, and D. A. Schupke, ``Power consumption analysis of NB-IoT technology for low-power aircraft applications," in \textit{Proc. IEEE WF-IoT’19,} Apr. 2019, pp. 719--723.

\bibitem{VullersSchaijk}
R. J. M. Vullers, R. V. Schaijk, H. J. Visser, J. Penders, and C. V. Hoof, ``Energy harvesting for autonomous wireless sensor networks," \textit{IEEE Solid-State Circuits Mag.}, vol. 2, no. 2, pp. 29--38, Jun. 2010.










\bibitem{BackComS2}
C. Xu, L. Yang, and P. Zhang, ``Practical backscatter communication systems for battery-free Internet of Things: A tutorial and survey of recent research," \textit{IEEE Signal Process. Mag.}, vol. 35, no. 5, pp. 16--27, Sept. 2018.


\bibitem{BackComS3}
A. Bletsas, P. N. Alevizos, and G. Vougioukas, ``The art of signal processing in backscatter radio for $\mu$W (or less) Internet of Things: Intelligent signal processing and backscatter radio enabling batteryless connectivity,"  \textit{IEEE Signal Process. Mag.}, vol. 35, no. 5, pp. 28--40, Sept. 2018.

\bibitem{WangZhangYang}
P. -H. P. Wang, C. Zhang, H. Yang, D. Bharadia, and P. P. Mercier, ``A 28$\mu$W IoT tag that can communicate with commodity WiFi transceivers via a single-side-band QPSK backscatter communication technique," in \textit{Proc.  IEEE ISSCC},  Apr. 2020, pp. 312--314.




\bibitem{LinAhmed}
L. Lin, K. A. Ahmed, P. S. Salamani, and M. Alioto, ``Battery-less IoT sensor node with PLL-less WiFi backscattering communications in a 2.5-$\mu$W peak power envelope," in \textit{Proc.  Symp. VLSI Circuits}, Jun. 2021, pp. 1-2.

\bibitem{BiBC}
J. Kimionis, A. Bletsas, and J. N. Sahalos, ``Increased range bistatic scatter radio," \textit{IEEE Trans. Commun.}, vol. 62, no. 3, pp. 1091--1104, Mar 2014.



\bibitem{modulation}
G. Khadka, M. Nemati, X. Zhou, and J. Choi, ``Index modulation in backscatter communication for IoT-sensor-based applications: A review," \textit{IEEE Sensors J.}, vol. 22, no. 22, pp. 21445--21461, Nov. 2022.

\bibitem{coding}
F. R. REZAEIDINDARLOO, D. Galappaththige, C. Tellambura, and S. Herath, ``Coding techniques for backscatter communications—A contemporary survey," \textit{IEEE Commun. Surveys Tut.}, vol. 25, no. 2, pp. 1020--1058, 2nd Quart. 2023.

\bibitem{mmwave}
W. Chen, ``Survey of Millimeter wave backscatter communition systems,'' \textit{arXiv preprint arXiv}:2305.10302, 2023.

\bibitem{Lora}
J. P. Shanmuga Sundaram, W. Du, and Z. Zhao, ``A survey on LoRa networking: Research problems, current solutions, and open Issues," \textit{IEEE Commun. Surveys  Tut.}, vol. 22, no. 1, pp. 371--388, 1st Quart. 2020.


\bibitem{AmBC}
N. Van Huynh, D. T. Hoang, X. Lu, D. Niyato, P. Wang, and D. I. Kim, ``Ambient backscatter communications: A contemporary survey," \textit{IEEE Commun. Surveys Tut.}, vol. 20, no. 4, pp. 2889--2922, 4th Quart. 2018.

\bibitem{bcoverview1}
J. -P. Niu and G. Y. Li, ``An overview on backscatter communications," \textit{J. Commun. Information Netw.}, vol. 4, no. 2, pp. 1--14, Jun. 2019.

 \bibitem{GIot}
W. Zhang \textit{et al.}, ``A green paradigm for Internet of Things: Ambient backscatter communications,''  \textit{China Commun.}, vol. 16, no. 7, pp. 109--119, Jul. 2019.

 \bibitem{bcss3}
M. L. Memon, N. Saxena, A. Roy, and D. R. Shin, ``Backscatter communications: Inception of the battery-free eraa comprehensive survey,'' \textit{Electronics}, vol. 8, no. 2, p. 129, Feb. 2019.

 \bibitem{bcss1}
C. Yao, Y. Liu, X. Wei, G. Wang, and F. Gao, ``Backscatter technologies and the future of Internet of Things: Challenges and opportunities," \textit{Intelligent and Converged Netw.}, vol. 1, no. 2, pp. 170--180, Sept. 2020.

\bibitem{bcss6}
W. Liu, K. Huang, X. Zhou, and S. Durrani, ``Next generation backscatter communication: Systems, techniques, and applications,'' \textit{EURASIP J. Wireless Commun. Netw.}, vol. 2019, no. 1, pp. 1--11, Mar. 2019.

 
 \bibitem{BTTN}
 M. Stanacevic, A. Athalye, Z. J. Haas, S. R. Das, and P. M. Djuric,
 ``Backscatter communications with passive receivers: From fundamentals
 to applications,''\textit{ ITU J. Future and Evolving Technologies},
 vol. 1, no. 1, Dec. 2020.
 
 \bibitem{bcss4}
 F. Rezaei, C. Tellambura, and S. Herath, ``Large-scale wireless-powered networks with backscatter communications—A comprehensive Survey,"  \textit{IEEE Open J. Commun. Society}, vol. 1, pp. 1100--1130, 2020.
 
\bibitem{bcss}
R. Torres \textit{et al.}, ``Backscatter communications,"  \textit{IEEE J. Microw.}, vol. 1, no. 4, pp. 864--878, Oct. 2021.

\bibitem{ACD}
C. Song \textit{et al.}, ``Advances in wirelessly powered backscatter communications: From antenna/RF circuitry design to printed flexible electronics," \textit{Proceedings of the IEEE}, vol. 110, no. 1, pp. 171--192, Jan. 2022.


\bibitem{bcss2}
U. S. Toro, K. Wu, and V. C. M. Leung, ``Backscatter wireless communications and sensing in green Internet of Things,"  \textit{IEEE Trans. Green Commun. Netw.}, vol. 6, no. 1, pp. 37--55, Mar. 2022.

\bibitem{BBWS}
U. S. Toro \textit{et al.}, ``Backscatter communication-based wireless sensing (BBWS): Performance enhancement and future applications,” \textit{J. Network and Computer Applications}, 2022, p. 103518.



\bibitem{bcss5}
D. A. Loku Galappaththige, F. Rezaei, C. Tellambura, and S. Herath, ``Link budget analysis for backscatter-based passive IoT,"  \textit{IEEE Access}, vol. 10, pp. 128890--128922, 2022.

\bibitem{bcs2}
W. Wu, X. Wang, A. Hawbani, L. Yuan, and W. Gong, ``A survey on ambient backscatter communications: Principles, systems, applications, and challenges,'' \textit{Computer Networks}, p. 109235, Oct., 2022.

\bibitem{bcss7}
F. Xu \textit{et al.}, ``The state of AI-empowered backscatter communications: A comprehensive survey," \textit{IEEE Internet  Things J.}, early access. 2023. doi: 10.1109/JIOT.2023.3299210.


\bibitem{BackComS1}
T. Jiang \textit{et al.}, ``Backscatter communication meets practical battery-free Internet of Things: A survey and outlook," \textit{IEEE Commun. Surveys Tut.}, vol. 25, no. 3, pp. 2021--2051, 3rd Quart. 2023.

{\color{black}\bibitem{photophone}
A. G. Bell, ``The photophone,'' \textit{Science }(80-.)., vol. 1, no. 11, pp. 130--134, 1880.

\bibitem{Lero}
P. Nikitin, ``Leon Theremin (Lev Termen),'' \textit{IEEE Antennas Propag.
Mag.}, vol. 54, no. 5, pp. 252--257, Oct. 2012.

\bibitem{Stockman}
H. Stockman, ``Communication by means of reflected power,'' in \textit{Proc. IRE}, vol. 36, no. 10, pp. 1196--1204, Oct. 1948.

\bibitem{harris}
D. B. Harris, ``Radio transmission systems with modulatable passive responder,'' 1960.

\bibitem{Boyer}
C. Boyer and S. Roy. ``Backscatter communication and RFID: Coding, energy,
and MIMO analysis.'' \textit{IEEE Trans. Commun.}, vol. 62, no. 3,  770-785, Mar. 2014.

\bibitem{Finkenzeller}
K. Finkenzeller, \textit{RFID Handbook}. Chichester, U.K.: Wiley, Mar. 2003.

\bibitem{WISP}
J. R. Smith, A. P. Sample, P. S. Powledge, S. Roy, and A. Mamishev,
``A wirelessly powered platform for sensing and computation,'' in \textit{Proc. 8th Int. Conf. Ubiquitous Comput.}, 2006, vol. 4206, pp. 495-506.

\bibitem{tunnelscatter}
A. Varshney, A. Soleiman, and T. Voigt, ``TunnelScatter: Low power
communication for sensor tags using tunnel diodes,'' in \textit{Proc. 25th
Annu. Int. Conf. Mobile Comput. Netw}.  Aug. 2019, pp. 1--17. 


\bibitem{quantum}
R. Jantti, R. Duan, J. Lietzen, H. Khalifa, and L. Hanzo, ``Quantum-enhanced microwave backscattering communications," \textit{IEEE Commun, Mag.}, vol. 58, no. 1, pp. 80--85, Jan. 2020.

\bibitem{symbotic}
R. Long, Y. -C. Liang, H. Guo, G. Yang, and R. Zhang, ``Symbiotic radio: A new communication paradigm for passive Internet of Things," \textit{IEEE Internet  Things J.}, vol. 7, no. 2, pp. 1350--1363, Feb. 2020.


\bibitem{RC1}
Y. Zhao and B. Clerckx, ``RIScatter: Unifying backscatter communication and reconfigurable intelligent surface,`` \textit{arXiv preprint arXiv}:2212.09121, Dec. 2022.

\bibitem{structure}
Z. Gong, L. Han, Z. An, L. Yang, S. Ding, and Y. Xiang, ``Empowering
smart buildings with self-sensing concrete for structural health monitoring,'' in \textit{Proc. ACM SIGCOMM Conf.}, Aug. 2022, pp. 560--575.




\bibitem{agriculture}
S. -N. Daskalakis, J. Kimionis, A. Collado, M. M. Tentzeris, and A. Georgiadis, ``Ambient FM backscattering for smart agricultural monitoring," in \textit{Proc. IEEE MTT-S  IMS},  2017, pp. 1339--1341.

\bibitem{transportation}
W. U. Khan, M. A. Javed, T. N. Nguyen, S. Khan, and B. M. Elhalawany, ``Energy-efficient resource allocation for 6G backscatter-enabled NOMA IoV networks," \textit{IEEE Trans. Intelligent Transportation Syst.}, vol. 23, no. 7, pp. 9775--9785, July 2022.

\bibitem{ehealth}
F. Jameel, R. Duan, Z. Chang, A. Liljemark, T. Ristaniemi, and R. Jantti, ``Applications of backscatter communications for healthcare networks,"  \textit{IEEE Netw.}, vol. 33, no. 6, pp. 50-57, Nov.--Dec. 2019.

\bibitem{tracking}
D. -T. Phan-Huy, D. Barthel, P. Ratajczak, R. Fara, M. d. Renzo, and J. d. Rosny, ``Ambient backscatter communications in mobile networks: Crowd-detectable zero-energy-devices," in \textit{Proc. IEEE RFID-TA}, 2021, pp. 81--84.

\bibitem{underwater}
 J. Jang and F. Adib, ``Underwater backscatter networking,’’ in \textit{Proc. ACM
Special Interest Group Data Commun.}, Aug. 2019, pp. 187--199.



\bibitem{rfeh}
X. Lu, P. Wang, D. Niyato, D. I. Kim, and Z. Han, ``Wireless networks with RF energy harvesting: A contemporary survey," \textit{IEEE Commun. Surveys  Tut.}, vol. 17, no. 2, pp. 757--789, 2nd Quart. 2015.



\bibitem{ehsen}
D. Khan \textit{et al.}, ``A design of ambient RF energy harvester with sensitivity of -21 dBm and power efficiency of a 39.3\% using internal threshold voltage compensation,'' \textit{Energies}, vol. 11, no. 5, p. 1258, May 2018.

\bibitem{ehce}
Y. Chen, K. T. Sabnis, and R. A. Abd-Alhameed, ``New formula for conversion efficiency of RF EH and its wireless applications,"  \textit{IEEE Trans. Veh. Tech.}, vol. 65, no. 11, pp. 9410--9414, Nov. 2016.

\bibitem{ehputput}
X. Gu, S. Hemour, and K. Wu, ``Far-field wireless power harvesting: Nonlinear modeling, rectenna design, and emerging applications," \textit{Proceedings of the IEEE}, vol. 110, no. 1, pp. 56--73, Jan. 2022.



\bibitem{ehmodel0}
G. Lu, L. Shi, and Y. Ye, ``Maximum throughput of TS/PS scheme in an AF relaying network with non-linear energy harvester," \textit{IEEE Access}, vol. 6, pp. 26617--26625, May 2018.

\bibitem{ehmodel1}
 E. Boshkovska, D. W. K. Ng, N. Zlatanov, and R. Schober, ``Practical non-linear energy harvesting model and resource allocation for SWIPT systems,'' \textit{IEEE Commun. Lett.}, vol. 19, no. 12, pp. 2082--2085, Dec. 2015.

\bibitem{ehmode2}
S. Wang, M. Xia, K. Huang, and Y. -C. Wu, ``Wirelessly powered two-way communication with nonlinear energy harvesting model: Rate regions under fixed and mobile relay,"  \textit{IEEE Trans. Wireless Commun.}, vol. 16, no. 12, pp. 8190--8204, Dec. 2017.


\bibitem{wifi1}
J. Zhao, W. Gong, and J. Liu, ``Spatial stream backscatter using commodity wifi,'' in \textit{Proc. of ACM MobiSys}, Jun. 2018. pp. 191--203.

\bibitem{lora1}
Y. Peng \textit{et al.}, ``PLoRa: A passive long-range data network from ambient LoRa transmissions,'' in \textit{Proc. Conf. Special Interest Group Data Commun.},  2018, pp. 147--160.

\bibitem{mbc1}
F. Amato, H. M. Torun, and G. D. Durgin, ``RFID backscattering in long-range scenarios," \textit{IEEE Trans. Wireless Commun.}, vol. 17, no. 4, pp. 2718--2725, Apr. 2018.

\bibitem{mbc2}
X. Wang \textit{et al.}, ``AllSpark: Enabling long-range backscatter for vehicle-to-infrastructure communication,"  \textit{IEEE Internet Things J.}, vol. 9, no. 24, pp. 25525--25537, Dec. 2022.

\bibitem{mbc3}
A. Eid, J. Rademacher, W. Akbar, P. Wang, A. Allam, and F. Adib, ``Enabling long-range underwater backscatter via Van Atta acoustic networks,'' in \textit{Proc. ACM SIGCOMM '23}, 2023, pp. 1--19.

\bibitem{tv}
V. Liu \textit{et al.}, ``Ambient backscatter: Wireless communication out of thin air,'' in \textit{Proc. ACM SIGCOMM}, Hong Kong, Aug. 2013, pp. 39--50.



\bibitem{fm}
A. Wang, V. Iyer, V. Talla, J. R. Smith, and S. Gollakota, ``FM backscatter: Enabling connected cities and smart fabrics,'' in \textit{Proc. 14th USENIX Symp. Netw. Syst. Design Implement. (NSDI)}, 2017, pp. 243--258.

\bibitem{lte}
Z. Chi, X. Liu, W. Wang, Y. Yao, and T. Zhu, ``Leveraging ambient LTE traffic for ubiquitous passive communication,'' in \textit{Proc. ACM SIGCOMM’20}, Aug. 2020, pp. 172--185.

\bibitem{wifi}
 P. Zhang, D. Bharadia, K. Joshi, and S. Katti, ``HitchHike:Practical backscatter using commodity WiFi,'' in \textit{Proc. 14th ACM Conf. Embedded Netw. Sensor Syst. CD-ROM}, Nov. 2016, pp. 259--271.

\bibitem{FR}
P. Zhang, C. Josephson, D. Bharadia, and S. Katti, ``Freerider: Backscatter communication using commodity radios,'' in \textit{Proc. CoNEXT}, 2017, pp. 389--401.


\bibitem{zigbee}
Wang, Shixin, Zhaoyuan Xu, and Wei Gong, ``Poster: Enhanced ZigBee backscatter communication using fine-grained chip-level modulation." in \textit{Proc. MobiSys'23.} Jun. 2023. pp. 565--566.

\bibitem{tv1}
A. N. Parks, A. Liu, S. Gollakota, and J. R. Smith, ``Turbocharging ambient backscatter communication,'' in \textit{Proc. ACM SIGCOMM Comput. Commun. Rev.}, vol. 44, no. 4, pp. 619–630, Oct. 2014.

\bibitem{fm4}
Z. Kapetanovic, A. Saffari, R. Chandra, and J. R. Smith, ``Glaze: Overlaying occupied spectrum with downlink iot transmissions,'' in \textit{Proc. ACM Interact. Mob. Wearable Ubiquitous Technol.}, vol. 3, no. 4, Dec. 2019.


\bibitem{fm2}
S. N. Daskalakis, J. Kimionis, A. Collado, G. Goussetis, M. M. Tentzeris, and A. Georgiadis, "Ambient backscatterers using FM broadcasting for low cost and low power wireless applications," \textit{IEEE Trans. Microw. Theory  Tech.}, vol. 65, no. 12, pp. 5251--5262, Dec. 2017.


\bibitem{fm5}
J. Hu, L. Zhong, T. Ma, Z. Ding, and Z. Xu, ``Long-range FM backscatter tag with tunnel diode," \textit{ IEEE Microw. Wireless Compon. Lett.}, vol. 32, no. 1, pp. 92--95, Jan. 2022.


\bibitem{lte2}
Y. Feng, S. Chen, W. Xi, S. Wang, J. Zhao, and W. Gong, ``Heartbeating with LTE networks for ambient backscatter," \textit{IEEE Trans. Mobile Comput.}, early access. Jun. 2023. doi: 10.1109/TMC.2023.3290298.




\bibitem{vms}
X. Liu, Z. Chi, W. Wang, Y. Yao, and T. Zhu, ``VMscatter: A versatile MIMO backscatter,'' in \textit{Proc. USENIX NSDI’20},  Feb. 2020, pp. 895–909.

\bibitem{wifiscatter}
M. Dunna, M. Meng, P.-H. Wang, C. Zhang, P. P. Mercier, and D. Bharadia, ``Syncscatter: Enabling wifi like synchronization and range for wifi backscatter communication.'' in \textit{Proc. of USENIX NSDI}, 2021.

\bibitem{wifiscatter1}
L. Yuan and W. Gong, ``SubScatter: Sub-symbol WiFi backscatter for high throughput," in \textit{Proc. IEEE ICNP},  2022, pp. 1--11.


\bibitem{bcmag1}
C. Xu, L. Yang, and P. Zhang, ``Practical backscatter communication systems for battery-free Internet of Things: A tutorial and survey of recent research," \textit{IEEE Signal Process. Mag.}, vol. 35, no. 5, pp. 16--27, Sept. 2018.





\bibitem{bluetooth}
J. Rosenthal and M. S. Reynolds, ``A 1.0-Mb/s 198-pJ/bit Bluetooth low-energy compatible single sideband backscatter uplink for the NeuroDisc brain–computer interface,"  \textit{IEEE Trans. Microw. Theory  Tech.}, vol. 67, no. 10, pp. 4015--4022, Oct. 2019.

\bibitem{blue1}
M. Zhang, J. Zhao, S. Chen, and W. Gong, ``Reliable backscatter with commodity ble,'' in \textit{Proc. IEEE INFOCOM 2020 - IEEE Conference on Computer Communications}, pp. 1291–1299, 2020.

\bibitem{blue3}
 W. Gong, L. Yuan, Q. Wang, and J. Zhao, ``Multiprotocol backscatter for personal iot sensors,'' in \textit{Proc. of ACM CONEXT}, Nov. 2020, pp. 261--273.
 
 \bibitem{blue4}
 J. Jung \textit{et al.}, ``Gateway over the air: Towards pervasive  internet connectivity for commodity iot,” in \textit{Proc. MobiSys ’20}, 2020, pp. 54--66.

\bibitem{blue2}
M. Zhang, S. Chen, J. Zhao, and W. Gong, ``Commodity-level BLE backscatter,'' in \textit{Proc. MobiSys’21}, Jun. 2021, pp. 402--414.




\bibitem{lora}
V. Talla \textit{et al.}, ``LoRa backscatter: Enabling the vision of ubiquitous connectivity,'' in \textit{Proc. ACM Interact. Mobile Wearable Ubiquitous Technol}., vol. 1, no. 3, Sep. 2017, Art. no. 105, pp 1--24.



\bibitem{lora2}
X. Guo, L. Shangguan, Y. He, J. Zhang, H. Jiang, A. A. Siddiqi, and Y. Liu, ‘‘Aloba: Rethinking ON-OFF keying modulation for ambient LoRa backscatter,’’ in \textit{Proc. 18th Conf. Embedded Networked Sensor Syst.} 2020, pp. 192--204.

\bibitem{lora3}
J. Jiang, Z. Xu, F. Dang, and J. Wang, ``Long-range ambient LoRa backscatter with parallel decoding,'' in \textit{Proc. 27th Annu. Int. Conf. Mobile Comput. Netw.}, 2021, pp. 684--696.

\bibitem{lora4}
X. Guo, L. Shangguan, Y. He, N. Jing, J. Zhang, H. Jiang, and Y. Liu, “Saiyan: Design and implementation of a low-power demodulator for lora backscatter systems,'' in \textit{Proc. of USENIX NSDI},  Apr. 2022, pp. 437--451.



}

\bibitem{TC}
A. Bletsas, S. Siachalou, and J. N. Sahalos, ``Anti-collision backscatter sensor networks," \textit{IEEE Trans. Wireless Commun.}, vol. 8, no. 10, pp. 5018--5029, Oct. 2009.


%
%



\bibitem{SIC1}
S. Zhou, W. Xu, K. Wang, C. Pan, M. -S. Alouini, and A. Nallanathan, ``Ergodic rate analysis of cooperative ambient backscatter communication," \textit{IEEE Wireless Commun. Lett.}, vol. 8, no. 6, pp. 1679--1682, Dec. 2019.


\bibitem{SIC3}
B. Gu, H. Xie, and D. Li, ``Act before another is aware: Safeguarding backscatter systems with covert communications,"  \textit{IEEE Wireless Commun. Lett.}, vol. 12, no. 6, pp. 1106--1110, Jun. 2023.

\bibitem{SIC4}
H. Yang, H. Ding, M. Elkashlan, H. Li, and K. Xin, ``A novel symbiotic backscatter-NOMA system," \textit{IEEE Trans. Veh.Tech.}, vol. 72, no. 8, pp. 11006--11011, Aug. 2023.






\bibitem{ESIC0a}
T. Brauner and X. Zhao, ``A novel carrier suppression method for RFID," \textit{IEEE Microw. Wireless Compon. Lett.}, vol. 19, no. 3, pp. 128--130, Mar. 2009.

\bibitem{ESIC0}
D. P. Villame and J. S. Marciano, ``Carrier suppression locked loop mechanism for UHF RFID readers," in \textit{Proc.  IEEE RFID}, 2010, pp. 141-145.


\bibitem{ESIC1}
Q. Tao, Y. Li, C. Zhong, S. Shao, and Z. Zhang, ``A novel interference cancellation scheme for bistatic backscatter communication systems," \textit{IEEE Commun. Lett.}, vol. 25, no. 6, pp. 2014--2018, Jun. 2021.


\bibitem{ESIC2}
R. Luo, H. Yang, C. Meng, and X. Zhang, ``A novel SR-DCSK-based ambient backscatter communication system," \textit{IEEE Trans. Circuits Syst. II, Exp. Briefs}, vol. 69, no. 3, pp. 1707--1711, Mar. 2022.

\bibitem{ESIC2a}
S. Li, H. Zheng, C. Zhang, Y. Song, S. Yang, M. Chen, L. Lu, and M. Li, ``Passive DSSS: Empowering the downlink communication for backscatter systems,'' in \textit{Proc. of USENIX NSDI}, 2022. pp. 913-928.

\bibitem{ESIC3}
W. Guo, H. Zhao, C. Song, S. Shao, and Y. Tang, ``Direct-link interference cancellation design for backscatter communications over ambient DVB signals," \textit{IEEE Trans. Broadcasting}, vol. 68, no. 2, pp. 317--330, Jun. 2022.

\bibitem{OSIC}
D. Li, H. Zhang, and L. Fan, ``Adaptive mode selection for backscatter-assisted communication systems with opportunistic SIC," \textit{IEEE Trans. Veh. Tech.}, vol. 69, no. 2, pp. 2327--2331, Feb. 2020.



\bibitem{OFDM0}
J. D. Rosenthal and M. S. Reynolds, ``Hardware-efficient all-digital architectures for OFDM backscatter modulators," \textit{IEEE Trans. Microw. Theory  Techn.}, vol. 69, no. 1, pp. 803--811, Jan. 2021.

\bibitem{OFDM1}
G. Yang, Y.-C. Liang, R. Zhang, and Y. Pei, ``Modulation in the air: Backscatter communication over ambient OFDM carrier,’’ \textit{IEEE Trans. Commun.}, vol. 66, no. 3, pp. 1219--1233, Mar. 2018.

\bibitem{OFDM1a}
G. Yang, Q. Zhang, and Y. -C. Liang, ``Cooperative ambient backscatter communications for green Internet-of-Things," \textit{IEEE Internet Things J.}, vol. 5, no. 2, pp. 1116--1130, Apr. 2018.


\bibitem{OFDM2}
M. A. ElMossallamy, M. Pan, R. Jäntti, K. G. Seddik, G. Y. Li, and Z. Han, ``Noncoherent backscatter communications over ambient OFDM signals,’’ \textit{IEEE Trans. Commun.}, vol. 67, no. 5, pp. 3597--3611, May 2019.




\bibitem{FS1}
P. Zhang, M. Rostami, P. Hu, and D. Ganesan, ``Enabling practical backscatter communication for on-body sensors,'' in \textit{Proc. ACM SIGCOMM},  Aug. 2016, pp. 370--383.

\bibitem{FS2}
D. Li, ``Capacity of backscatter communication with frequency shift in Rician fading channels," \textit{IEEE Wireless Commun. Lett.}, vol. 8, no. 6, pp. 1639--1643, Dec. 2019.



\bibitem{FS4a}
D. Li and Y.-C. Liang, ``Price-based bandwidth allocation for backscatter communication with bandwidth constraints,'' \textit{IEEE Trans. Wireless Commun.}, vol. 18, no. 11, pp. 5170--5180, Nov. 2019.

\bibitem{FS5}
Y. Ding, R. Lihakanga, R. Correia, G. Goussetis, and N. B. Carvalho, ``Harmonic suppression in frequency shifted backscatter communications," \textit{IEEE Open J. Commun. Society}, vol. 1, pp. 990--999, 2020.

\bibitem{FS6}
M. Rostami, K. Sundaresan, E. Chai, S. Rangarajan, and D. Ganesan, ``Redefining passive in backscattering with commodity devices,'' in \textit{Proc. 26th Annu. Int. Conf. Mobile Comput. Netw.}, 2020, pp. 1--13.




%





\bibitem{PC2}
R. Fara, D.-T. Phan-Huy, A. Ourir, Y. Kokar, J.-C. Prevotet, M. Helard, M. Di Renzo, and J. De Rosny, ``Polarization-based reconfigurable tags for robust ambient backscatter communications,'' \textit{IEEE Open J. Commun. Soc.}, vol. 1, pp. 1140--1152, 2020.

\bibitem{PC3}
J. Lietzén, A. Liljemark, R. Duan, R. Jäntti, and V. Viikari, ``Polarization conversion-based ambient backscatter system," \textit{IEEE Access}, vol. 8, pp. 216793--216804, 2020.


\bibitem{MAR0}
A. N. Parks, A. Liu, S. Gollakota, and J. R. Smith, ``Turbocharging ambient backscatter communication,’’ in \textit{Proc. ACM SIGCOMM}, Aug. 2014, pp. 619--630.

\bibitem{MAR1}
H. Guo, Q. Zhang, S. Xiao, and Y. -C. Liang, ``Exploiting multiple antennas for cognitive ambient backscatter communication,"  \textit{IEEE Internet  Things J.}, vol. 6, no. 1, pp. 765--775, Feb. 2019.

\bibitem{MAR2}
R. Duan, E. Menta, H. Yigitler, R. Jantti, and Z. Han, ``Hybrid beamformer design for high dynamic range ambient backscatter receivers,'' in \textit{Proc. IEEE ICC Workshops},  May 2019, pp. 1--6.

\bibitem{MAR2a}
D. Li, ``Backscatter communication powered by selective relaying,"  \textit{IEEE Trans. Veh. Technol.}, vol. 69, no. 11, pp. 14037--14042, Nov. 2020.

\bibitem{MAR2b}
K. Ahmet, J. Vieira, and E. G. Larsson. ``Direct link interference suppression for bistatic backscatter communication in distributed MIMO,"  \textit{IEEE Trans. Wireless Commun.}, early access. 2023. doi: 10.1109/TWC.2023.3285250.

\bibitem{OFDMA0}
F. Zhu, Y. Feng, Q. Li, X. Tian, and X. Wang, ``DigiScatter: Efficiently prototyping large-scale OFDMA backscatter networks,'' in \textit{Proc. ACM MobiSys}, Jun. 2020, pp. 42--53.

\bibitem{OFDMA1}
P. X. Nguyen \textit{et al.}, ``Backscatter-assisted data offloading in OFDMA-based wireless-powered mobile edge computing for IoT networks," \textit{IEEE Internet  Things J.}, vol. 8, no. 11, pp. 9233--9243, Jun. 2021.

\bibitem{OFDMA2}
Y. Xu, B. Gu, R. Q. Hu, D. Li, and H. Zhang, ``Joint computation offloading and radio resource allocation in MEC-based wireless-powered backscatter communication networks,"  \textit{IEEE Trans. Veh. Tech.}, vol. 70, no. 6, pp. 6200--6205, Jun. 2021.

\bibitem{OFDMA3}
F. Zhu \textit{et al.}, ``Enabling OFDMA in Wi-Fi backscatter,"  \textit{IEEE/ACM Trans. Netw.}, early access. 2023. doi: 10.1109/TNET.2023.3290370.





\bibitem{TDMA4}
Y. Xu, B. Gu, and D. Li, ``Robust energy-efficient optimization for secure wireless-powered backscatter communications with a non-linear EH model,"  \textit{IEEE Commun. Lett.}, vol. 25, no. 10, pp. 3209--3213, Oct. 2021.

\bibitem{TDMA5}
Z. Ling, F. Hu, Y. Zhang, L. Fan, F. Gao, and Z. Han, ``Distributionally robust chance-constrained backscatter communication-assisted computation offloading in WBANs," \textit{IEEE Trans.  Commun.}, vol. 69, no. 5, pp. 3395--3408, May 2021.

\bibitem{TDMA6}
Y. Ye, L. Shi, X. Chu, G. Lu, and S. Sun, ``Mutualistic cooperative ambient backscatter communications under hardware impairments,"  \textit{IEEE Trans. Commun.}, vol. 70, no. 11, pp. 7656--7668, Nov. 2022.

\bibitem{SDMA1}
C. Psomas and I. Krikidis, ``Collision avoidance in wireless powered sensor networks with backscatter communications," in \textit{Proc. IEEE SPAWC}, Dec. 2017, pp. 1--5.

\bibitem{MSMA1}
Y. Igarashi, Y. Sato, Y. Kawakita, J. Mitsugi, and H. Ichikawa, ``A feasibility study on simultaneous data collection from multiple sensor RF tags with multiple subcarriers," in \textit{Proc. IEEE RFID},  May. 2014, pp. 141-146.

\bibitem{MSMA2}
N. Rajoria, H. Kamei, J. Mitsugi, Y. Kawakita, and H. Ichikawa, ``Multi-carrier backscatter communication system for concurrent wireless and batteryless sensing," in Proc. IEEE WiSPNET, Feb. 2018, pp. 1078-1082.


\bibitem{MSMA3}
J. Mitsugi, Y. Kawakita, K. Egawa, and H. Ichikawa, ``Perfectly synchronized streaming from digitally modulated multiple backscatter sensor tags," in \textit{Proc. IEEE RFID-TA}, Dec. 2018, pp. 1-6.






\bibitem{TS1}
D. Li, ``Fairness-based multiuser scheduling for ambient backscatter communication systems," \textit{IEEE Wireless Commun. Lett.}, vol. 9, no. 8, pp. 1150--1154, Aug. 2020.

\bibitem{TS2}
Y. H. Al-Badarneh, M. -S. Alouini, and C. N. Georghiades, ``Performance analysis of monostatic multi-tag backscatter systems with general order tag selection," \textit{IEEE Wireless Commun. Lett.}, vol. 9, no. 8, pp. 1201--1205, Aug. 2020.

\bibitem{TS3}
Y. Liu, Y. Ye, and R. Q. Hu, ``Secrecy outage probability in backscatter communication systems with tag selection," \textit{IEEE Wireless Commun. Lett.}, vol. 10, no. 10, pp. 2190--2194, Oct. 2021.


\bibitem{TS4}
D. Deng, X. Li, S. Dang, and K. Rabie, ``Outage analysis for tag selection in reciprocal backscatter communication systems," \textit{IEEE Wireless Commun. Lett.}, vol. 11, no. 2, pp. 210--214, Feb. 2022.

\bibitem{UC1}
C. Yang, X. Wang, and K. -W. Chin, ``On max–min throughput in backscatter-assisted wirelessly powered IoT," \textit{IEEE Internet of Things J.}, vol. 7, no. 1, pp. 137--147, Jan. 2020.



\bibitem{UC2}
D. Han and H. Minn, ``Coverage probability analysis under clustered ambient backscatter nodes,"  \textit{IEEE Wireless Commun. Lett.}, vol. 8, no. 6, pp. 1713--1717, Dec. 2019.

\bibitem{UC3}
Q. Wang, Y. Zhou, H. -N. Dai, G. Zhang, and W. Zhang, ``Performance on cluster backscatter communication networks with coupled interferences," \textit{IEEE Internet Things J.}, vol. 9, no. 20, pp. 20282--20294, Oct. 2022.




\bibitem{MS1}
A. Almaaitah, H. S. Hassanein, and M. Ibnkahla, ``Tag modulation silencing: Design and application in RFID anti-collision protocols,'' \textit{IEEE Trans. Commun.}, vol. 62, no. 11, pp. 4068–4079, Nov. 2014.





\bibitem{CS1}
J. Wang, H. Hassanieh, D. Katabi, and P. Indyk, ``Efficient and reliable low-power backscatter networks,'' in \textit{Proc. ACM SIGCOMM}, 2012, pp. 61--72.

\bibitem{LF}
P. Hu, P. Zhang, and D. Ganesan, ``Laissez-faire: Fully asymmetric backscatter communication,'' in \textit{Proc. ACM SIGCOMM}, 2015, pp. 255--267. 

\bibitem{BST}
P. Hu, P. Zhang, and D. Ganesan, ``Leveraging interleaved signal edges for concurrent backscatter,'' in \textit{Proc. SIGMOBILE Mobile Comput. Commun. Rev.}, vol. 18, no. 3, Jan. 2015, pp. 26--31.

\bibitem{Bigroup}
J. Ou, M. Li, and Y. Zheng, ``Come and be served: Parallel decoding for cots RFID tags,'' in \textit{Proc. Annu. Int. Conf. Mobile Comput. Netw.}, 2015, pp. 500--511.

\bibitem{Fliptracer}
M. Jin, Y. He, X. Meng, Y. Zheng, D. Fang, and X. Chen, ``Fliptracer: Practical parallel decoding for backscatter communication,'' in \textit{Proc. 23rd Annu. Int. Conf. Mobile Computing and Networking}, 2017, pp. 275--287.

\bibitem{Hubble}
M. Jin, Y. He, X. Meng, D. Fang, and X. Chen, ``Parallel backscatter in the wild: When burstiness and randomness play with you,'' in \textit{Proc.  Annu. Int. Conf. Mobile Comput. Netw.}, Oct. 2018, pp. 471--485.

\bibitem{THSS1}
W. Liu, K. Huang, X. Zhou, and S. Durrani, ``Full-duplex backscatter interference networks based on time-hopping spread spectrum,"  \textit{IEEE Trans. Wireless Commun.}, vol. 16, no. 7, pp. 4361--4377, Jul. 2017.


\bibitem{RA1}
J. Guo, S. Durrani, and X. Zhou, ``Monostatic backscatter system with multi-tag to reader communication,"  \textit{IEEE Trans. Veh.Tech.}, vol. 68, no. 10, pp. 10320--10324, Oct. 2019.

\bibitem{RA2}
X. Cao, Z. Song, B. Yang, M. A. Elmossallamy, L. Qian, and Z. Han, ``A distributed ambient backscatter MAC protocol for Internet-of-Things networks," \textit{IEEE Internet Things J.}, vol. 7, no. 2, pp. 1488--1501, Feb. 2020.

\bibitem{RA3}
B. Gu, H. Xie, D. Li, Y. Liu, and Y. Xu, "Exploiting hybrid active and passive multiple access via slotted ALOHA-driven backscatter communications," \textit{arXiv preprint arXiv}:2209.12526, 2022.

\bibitem{CCH}
H. Ding, M. Elkashlan, H. Yang, H. Li, and K. Xin, ``Symbiotic backscatter system over cascaded fading channels," in \textit{Proc. IEEE VTC2022-Fall}, 2022, pp. 1--7.


\bibitem{APAS1}
D. Li, ``Hybrid active and passive antenna selection for backscatter-assisted MISO systems,"  \textit{IEEE Trans. Commun.}, vol. 68, no. 11, pp. 7258--7269, Nov. 2020.

\bibitem{APAS1a}
A. E. Mostafa and V. W. S. Wong, ``Transmit or backscatter: Communication mode selection for narrowband IoT systems," \textit{IEEE Trans. Veh. Tech.}, vol. 71, no. 5, pp. 5477--5491, May 2022.

\bibitem{APAS2}
B. Gu, H. Xie, and D. Li, "Computation-Efficient Backscatter-Blessed MEC with User Reciprocity," \textit{IEEE Trans. Veh. Tech.}, early access. Dec. 2023. doi: 10.1109/TVT.2023.3348200.



\bibitem{WPBCN2}
D. Li, W. Peng, and Y. -C. Liang, ``Hybrid ambient backscatter communication systems with harvest-then-transmit protocols,"  \textit{IEEE Access}, vol. 6, pp. 45288--45298, 2018.



\bibitem{WPBCN4a}
D. Li, ``Two birds with one stone: Exploiting decode-and-forward relaying for opportunistic ambient backscattering," \textit{IEEE Trans. Commun.}, vol. 68, no. 3, pp. 1405--1416, Mar. 2020.

\bibitem{WPBCN5}
B. Gu, Y. Xu, C. Huang, and R. Q. Hu, ``Energy-efficient resource allocation for OFDMA-based wireless-powered backscatter communications,” in \textit{Proc. IEEE ICC},  2021, pp. 1--6.

\bibitem{WPBCN6}
L. Shi, R. Q. Hu, J. Gunther, Y. Ye, and H. Zhang, ``Energy efficiency for RF-powered backscatter networks using HTT protocol," \textit{IEEE Trans. Veh. Tech.}, vol. 69, no. 11, pp. 13932--13936, Nov. 2020.

\bibitem{WPBCN7}
Y. Zhuang, X. Li, H. Ji, H. Zhang, and V. C. M. Leung, ``Optimal resource allocation for RF-powered underlay cognitive radio networks with ambient backscatter communication," \textit{IEEE Trans.  Veh. Tech.}, vol. 69, no. 12, pp. 15216--15228, Dec. 2020.




\bibitem{MAS0}
N. Deepan and B. Rebekka, ``Backscatter-assisted wireless powered communication networks with multiple antennas," in \textit{Proc. WiSPNET}, 2020, pp. 135--138.

\bibitem{MAS1}
R. Long, G. Yang, Y. Pei, and R. Zhang, ``Transmit beamforming for cooperative ambient backscatter communication systems," in \textit{Proc. GLOBECOM}, 2017, pp. 1--6.

\bibitem{MAS1a}
W. Zang, F. Sun, Y. Cai and, Y. Li, ``Transmit beamforming for ambient backscatter communication enabled wireless body area network in multiuser MISO system,"  \textit{IEEE Trans. Cognitive Commun. Netw.}, vol. 8, no. 4, pp. 1839--1847, Dec. 2022.

\bibitem{MAS2}
C. Zhou \textit{et al.}, ``Energy-efficient maximization for RIS-aided MISO symbiotic radio systems,"  \textit{IEEE Trans. Veh. Tech.}, vol. 72, no. 10, pp. 13689--13694, Oct. 2023.


\bibitem{MAT1}
C. He and Z. J. Wang, ``Closed-form BER analysis of non-coherent FSK in MISO double Rayleigh fading/RFID channel," \textit{IEEE Commun. Lett.}, vol. 15, no. 8, pp. 848--850, Aug. 2011.




\bibitem{MAT1a}
J. D. Griffin and G. D. Durgin, ``Multipath fading measurements at 5.8 GHz for backscatter tags with multiple antennas," \textit{IEEE Trans. Antennas  Propag.}, vol. 58, no. 11, pp. 3693--3700, Nov. 2010.

\bibitem{MAT2}
C.-H. Kang, W.-S. Lee, Y.-H. You, and H.-K. Song, ``Signal detection scheme in ambient backscatter system with multiple antennas,'' \textit{IEEE Access}, vol. 5, pp. 14543--14547, 2017.

\bibitem{MAT3}
J. Liu, J. Yu, D. Niyato, R. Zhang, X. Gao, and J. An, ``Covert ambient backscatter communications with multi-antenna tag," \textit{IEEE Trans. Wireless Commun.},vol. 22, no. 9, pp. 6199--6212, Sept. 2023.


\bibitem{MAR4}
J. Zhao, W. Gong, and J. Liu, ``Spatial stream backscatter using commodity WiFi,'' in \textit{Proc. of ACM MobiSys}, 2018, pp. 191-203.

\bibitem{MAR5}
D. Li and Y. -C. Liang, ``Adaptive ambient backscatter communication systems with MRC," \textit{IEEE Trans. Veh. Tech.}, vol. 67, no. 12, pp. 12352--12357, Dec. 2018.




\bibitem{MAR9}
D. Lin, K. Cumanan, and Z. Ding, ``Beamforming design for BackCom assisted NOMA systems," \textit{IEEE Wireless Commun. Lett.}, vol. 12, no. 9, pp. 1494--1498, Sept. 2023.




\bibitem{MIMO2}
C. He, S. Chen, H. Luan, X. Chen, and Z. J. Wang, ``Monostatic MIMO backscatter communications," \textit{IEEE J. Selected Areas Commun.}, vol. 38, no. 8, pp. 1896--1909, Aug. 2020.

\bibitem{MIMO3}
C. He, H. Luan, X. Li, C. Ma, L. Han, and Z. Jane Wang, "A simple, high-performance space–time code for MIMO backscatter communications,"  \textit{IEEE Internet Things J.}, vol. 7, no. 4, pp. 3586--3591, Apr. 2020.

\bibitem{MIMO4}
H. Luan, X. Xie, L. Han, C. He, and Z. J. Wang, ``A better than alamouti OSTBC for MIMO backscatter communications," \textit{IEEE Trans. Wireless Commun.}, vol. 21, no. 2, pp. 1117--1131, Feb. 2022.

\bibitem{MIMO5}
X. Wang, H. Yiğitler, and R. Jäntti, ``Gaining from multiple ambient sources: signaling matrix for multi-antenna backscatter devices," \textit{IEEE Wireless Commun. Lett.}, vol. 12, no. 3, pp. 491--495, Mar. 2023.

\bibitem{MSEH0}
U. S. Toro, S. Khan, U. Aslam, B. M. ElHalawany, and K. Wu, ``Enhancing the sustainability of acoustic backscatter communication with multi-source energy harvesting," \textit{IEEE Commun. Mag.}, vol. 61, no. 10, pp. 116--120, Oct. 2023.

\bibitem{MSEH1}
A. Khaleghi, A. Hasanvand, and I. Balasingham, ``Wireless backscatter communication using multiple transmitter scheme,'' in \textit{Proc. IEEE EuCAP}, 2016, pp. 1--4.


\bibitem{MSEH2}
C. Yang, J. Gummeson, and A. Sample, ``Riding the airways: Ultra-wideband ambient backscatter via commercial broadcast systems," in \textit{Proc. IEEE INFOCOM}, 2017, pp. 1--9.


\bibitem{MSEH2a}
A. Galisteo, A. Varshney, and D. Giustiniano, ``Two to tango: Hybrid light and backscatter networks for next billion devices,'' in \textit{Proc. 18th International Conference on Mobile Systems, Applications, and Services}, Jun. 2020. pp. 80--93.


\bibitem{MSEH3}
S. H. Kim and D. I. Kim, ``Hybrid backscatter communication for wireless-powered heterogeneous networks,'' \textit{IEEE Trans. Wireless Commun.}, vol. 16, no. 10, pp. 6557--6570, 2017.

\bibitem{MSEH4}
S. H. Kim, S. Y. Park, K. W. Choi, T. -J. Lee, and D. I. Kim, ``Backscatter-aided cooperative transmission in wireless-powered heterogeneous networks," \textit{IEEE Trans. Wireless Commun.}, vol. 19, no. 11, pp. 7309--7323, Nov. 2020.

\bibitem{MSEH5}
M. Katanbaf, A. Saffari, and J. R. Smith, ``MultiScatter: Multistatic backscatter networking for battery-free sensors,'' in \textit{Proc. ACM Conf. Embedded Netw. Sens. Syst.}, 2021, pp. 69--83.

\bibitem{MSEH6}
L. Yuan, Q. Wang, J. Zhao, and W. Gong, ``Multiprotocol backscatter with commodity radios for personal IoT sensors," \textit{IEEE/ACM Trans. Netw.}, vol. 31, no. 3, pp. 1132--1144, Jun. 2023.

\bibitem{BER1}
R. Zhang, X. Kang, and Y. -C. Liang, ``Minimum throughput maximization for peer-assisted NOMA-plus-TDMA symbiotic radio networks," \textit{IEEE Wireless Commun. Lett.}, vol. 10, no. 9, pp. 1847--1851, Sept. 2021.

\bibitem{BER2}
B. Gu, D. Li, Y. Xu, C. Li, and S. Sun, ``Many a little makes a mickle: Probing backscattering energy recycling for backscatter communications," \textit{IEEE Trans. Veh. Tech.}, vol. 72, no. 1, pp. 1343--1348, Jan. 2023.

\bibitem{UAV1}
J. Hu, X. Cai, and K. Yang, ``Joint trajectory and scheduling design for UAV aided secure backscatter communications," \textit{IEEE Wireless Commun. Lett.}, vol. 9, no. 12, pp. 2168--2172, Dec. 2020.

\bibitem{UAV2}
 D. Nagarajan, D. N. K. Jayakody, and R. Balakrishnan, ``Performance analysis of UAV-enabled backscatter wireless communication network,'' in \textit{Proc. DroneCom}.  2020, pp. 20--24.
 
 \bibitem{UAV3}
 G. Yang, R. Dai, and Y. -C. Liang, ``Energy-efficient UAV backscatter communication with joint trajectory design and resource optimization," \textit{IEEE Trans. Wireless Commun.}, vol. 20, no. 2, pp. 926--941, Feb. 2021.
  
 \bibitem{UAV4}
 Y. Zhang, Z. Mou, F. Gao, L. Xing, J. Jiang, and Z. Han, ``Hierarchical deep reinforcement learning for backscattering data collection with multiple UAVs," \textit{IEEE Internet Things J.}, vol. 8, no. 5, pp. 3786--3800, Mar. 2021.
 
 \bibitem{UAV5}
 H. Yang, Y. Ye, X. Chu, and S. Sun, ``Energy efficiency maximization for UAV-enabled hybrid backscatter-harvest-then-transmit communications," \textit{IEEE Trans. Wireless Commun.}, vol. 21, no. 5, pp. 2876--2891, May 2022.
 
 \bibitem{UAV6}
 Y. Du, Z. Chen, J. Hao, and Y. Guo, ``Joint optimization of trajectory and communication in multi-UAV assisted backscatter communication networks," \textit{IEEE Access}, vol. 10, pp. 40861--40871, 2022.
 
 \bibitem{UAV7}
 L. Bai, Q. Chen, T. Bai, and J. Wang, ``UAV-enabled secure multiuser backscatter communications with planar array," \textit{IEEE J. Selected Areas Commun.}, vol. 40, no. 10, pp. 2946--2961, Oct. 2022.

\bibitem{ARelay1}
B. Lyu, Z. Yang, H. Guo, F. Tian, and G. Gui, ``Relay cooperation enhanced backscatter communication for Internet-of-Things," \textit{IEEE Internet Things J.}, vol. 6, no. 2, pp. 2860--2871, Apr. 2019.

\bibitem{ARelay2}
A. E. Mostafa and V. W. S. Wong, ``Connection density enhancement of backscatter communication systems with relaying," in \textit{Proc. IEEE GLOBECOM},  2020, pp. 1--6.

\bibitem{ARelay3}
W. -J. Wang, K. Xu, Y. Yan, and L. Chen, ``Relay selection-based cooperative backscatter transmission with energy harvesting: Throughput maximization,"  \textit{IEEE Wireless Commun. Lett.}, vol. 11, no. 7, pp. 1533--1537, Jul. 2022.

\bibitem{PRelay1}
S. Gong, X. Huang, J. Xu, W. Liu, P. Wang, and D. Niyato, ``Backscatter relay communications powered by wireless energy beamforming," \textit{IEEE Trans. Commun.}, vol. 66, no. 7, pp. 3187--3200, July 2018.

\bibitem{PRelay2}
S. Gong, L. Gao, J. Xu, Y. Guo, D. T. Hoang, and D. Niyato, ``Exploiting backscatter-aided relay communications with hybrid access model in device-to-device networks," \textit{IEEE Trans. Cognitive Commun. Netw.}, vol. 5, no. 4, pp. 835--848, Dec. 2019.

\bibitem{PRelay3}
J. Xu, J. Li, S. Gong, K. Zhu, and D. Niyato, ``Passive relaying game for wireless powered Internet of Things in backscatter-aided hybrid radio networks," \textit{IEEE Internet Things J.}, vol. 6, no. 5, pp. 8933--8944, Oct. 2019.

\bibitem{PRelay4}
X. Gao, D. Niyato, K. Yang, and J. An, ``Cooperative scheme for backscatter-aided passive relay communications in wireless-powered D2D networks," \textit{IEEE Internet Things J.}, vol. 9, no. 1, pp. 152--164, 1 Jan.1, 2022.

\bibitem{HRelay1}
C. Zheng, W. Cheng, and H. Zhang, ``Optimal resource allocation for two-user and single-DF-relay network with ambient backscatter," \textit{IEEE Access}, vol. 7, pp. 91375--91389, 2019.

\bibitem{HRelay2}
Y. He, X. Wu, Z. He, and M. Guizani, ``Energy efficiency maximization of backscatter-assisted wireless-powered MEC with user cooperation," \textit{IEEE Trans. Mobile Comp.}, eraly access. 2023. doi: 10.1109/TMC.2023.3243161.


\bibitem{ARelay4}
D. Li, ``Backscatter communication via harvest-then-transmit relaying,"  \textit{IEEE Trans. Veh. Tech.}, vol. 69, no. 6, pp. 6843--6847, Jun. 2020.

\bibitem{ARelay5}
B. Lyu, D. T. Hoang, and Z. Yang, ``Backscatter then forward: A relaying scheme for batteryless IoT networks," \textit{IEEE Wireless Commun. Lett.}, vol. 9, no. 4, pp. 562--566, Apr. 2020.

\bibitem{ARelay6}
Y. Zhuang, X. Li, H. Ji, and H. Zhang, ``Exploiting hybrid SWIPT in ambient backscatter communication-enabled relay networks: Optimize power allocation and time scheduling," \textit{IEEE Internet  Things J.}, vol. 9, no. 24, pp. 24655--24668, Dec. 2022.

\bibitem{PRelay5}
J. Zhao, W. Gong, and J. Liu, ``X-Tandem: Towards MultiHop Backscatter Communication with Commodity WiFi,'' in \textit{Proc. 24th Annual ACM Int’l. Conf. Mobile Computing and
	Networking}, 2018, pp. 497--511.

\bibitem{PRelay6}
A. Y. Majid, M. Jansen, G. O. Delgado, K. S. Yildirim, and P. Pawełłzak, ``Multi-hop backscatter tag-to-tag networks," in \textit{Proc. IEEE INFOCOM}, 2019, pp. 721--729.

\bibitem{PRelay7}
J. Zhao, W. Gong, and J. Liu, ``Towards scalable backscatter sensor mesh with decodable relay and distributed excitation,'' in \textit{Proc. 24th Annual International Conference on Mobile Computing and Networking}, 2018, pp. 77--79.

\bibitem{PRelay8}
D. Piumwardane, C. Rohner, and T. Voigt, ``Reliable flooding in dense backscatter-based tag-to-tag networks," in \textit{Proc. IEEE RFID},  2021, pp. 1--8.


\bibitem{URelay1}
S. Yang, Y. Deng, X. Tang, Y. Ding, and J. Zhou, "Energy efficiency optimization for UAV-assisted backscatter communications,"  \textit{IEEE Commun. Lett.}, vol. 23, no. 11, pp. 2041--2045, Nov. 2019.

\bibitem{URelay2}
M. Z. Hassan, M. J. Hossain, J. Cheng, and V. C. Leung, ``Statisticalqos guarantee for IoT network driven by laser-powered UAV relay and RF backscatter communications,'' \textit{IEEE Trans.  Green Commun. Netw.}, vol. 5, no. 1, pp. 406--425, Mar. 2020.

\bibitem{URelay3}
M. Hua, L. Yang, C. Li, Q. Wu, and A. L. Swindlehurst, ``Throughput maximization for UAV-aided backscatter communication networks," \textit{IEEE Trans. Commun.}, vol. 68, no. 2, pp. 1254--1270, Feb. 2020.

\bibitem{URelay4}
M. Z. Hassan, M. J. Hossain, J. Cheng, and V. C. M. Leung, ``Statistical-QoS guarantee for IoT network driven by laser-powered UAV relay and RF backscatter communications," \textit{IEEE Trans. Green Commun. Netw.}, vol. 5, no. 1, pp. 406--425, Mar. 2021.

\bibitem{URelay5}
R. Han, L. Bai, Y. Wen, J. Liu, J. Choi, and W. Zhang, ``UAV-aided backscatter communications: Performance analysis and trajectory optimization," \textit{IEEE J. Selected Areas Commun.}, vol. 39, no. 10, pp. 3129--3143, Oct. 2021.

\bibitem{RIS1}
D. Li, ``Bound analysis of number configuration for reflecting elements in IRS-assisted D2D communications,'' \textit{IEEE Wireless Commun. Lett.}, vol. 11, no. 10, pp. 2220--2224, Oct. 2022.


\bibitem{RIS2}
D. Li, ``How many reflecting elements are needed for energy- and spectral-efficient intelligent reflecting surface-assisted communication,"  \textit{IEEE Trans. Commun.}, vol. 70, no. 2, pp. 1320--1331, Feb. 2022.


\bibitem{RIS3}
Y. Xu, B. Gu, Z. Gao, D. Li, Q. Wu,A and C. Yuen, ``Applying RIS in multi-user SWIPT-WPCN systems: A robust and environmentally-friendly design,"  \textit{IEEE Trans. Cogn. Commun. Netw.}, early access. Oct. 2023. doi: 10.1109/TCCN.2023.3324636.


\bibitem{RIS4}
D. Li, ``Fairness-aware multiuser scheduling for finite-resolution intelligent reflecting surface-assisted communication,'' \textit{IEEE Commun. Lett.}, vol. 25, no. 7, pp. 2395--2397, Jul. 2021.


\bibitem{RIS5}
H. Xie, B. Gu, D. Li, Z. Lin, and Y. Xu, ``Gain without pain: Recycling reflected energy from wireless powered RIS-aided communications," \textit{IEEE Internet Things J.}, vol. 10, no. 15, pp. 13264--13280, Aug. 2023.


\bibitem{RABC1}
Y. -C. Liang, Q. Zhang, J. Wang, R. Long, H. Zhou, and G. Yang, ``Backscatter communication assisted by reconfigurable intelligent surfaces," \textit{Proceedings of the IEEE}, vol. 110, no. 9, pp. 1339--1357, Sept. 2022.



\bibitem{RC2}
S. Y. Park and D. In Kim, ``Intelligent reflecting surface-aided phase-shift backscatter communication," \textit{IEEE IMCOM,}  2020, pp. 1--5.




\bibitem{RC6}
S. Li, L. Bariah, S. Muhaidat, A. Wang, and J. Liang, ``Outage analysis of NOMA-enabled backscatter communications with intelligent reflecting surfaces,"  \textit{IEEE Internet  Things J.}, vol. 9, no. 16, pp. 15390--15400, Aug. 2022.

\bibitem{RC7}
S. Xu, J. Liu, N. Kato, and Y. Du, ``Intelligent reflecting surface backscatter enabled multi-tier computing for 6G Internet of Things," \textit{IEEE J. Selected Areas  Commun.}, vol. 41, no. 2, pp. 320--333, Feb. 2023.

\bibitem{RC8}
H. Yang, H. Ding, K. Cao, M. Elkashlan, H. Li, and K. Xin, ``A RIS-segmented symbiotic ambient backscatter communication system," \textit{IEEE Trans. Veh. Tech.}, early access. 2023. doi: 10.1109/TVT.2023.3306037.


\bibitem{RAB0a}
J. Zuo, Y. Liu, L. Yang, L. Song, and Y. -C. Liang, ``Reconfigurable intelligent surface enhanced NOMA assisted backscatter communication system," \textit{IEEE Trans. Veh. Tech.}, vol. 70, no. 7, pp. 7261--7266, Jul. 2021.



\bibitem{RAB2}
S. Mao, N. Zhang, J. Hu, and K. Yang, ``Intelligent reflecting surface-assisted over-the-air computation for backscatter sensor networks," \textit{IEEE Trans. Veh. Tech.}, vol. 72, no. 5, pp. 6839--6843, May 2023.

\bibitem{RAB3}
X. Jia, X. Zhou, D. Niyato, and J. Zhao, ``Intelligent reflecting surface-assisted bistatic backscatter networks: Joint beamforming and reflection design," \textit{IEEE Trans, Green Commun. Netw.}, vol. 6, no. 2, pp. 799--814, Jun. 2022.

\bibitem{RAB4}
D. L. Galappaththige, F. Rezaei, C. Tellambura, and S. Herath, ``RIS-empowered ambient backscatter communication systems," \textit{IEEE Wireless Commun. Lett.}, vol. 12, no. 1, pp. 173--177, Jan. 2023.

\bibitem{RAB5}
Q. Liu, M. Fu, W. Li, J. Xie, and M. Kadoch, ``RIS-assisted ambient backscatter communication for SAGIN IoT," \textit{IEEE Internet  Things J.}, vol. 10, no. 11, pp. 9375--9384, Jun. 2023.



\bibitem{RAB6}
Z. Yang, L. Feng, F. Zhou, X. Qiu, and W. Li, ``Analytical performance analysis of intelligent reflecting surface aided ambient backscatter communication network," \textit{IEEE Wireless Commun. Lett.}, vol. 10, no. 12, pp. 2732--2736, Dec. 2021.

\bibitem{RAB7}
Y. Chen, ``Performance of ambient backscatter systems using reconfigurable intelligent surface,"  \textit{IEEE Commun. Lett.}, vol. 25, no. 8, pp. 2536--2539, Aug. 2021.

\bibitem{RAB8}
A. Bhowal, S. Aïssa, and R. S. Kshetrimayum, ``RIS-assisted advanced spatial modulation techniques for ambient backscattering communications," \textit{IEEE Trans. Green Commun. Netw.}, vol. 5, no. 4, pp. 1684--1696, Dec. 2021.




\bibitem{URIS1}
S. Solanki, S. Gautam, S. K. Sharma, and S. Chatzinotas, ``Ambient backscatter assisted co-existence in aerial-IRS wireless networks," \textit{IEEE Open J. Commun. Society}, vol. 3, pp. 608--621, 2022.



\bibitem{OSB}
H. Yang, H. Ding, and M. Elkashlan, ``Opportunistic symbiotic backscatter communication systems," \textit{IEEE Commun. Lett.}, vol. 27, no. 1, pp. 100--104, Jan. 2023.

\bibitem{DL}
T. Oyedare, V. K. Shah, D. J. Jakubisin, and J. H. Reed, ``Interference suppression using deep learning: Current approaches and open challenges," \textit{IEEE Access}, vol. 10, pp. 66238--66266, 2022.




\bibitem{YuLi}
X. Yu and D. Li, ``Attention mechanism aided signal detection in backscatter communications with insufficient training data," \textit{IEEE Trans. Veh. Tech.}, early access. Dec. 2023. doi: 10.1109/TVT.2023.3346198.

\bibitem{CI1}
B. Gu, D. Li, Y. Liu, and Y. Xu, ``Exploiting constructive interference for backscatter communication systems," \textit{IEEE Trans. Commun.},  vol. 71, no. 7, pp. 4344--4359, Jul. 2023.


\bibitem{NOMA}
S. M. R. Islam, N. Avazov, O. A. Dobre, and K. S. Kwak, ``Power-domain non-orthogonal multiple access (NOMA) in 5G Systems: Potentials and challenges," \textit{ IEEE Commun. Surveys  Tut.}, vol. 19, no. 2, pp. 721--742, 2nd Quart. 2017.

%
%
\bibitem{NOMA3}
F. D. Ardakani, R. Huang, and V. W. S. Wong, ``Joint device pairing, reflection coefficients, and power control for NOMA backscatter systems," \textit{IEEE Trans. Veh. Technol.}, vol. 71, no. 4, pp. 4396--4411, Apr. 2022.

\bibitem{Quantum}
R. Jantti, R. Duan, J. Lietzen, H. Khalifa, and L. Hanzo, ``Quantum-enhanced microwave backscattering communications," \textit{ IEEE Commun. Mag.}, vol. 58, no. 1, pp. 80--85, Jan. 2020.

\bibitem{Quantum1}
C. Wang and A. Rahman, ``Quantum-enabled 6G wireless networks: Opportunities and challenges,"  \textit{IEEE Wireless Commun.}, vol. 29, no. 1, pp. 58--69, Feb. 2022.

\bibitem{DL1}
U. S. Toro, B. M. ElHalawany, A. B. Wong, L. Wang, and K. Wu, ``Machine-learning-assisted signal detection in ambient backscatter communication networks,"  \textit{IEEE Netw.}, vol. 35, no. 6, pp. 120--125, Nov. 2021.

\bibitem{Blockchain}
A. A. Monrat, O. Schelén, and K. Andersson, ``A survey of blockchain from the perspectives of applications, challenges, and opportunities," \textit{IEEE Access}, vol. 7, pp. 117134--117151, 2019.
%

%
%
%
%
%








\bibitem{CI2}
C. Pérez-Penichet, F. Hermans, and T. Voigt, ``On limits of constructive interference in backscatter systems," in \textit{Proc. GIoTS}, pp. 1-5, Aug. 2017.

\bibitem{CI3}
G. Zheng, M. Wen, Y. Chen, S. Zhao, and H. Du, ``Interference exploitation for ambient backscatter communication networks via symbol level precoding,"  \textit{IEEE Wireless Commun. Lett.}, vol. 11, no. 6, pp. 1166--1170, Jun. 2022.

\bibitem{Tunnel}
A. Varshney, A. Soleiman, and T. Voigt, ``TunnelScatter: Low power communication for sensor tags using tunnel diodes,'' in \textit{Proc. MobiCom}. 2019, pp. 1–17.




\bibitem{MA1}
L. Zhu, W. Ma, and R. Zhang, ``Movable antennas for wireless communication: Opportunities and challenges,`` \textit{arXiv prepint} arXiv:2306.02331, 2023.


\bibitem{MMwave1}
H. Lu, M. Mazaheri, R. Rezvani, and O. Abari, ``A millimeter wave backscatter network for Two-Way communication and localization,'' in \textit{Proc. ACM SIGCOMM'23}, Sept. 2023, pp. 49-61.

\bibitem{ARIS}
Z. Zhang \textit{et al.}, ``Active RIS vs. passive RIS: Which will prevail in 6G?" \textit{IEEE Trans. Commun.}, vol. 71, no. 3, pp. 1707--1725, Mar. 2023.

%

%

\end{thebibliography}
\end{document}